\documentclass[preprint]{aastex}
\usepackage{graphicx} 
\usepackage{figcaps}
\usepackage{balance}

\usepackage{times} 
\usepackage{amsmath}

\def\bhat{\bf\skew{-5}\hat b} % Boldface b with a hat.

\figcapsoff

\shortauthors{Winebarger et al.}
\shorttitle{IDENTIFYING OBSERVABLES THAT CAN DIFFERENTIATE BETWEEN IMPULSIVE AND FOOTPOINT HEATING}

\begin{document}

%%------------------------------------------------------------------------
%%
%%						TITLE
%%
%%------------------------------------------------------------------------

\title{IDENTIFYING OBSERVABLES THAT CAN DIFFERENTIATE BETWEEN IMPULSIVE AND FOOTPOINT HEATING: TIME LAGS AND INTENSITY RATIOS}
\author{Amy R.\ Winebarger,}
\affil{NASA Marshall Space Flight Center, ZP 13, Huntsville, AL 35812}
\email{amy.r.winebarger@nasa.gov}
\and
\author{Roberto Lionello, Cooper Downs, Zoran Miki\'c, Jon Linker}
\affil{Predictive Science, Inc.,  9990 Mesa Rim Rd., Ste. 170, San Diego, CA
92121-2910}
\email{\{lionel, cdowns, mikicz, linkerj\}@predsci.com}

%%------------------------------------------------------------------------
%%
%%					ABSTRACT
%%
%%------------------------------------------------------------------------

\begin{abstract}
Observations of coronal loops have identified several common loop characteristics, including that loops appear to cool and have higher than expected densities.  Two potential heating scenarios have been suggested to explain these observations.  One scenario is that the loops are heated by a series of small-scale impulsive heating events, or nanoflares.   Another hypothesis is that the heating is quasi-steady and highly-stratified, i.e., ``footpoint heating'.   The goal of this paper is to identify observables that can be used to differentiate between these two heating scenarios.  For footpoint heating, we vary the heating magnitude and stratification, for impulsive heating, we vary the heating magnitude.  We use one-dimensional hydrodynamic codes to calculate the resulting temperature and density evolution and expected lightcurves in four channels of AIA and one channel of XRT.    We consider two principal diagnostics: the time lag between the appearance of the loop in two different channels, and the ratio of the peak intensities of the loop in the two channels.    We find that 1) footpoint heating can predict longer time lags than impulsive heating, 2) footpoint heating can predict zero or negative time lags, 3) the intensity ratio expected from impulsive heating is confined to a narrow range, while footpoint heating predicts a wider range of intensity ratios, and 4) the range of temperatures expected in impulsive heating is broader than the range of temperatures expected in footpoint heating.  This preliminary study identifies observables that may be useful in discriminating between heating models in future work.
\end{abstract}

%%------------------------------------------------------------------------
%%
%%					Introduction
%%
%%------------------------------------------------------------------------

\section{INTRODUCTION} 

Solar coronal loops, first observed in soft X-ray wavelengths, were originally found to be long lived and to have densities and temperatures consistent with steady, uniform heating  (e.g., \citealt{rosner1978, porter1995}). This picture, however, became much more complicated with new observations made in the narrow passbands of extreme-ultraviolet (EUV) imagers. The properties of the EUV loops were fundamentally different than those of the X-ray loops; their temperatures and densities did not agree with predictions of steady heating models (e.g. \citealt{lenz1999,aschwanden2000a,winebarger2003}).  Instead, the loops were observed to ``cool,'' meaning appear in higher-temperature channels before appearing in lower-temperature channels, though they still persisted longer than expected in a single channel based on the calculated cooling time (e.g., \citealt{winebarger2003a,winebarger2005,ugarte2006,mulu2011,viall2012}). Significant research of coronal loops has taken place over the last three decades (see review by \citealt{reale2014}).  

Almost immediately it was suggested that these loops were the result of a short bursts of heating occurring on sub-resolution strands 
\citep{warren2002b,warren2003}; such a heating scenario had previously been investigated \citep{cargill1995,cargill1997,cargill2004} as the natural outcome of the small-scale reconnection events suggested by \cite{parker1972,parker1983b}.  This heating scenario is now commonly called a ``short nanoflare storm'' \citep{klimchuk2006}.  In the original short nanoflare storm model, an observed loop is a bundle of strands.  Each strand is heated impulsively and independently and then allowed to evolve with no additional heating events.  Because the heating window is relatively narrow, the loop will appear to evolve as the sub-resolution strands evolve.  In this original heating scenario,  the strands were heated at low frequency, meaning that time between heating events was at least as long as the cooling time of the plasma (e.g., \citealt{warren2002b}).  More recently, several studies have considered the impact of the frequency of heating events in individual strands \citep{bradshaw2012,bradshaw2015,bradshaw2016,cargill2014,cargill2015}.

Qualitatively, such low-frequency impulsive heating explains the characteristics of the observed loops well.  However, quantitative comparisons between the observations and models are lacking.  In one such comparison, \cite{winebarger2003a} measured the light curve of a short isolated loop in two TRACE EUV channels (195\,\AA, sensitive to 1.5\,MK plasma, and 171\,\AA, sensitive to 1\,MK plasma).   \cite{warren2003} demonstrated that the light curve of this loop was well matched by the short nanoflare storm model (see Figure~9 of \citealt{warren2003}) with one exception.  The modeled 171\,\AA\ channel intensities were too bright by roughly 50\%, implying that the rate at which the density decreased in the model was too slow.  In another attempt to model the loop evolution, \cite{ugarte2006} followed a single loop from X-ray to EUV  images.  They attempted to model the evolution with many different assumptions of loop length, radiative loss functions, and heating functions.  Regardless of the assumptions, they found they could not match the evolution of the loop in both the X-ray and EUV channels; the observed loops cooled too slowly.  \cite{warren2010} found the same discrepancy when modeling a different loop observed with many instruments.  Finally, a recent study of the evolution of loops in AIA found light curves that suggested that the plasma is cooling to roughly 1\,MK, then being re-heated to higher temperatures \citep{kamio2011}.

Subsequently, quasi-steady, highly-stratified heating, similar to what would be expected for dissipation of  Alfv{\'e}n waves \citep{asgaritarghi2012},
 was investigated to determine if it could explain the properties of EUV loops \citep{klimchuk2010,mikic2013,lionello2013}.  Such heating profiles may not produce steady state solutions; instead, it is found that the heating can drive thermal non-equilibrium solutions to the hydrodynamic equations \citep{1982A&A...108L...1K,1983A&A...123..216M,1991ApJ...378..372A,1999ApJ...512..985A,2001ApJ...553L..85K,2003ApJ...593.1187K,2003A&A...411..605M,2004A&A...424..289M,2006ApJ...637..531K}.  Depending on the heating parameters and geometry of the loop, cold, dense condensations can be periodically formed in the corona, subsequently sliding down the field lines into the photosphere; such a phenomenon has been observed and is called coronal rain (e.g., \citealt{schrijver2001,antolin2010,antolin2012,antolin2015}).  \cite{klimchuk2010} argued against the possibility that highly-stratified heating could explain EUV loop observations on the basis of 1D plasma simulations, stating that it fails to reproduce the observed properties mentioned above.  However, \cite{2008ApJ...679L.161M,mok2016} and \cite{mikic2013} presented 3D and 1D plasma simulations showing that quasi-steady, footpoint heating can lead to the formation of realistic looking loops. \cite{lionello2013} showed that the properties of such simulated 3D loops match the observed properties of EUV loops.  
 Additionally, \cite{peter2012} found that the light curves shown in  \cite{kamio2011} can be explained by highly-stratified, footpoint heating.   \cite{froment2018} completed an extensive parameter space study on three loop geometries and determined that only specific combinations of geometry and heating profiles can drive thermal non-equilibrium;t in those cases, loops appear and disappear in ways that are similar to observations.  
 In cases where highly-stratified heating drives thermal non-equilibrium, the loops would experience repeated thermal cycles if the heating and magnetic field remain unchanged for many hours; such long-period ``pulsations" have been observed \citep{auchere2014,auchere2016,froment2015,froment2017} in isolated regions within solar active regions.

Both low-frequency impulsive and steady footpoint heating scenarios (and possibly many others) can explain some of the key observed properties of coronal loops.  In this paper, we investigate which observables may be used to discriminate between these two heating scenarios.  We use a single loop geometry, taken from \cite{mikic2013}, and vary the magnitude and stratification of the footpoint heating functions.  We also generate analogous low-frequency impulsive heating solutions with similar heating magnitudes.  We calculate the time lags and the relative intensities predicted for several EUV and one X-ray channel combinations.  From these simulations, we find that 1) footpoint heating can predict longer time lags than impulsive heating in some channel pairs, 2) footpoint heating can predict zero or negative time lags in some channel pairs, 3) the intensity ratio expected from impulsive heating is confined to a narrow range, while footpoint heating predicts a wider range of intensity ratios, and 4) the range of temperatures expected in impulsive heating is broader than the range of temperatures expected in footpoint heating.  This is a preliminary study, using only a single loop geometry.  In a future paper, we will consider how these conclusions are impacted when many loops are included along the line of sight and we will include the predicted differential emission measure (DEM) as an observable.    

In Section 2, we describe the equations, assumptions, and geometry of all the simulations shown in this paper.  In Section 3, we compare a single footpoint-heated solution to an analogous impulsively-heated solution.  In Section 4, we compare the observables for footpoint-heated solutions of different heating magnitudes to analogous impulsively-heated solutions.  In Section 5, we investigate how varying the stratification of the heating impacts the footpoint-heated solutions and observables.  In Section 6, we discuss these results.  

%%------------------------------------------------------------------------
%%
%%				Time-lag analysis
%%
%%------------------------------------------------------------------------

\section{DESCRIPTION OF SIMULATIONS}

To calculate the evolution of these loops we rely on the Predictive Science one-dimensional hydrodynamic model (\citealt{mikic2013}).  This model solves the following 1D hydrodynamic equations along the length of a loop:

\begin{equation}
\frac{\partial \rho}{\partial t}
+\frac{1}{A}\frac{\partial}{\partial s}\left(A\rho v\right)=0, 
\label{eq-mass}
\end{equation}
\begin{equation}
\rho\left(\frac{\partial v}{\partial t}+v\frac{\partial v}{\partial s}\right)= 
 -\frac{\partial p}{\partial s}+\rho g_s 
 +\frac{1}{A}\frac{\partial}{\partial s}\left(A\nu\rho
\frac{\partial v}{\partial s}\right),
\label{eq-mom}
\end{equation}
\begin{multline}
\frac{\partial T}{\partial t}
+\frac{1}{A}\frac{\partial}{\partial s}\left(ATv\right)=
-(\gamma-2)\frac{T}{A}\frac{\partial}{\partial s}\left(Av\right)+ \\
\frac{(\gamma-1)T}{p}
\left[\frac{1}{A}\frac{\partial}{\partial s}\left(A
\kappa_\parallel\frac{\partial T}{\partial s}\right)
-n_en_p{Q(T)}+{H(s,t)}\right],
\label{eq-energy}
\end{multline}
where $s$ is the length along the loop, $T$, $p$, and $v$ are the plasma temperature, pressure, and velocity, $n_e$ and $n_p$ are the electron and proton number density (assumed to be equal), and $\gamma=5/3$ is the ratio of specific heats. The mass density for the plasma is $\rho=m_pn_e$, where $m_p$ is the proton mass.    The plasma pressure is $p=2n_ekT$; where $k$ is Boltzmann's constant.  The gravitational acceleration projected along the loop is $g_s=\bhat\cdot{\bf g}$, where $\bhat$ is the unit vector along the magnetic field ${\bf B}$. The Spitzer thermal conductivity coefficient is   $\kappa_\parallel(T)=C_0 T^{5/2}\,\text{[erg/cm/s/K]}$, where $C_0 = 9 \times 10^{-7}$ in cgs units.

In the above equations, $Q(T)$ represents the radiative loss function.  We use a the radiative loss function derived from the CHIANTI (version~7.1)  atomic database \citep{1997A&AS..125..149D,2013ApJ...763...86L} with coronal abundances \citep[\texttt{sun\_coronal\_ext.abund},][]{1992ApJS...81..387F,2002ApJS..139..281L,1998SSRv...85..161G} and the CHIANTI collisional ionization equilibrium model \citep{2009A&A...498..915D}. Radiation is turned off at chromospheric temperatures.  

A uniform kinematic viscosity $\nu$ is used to damp out waves, and is made very small, corresponding to a diffusion time $L^2/\nu \sim 2{,}000\,\text{hours}$.  The small values of $\kappa_\parallel$ at low temperatures produce very steep gradients in temperature in the transition region that are difficult to resolve in numerical calculations. In order to minimize the mesh resolution requirements, we have developed a special treatment of the thermal conduction and radiative loss function to artificially broaden the transition region at low temperatures \citep{2009ApJ...690..902L} without significantly affecting the coronal solution \citep{mikic2013}.    The broadening has been applied for temperatures less than 0.25 MK.
 
\begin{figure*}[t!]
\begin{center}
\resizebox{.49\textwidth}{!}{\rotatebox{0}{\includegraphics{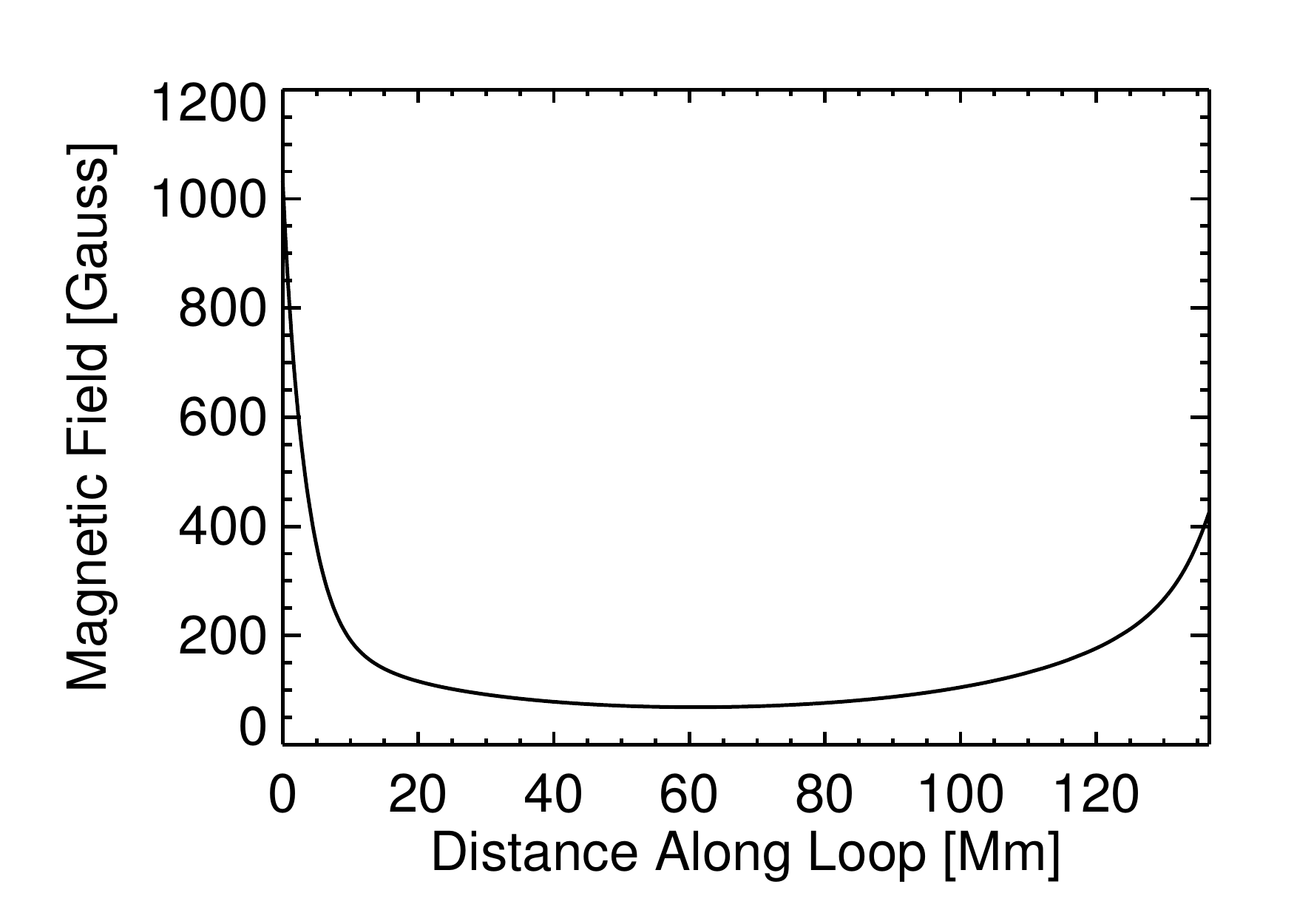}}}
\resizebox{.49\textwidth}{!}{\rotatebox{0}{\includegraphics{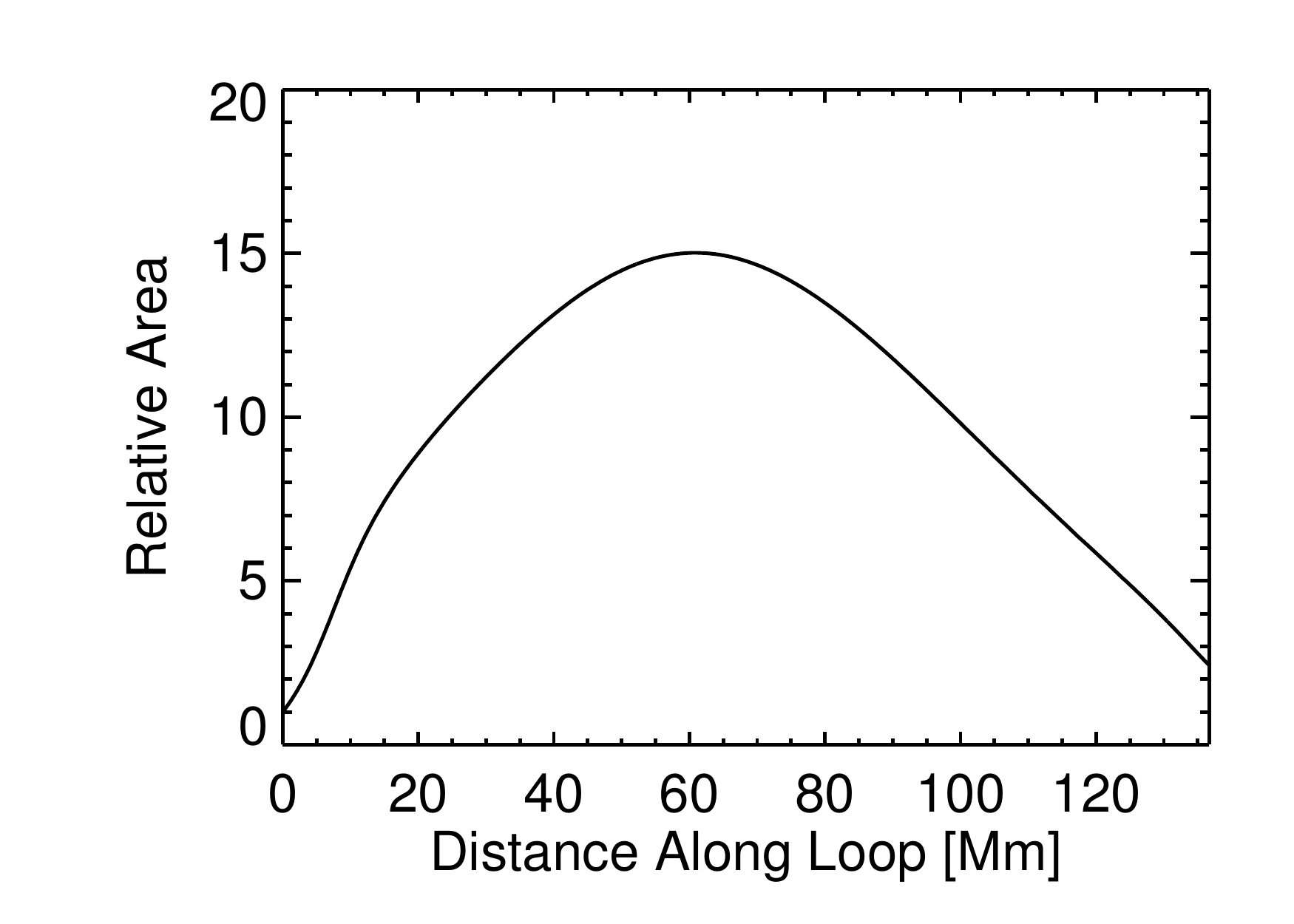}}}
\resizebox{.49\textwidth}{!}{\rotatebox{0}{\includegraphics{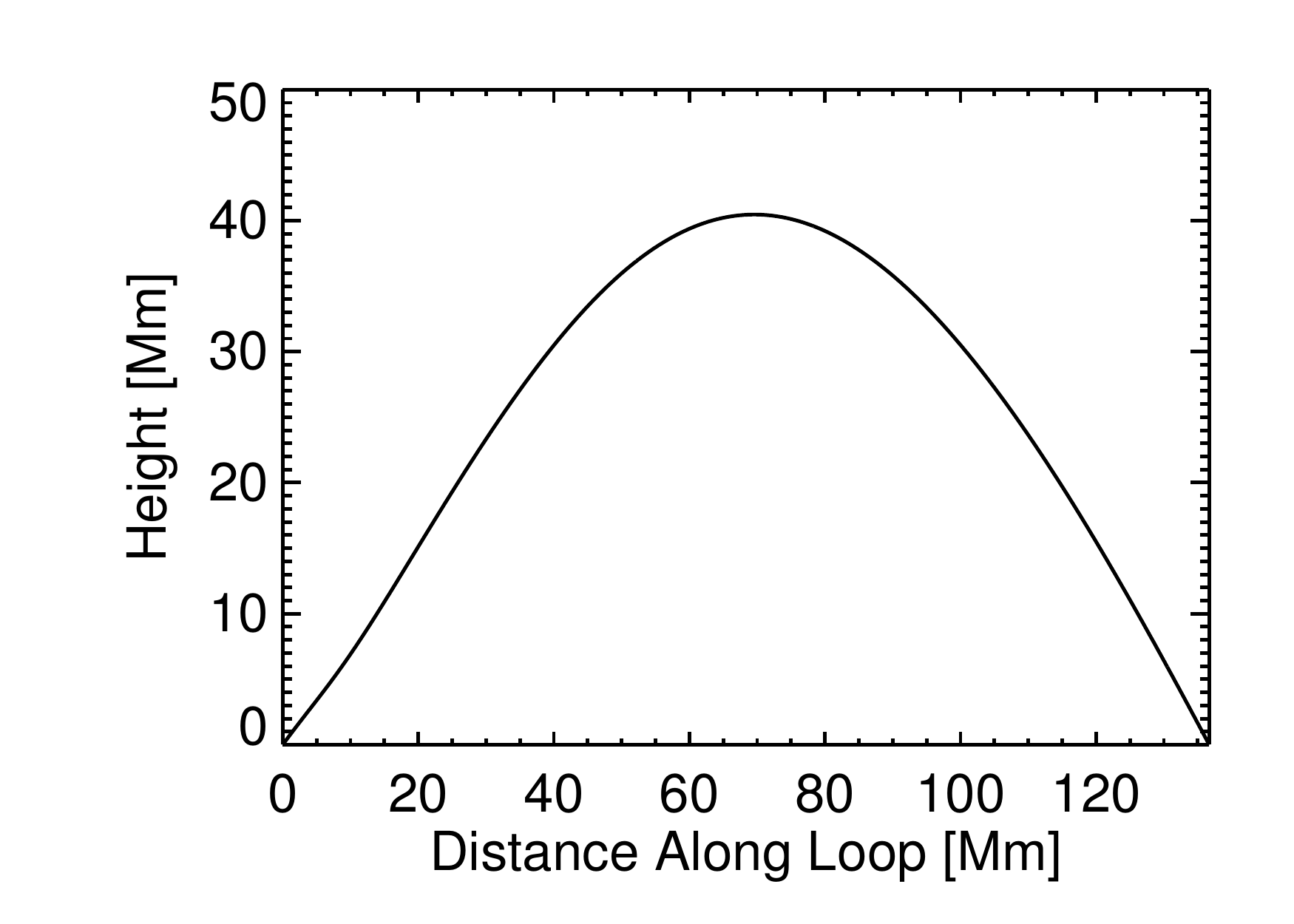}}}
\resizebox{.49\textwidth}{!}{\rotatebox{0}{\includegraphics{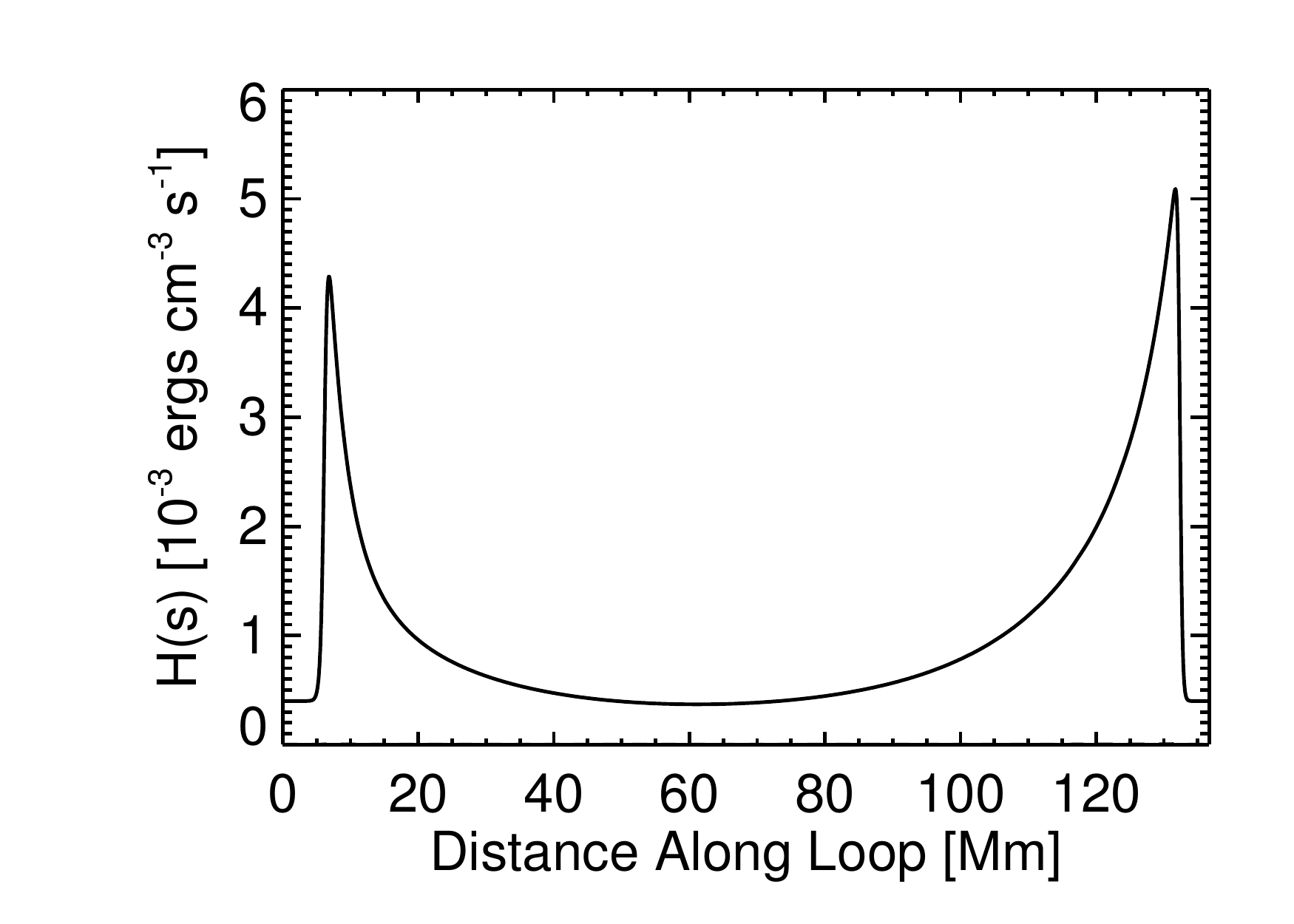}}}
\caption{The magnetic field (upper left), relative area (upper right) and height (lower left) as a function of loop length define the geometry of the loop considered in this study.  The initial stratified heating function used in \cite{mikic2013} is shown in the lower right panel.}
\label{fig:b_a_h}
\end{center}
\end{figure*}

The loop geometry we consider in this research is extracted from the 3D simulation of \citet{2005ApJ...621.1098M,2008ApJ...679L.161M,mok2016} and used in Section 7 of \cite{mikic2013}.  The loop is extracted from a non-linear force-free field extrapolation of Active Region 7986 observed in late August 1996 (see Figures 15 and 16 in \citet{mikic2013} for an image of this loop).  The loop is 137 Mm long.  The magnetic field strength, relative cross sectional area (proportional to the inverse of the magnetic field strength), and height as a function of position along the loop are shown in Figure \ref{fig:b_a_h}.  Note the loop is slightly asymmetric.

In Equation \ref{eq-energy}, $H(s,t)$ is the coronal heating function.  In this investigation, we use two different heating functions, a steady, highly-stratified heating function and an impulsive, uniform heating function. For our initial steady heating function, we use the heating function of \cite{mikic2013}, which is shown in the lower right panel of Figure~\ref{fig:b_a_h}.   For subsequent runs of footpoint heating, we change the magnitude or stratification of this heating function.  

We define the impulsive heating function as
\begin{equation}
H_{\rm imp}(t) = H_{\rm imp0} \sum_{n=0}^{n_{\rm pulse}} g(t-n\tau) + H_{\rm back}
\label{eq:imp}
\end{equation}
where $H_{\rm imp0}$ is the magnitude of the impulsive heating that is uniformly applied along the loop over a short time and $H_{\rm back}$ is a small background heating, also applied uniformly to insure that the temperature never goes to 0.  In Equation \ref{eq:imp}, $\tau$ is the time between heating events and $n_{\rm pulse}$ is the number of pulses in the simulation.  We chose the heating events to be evenly spaced in time, so the total number of pulses is the total simulation time divided by $\tau$.  The time dependence of the impulsive heating event, $g(t)$, is a triangular pulse defined as 
\begin{eqnarray}
g(t) = t/\delta  \hspace{0.5in} 0&<t\le&\delta \\
g(t) = (2\delta - t)/\delta \hspace{0.5in}\delta &< t\le&2\delta\\
g(t) = 0 \hspace{0.5in} t < 0 & or & t> 2\delta\
\label{eq:gt}
\end{eqnarray}
where $2\delta$ is the total duration of the impulsive heating.   The duration of the impulsive heating event does not greatly impact the expected intensities in the channels we are simulating as long as it is much less than the cooling time of the plasma.  For all impulsive heating simulations in this paper, $2\delta = 174$\,s and $H_{\rm back} = 1.6 \times 10^{-5}$ ergs cm$^{-3}$ s$^{-1}$.     The average heating per pulse over each period, $\tau$, is thus $H_{\rm back}  + \delta H_{\rm imp0}/\tau$.

\section{EXAMPLE SIMULATIONS AND OBSERVABLES}

In this section, we present an example of a highly-stratified simulation (shown previously in \citealt{mikic2013}) and a comparable impulsive heating simulation.  We then use these two example simulations to introduce the observables that will be measured for the larger parameter study.   

\subsection{Footpoint Heating}

\begin{figure*}[t!]
\begin{center}
\resizebox{.32\textwidth}{!}{\rotatebox{0}{\includegraphics{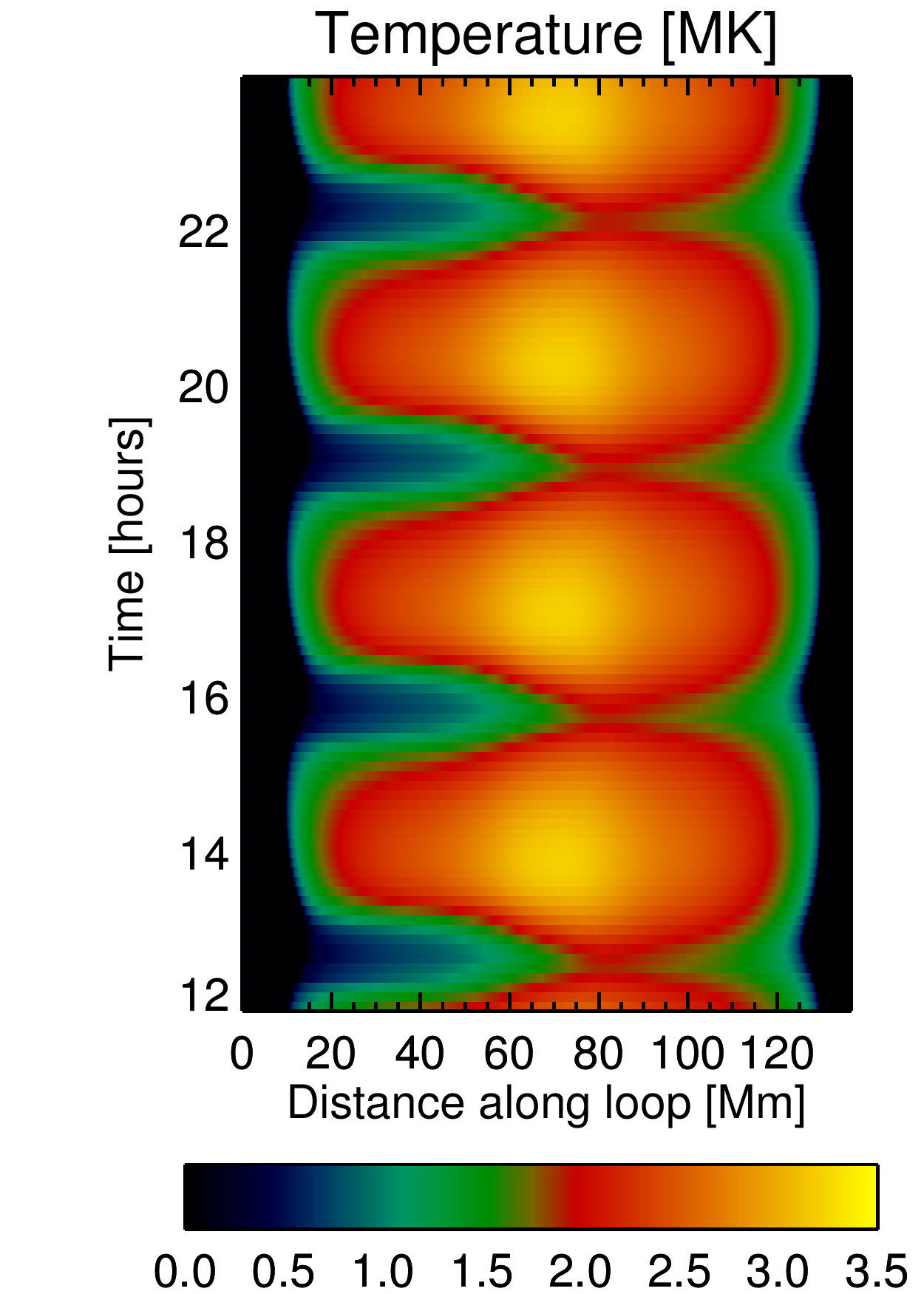}}}
\resizebox{.32\textwidth}{!}{\rotatebox{0}{\includegraphics{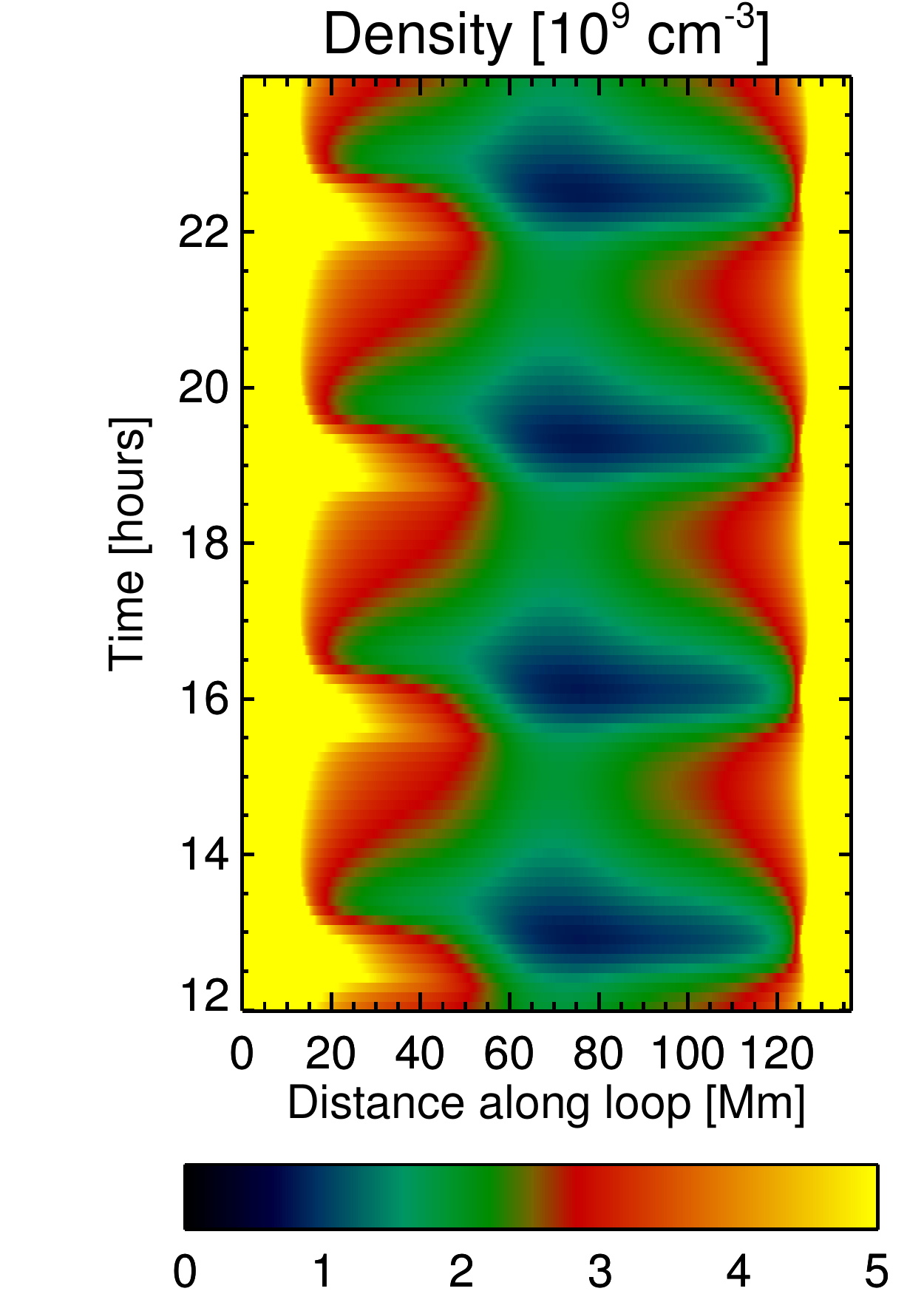}}}
\caption{The temperature and density evolution for the example footpoint heating simulation.}
\label{fig:example_map}
\end{center}
\end{figure*}

The simulation originally shown in \cite{mikic2013} forms the basis for much of the subsequent simulations in this paper.  The geometry and heating function are shown in Figure~\ref{fig:b_a_h}.  The temperature and density as a function of time and distance along the loop that result from this heating are shown in Figure~\ref{fig:example_map}.  The temperature and density evolution imply that the loop is in thermal non-equilibrium (TNE).  We focus on the simulation between 12 and 24 hours, which allows for enough time after the start of the simulations for the initial conditions to not impact our conclusions.
 
\begin{figure*}[t!]
\begin{center}
\resizebox{.49\textwidth}{!}{\rotatebox{0}{\includegraphics{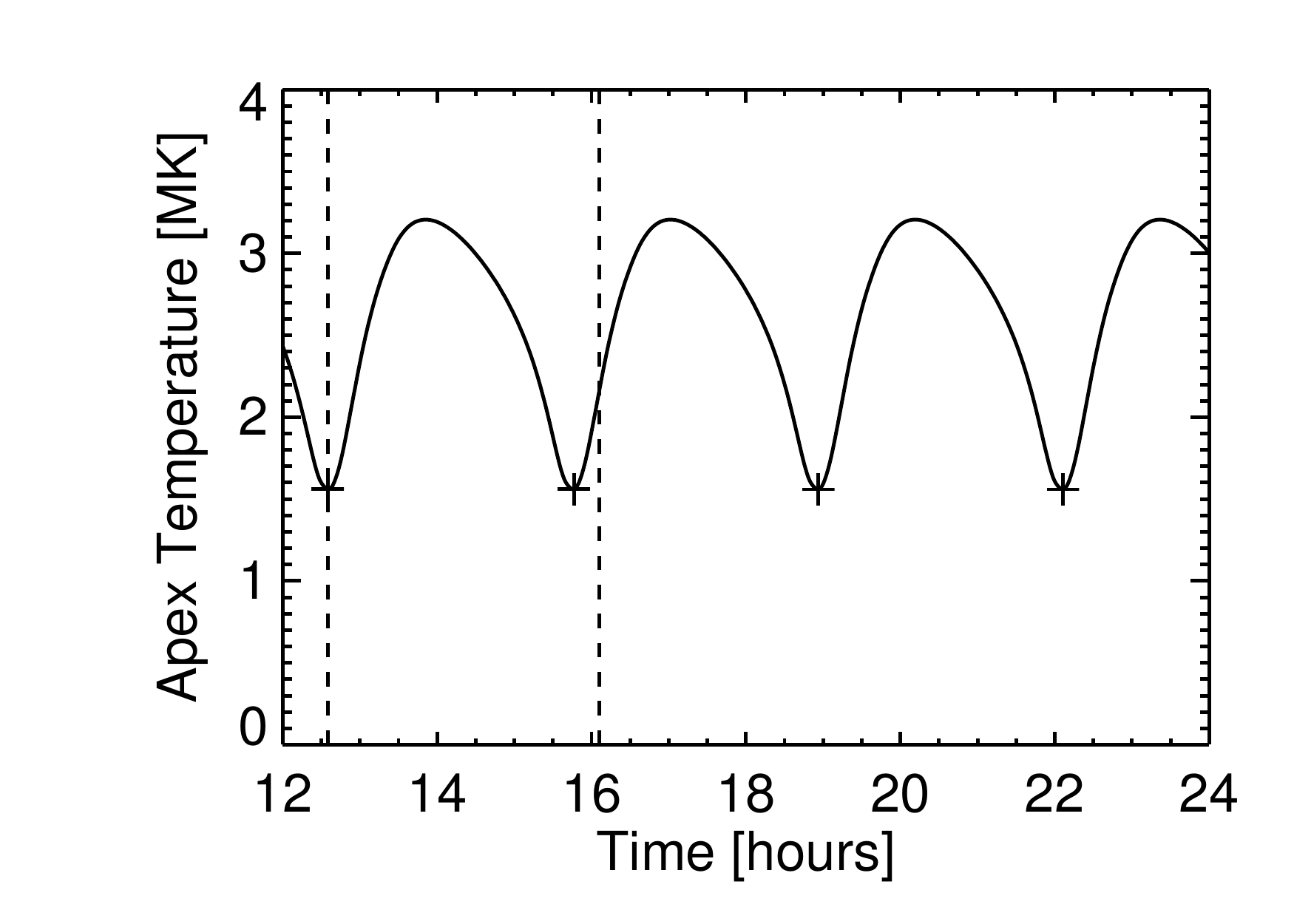}}}
\resizebox{.49\textwidth}{!}{\rotatebox{0}{\includegraphics{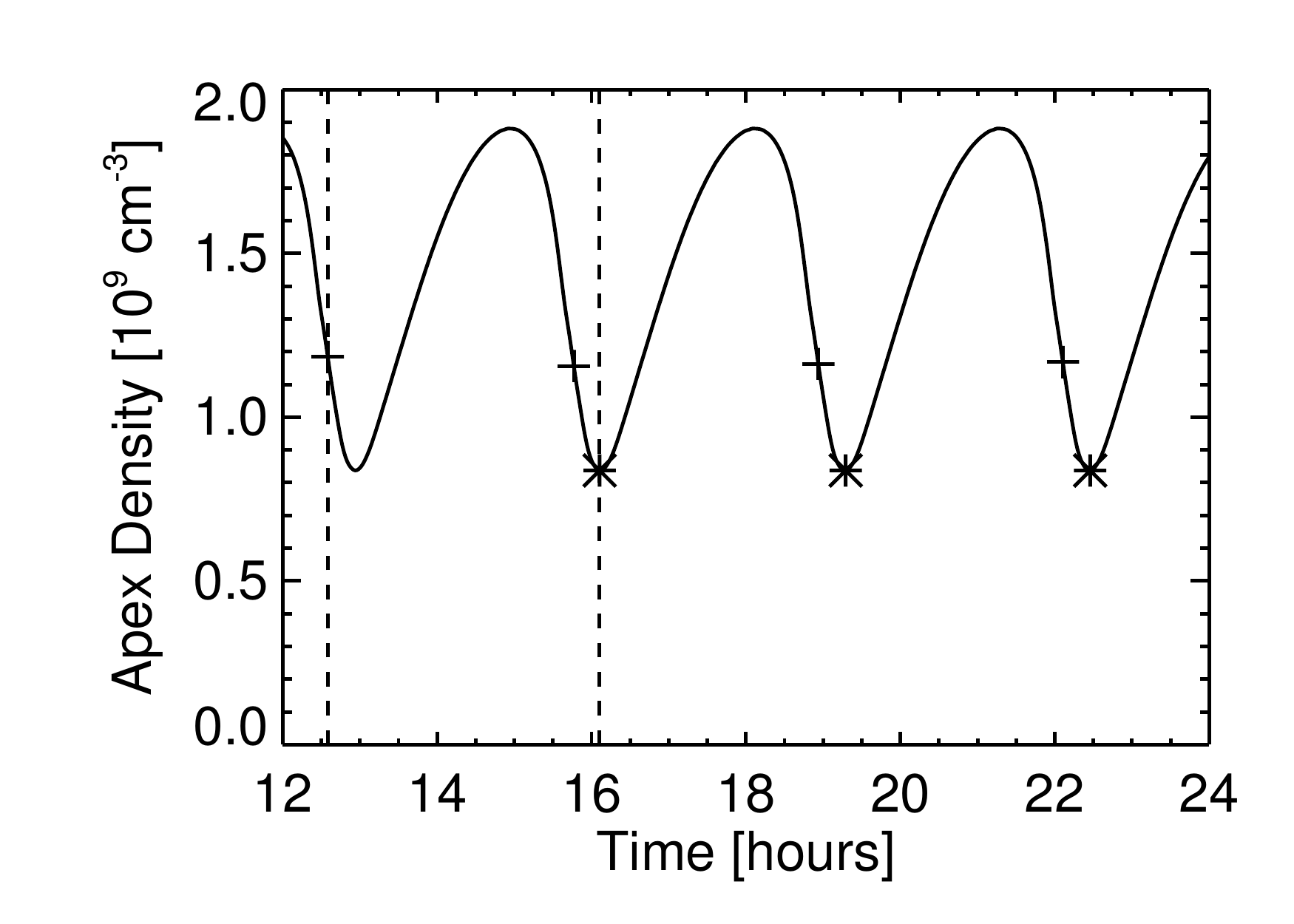}}}
\caption{The apex temperature and density for the example footpoint heating simulation.   The dashed lines show slightly more than a single cycle, from a minimum in the temperature evolution to the minimum in the density evolution.  This cycle is compared to the impulsive heating cycle in Figure~\ref{fig:norm}.  Crosses show the time when the temperature is at a minimum, asterisks show the time the apex density is at a minimum.}
\label{fig:example_apex}
\end{center}
\end{figure*}

We find the average apex values of the temperature and density by averaging over the top 10\% of the loop.  These values are shown in Figure~\ref{fig:example_apex} for hours 12 through 24 of the simulation.  The maximum and minimum apex temperatures of this simulation are 3.2 MK and 1.6 MK.  The maximum and minimum apex densities are 1.9$\times 10^9$ cm$^{-3}$ and 8.4$\times 10^8$\,cm$^{-3}$.  

Another concept we will discuss in this paper is the cycle time of the footpoint heating simulations.  For thermal non-equilibrium solutions, we expect the solution to cycle, or repeat, at some time scale depending on the heating profile and geometry of the loop.  We define the cycle time as the average time between local minima in the apex temperature.  The local minima are shown in Figure~\ref{fig:example_apex} as crosses.  The average cycle time of this simulations is 3.17 hours.  Later, we compare this cycle time to the cooling time of impulsive heating simulations.  Note that there is an offset between the minima in the apex temperature and the minima in the apex density.  In the right panel of Figure \ref{fig:example_apex}, the time of local minima in temperature, shown as crosses, are offset from the time of the local minima in density, shown as asterisks.   In Figure~\ref{fig:norm}, we show the evolution of the apex temperature and density between the two dashed lines in Figure \ref{fig:example_apex}.  This is from the minimum temperature in one cycle to the density minimum in the next cycle.  The total time of the simulation shown in Figure \ref{fig:norm} is slightly longer than one cycle.  

\subsection{Impulsive Heating}

Next, we want to complete an impulsive heating simulation that has the same average heating rate as the footpoint heating simulation for comparison.   There are many ways to choose the parameters of the impulsive heating solution.  This is not meant to be a full search of the possible parameter space; we pick one (typical) solution that is comparable to the footpoint heated solution.

The power in the footpoint heating  function, $P_{\rm FP}$, can be found from 
\begin{equation}
P_{\rm FP} = \int_0^{\rm{s_{max}}} H_{\rm FP0}(s) A(s) ds \hspace{0.5in} \rm{[ergs\ s^{-1}]}
\end{equation}
where $H_{\rm FP0}(s)$ is the initial volumetric heating rate as a function of position along the loop and $A(s)$ is the cross sectional area; both are shown in Figure~\ref{fig:b_a_h}.  For impulsive heating,  the total energy input of a single heating pulse can be found from
\begin{eqnarray}
E_{\rm imp} &  = & \int_0^{2\delta} \int_0^{\rm{s_{max}}} H_{\rm imp}(t) A(s) ds dt \hspace{0.5in} \rm{[ergs]} \\
E_{\rm imp} & = & H_{\rm imp0} \delta  \int_0^{\rm{s_{max}}} A(s) ds
\end{eqnarray}
where $H_{\rm imp}(t)$ is defined by Equations~\ref{eq:imp} and \ref{eq:gt} and we have ignored the small contribution of the background heating.  

Next we want to find the $H_{\rm imp0}$ so that the total power in both the impulsive and footpoint simulations are the same.  We want the impulsive heating to be infrequent or sporadic, such that the loop has an opportunity to fully cool to background levels (e.g., 0.5 MK).  We define the time it takes from the initial heating event until the loop returns to background levels as $\tau$, or the cooling time.  We can then relate the power in the footpoint simulation to the energy of the impulsive heating simulation through
\begin{equation}
P_{\rm FP} = E_{\rm imp}/\tau,
\end{equation}
which can be written as
\begin{equation}
H_{\rm imp0}/ \tau  = \frac{ \int_0^{\rm{s_{max}}} H_{\rm FP0}(s) A(s) ds }{\delta  \int_0^{\rm{s_{max}}} A(s) ds}.
\end{equation}
In this equation, the terms on the right-hand side except for $\delta$ are the footpoint heating function and the geometry of the loop, defined in our initial simulation.  The duration of the heating is chosen as $2\delta = 174$\,s.   The terms on the left-hand side of the equation are free parameters.  The cooling time, $\tau$, will depend on the impulsive heating magnitude $H_{\rm imp}$.   We iteratively adjusted $H_{\rm imp0}$ and $\tau$ until the minimum apex temperature of the solution was approximately 0.5\,MK, the apex temperature associated with the background heating rate.  For the original footpoint heating function, the values of $H_{\rm imp0}$ and $\tau$ are 3.16$\times 10^{-2}$ ergs cm$^{-3}$ s$^{-1}$ and 1 hour, respectively.

\begin{figure*}[t!]
\begin{center}
\resizebox{.32\textwidth}{!}{\rotatebox{0}{\includegraphics{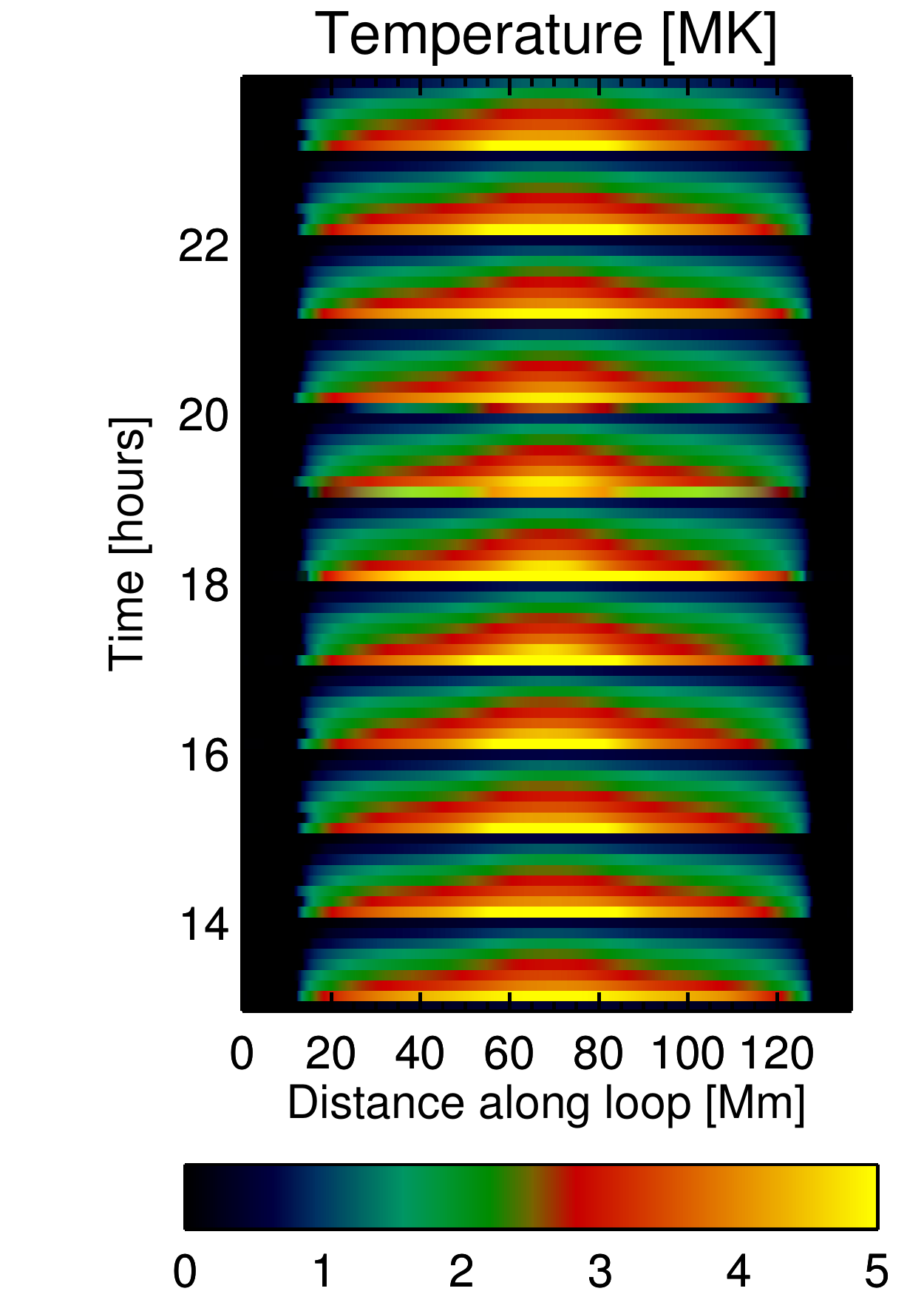}}}
\resizebox{.32\textwidth}{!}{\rotatebox{0}{\includegraphics{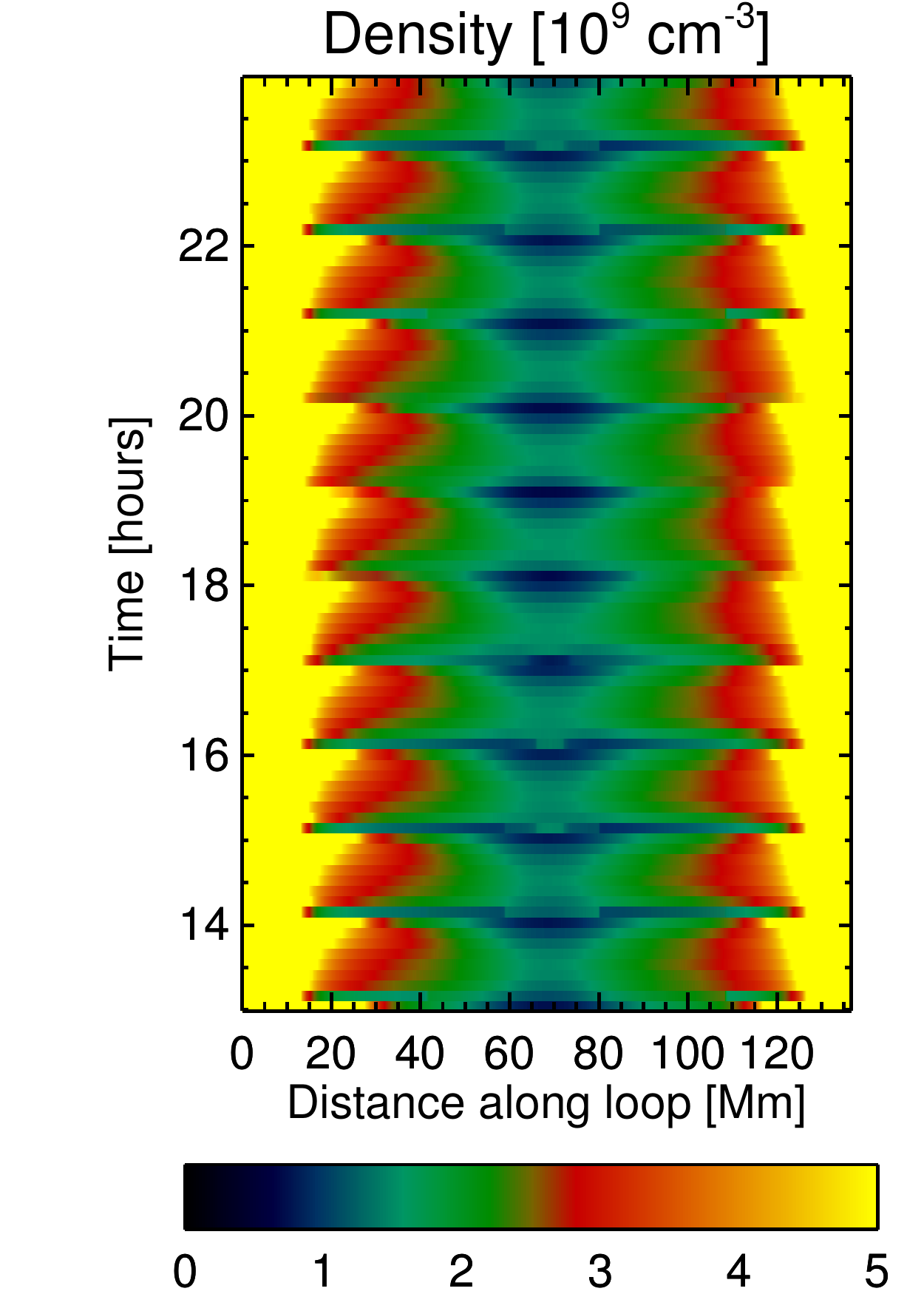}}}
\caption{The temperature and density evolution for the example impulsive heating simulation.}
\label{fig:example_map_imp}
\end{center}
\end{figure*}

\begin{figure*}[t!]
\begin{center}
\resizebox{.49\textwidth}{!}{\rotatebox{0}{\includegraphics{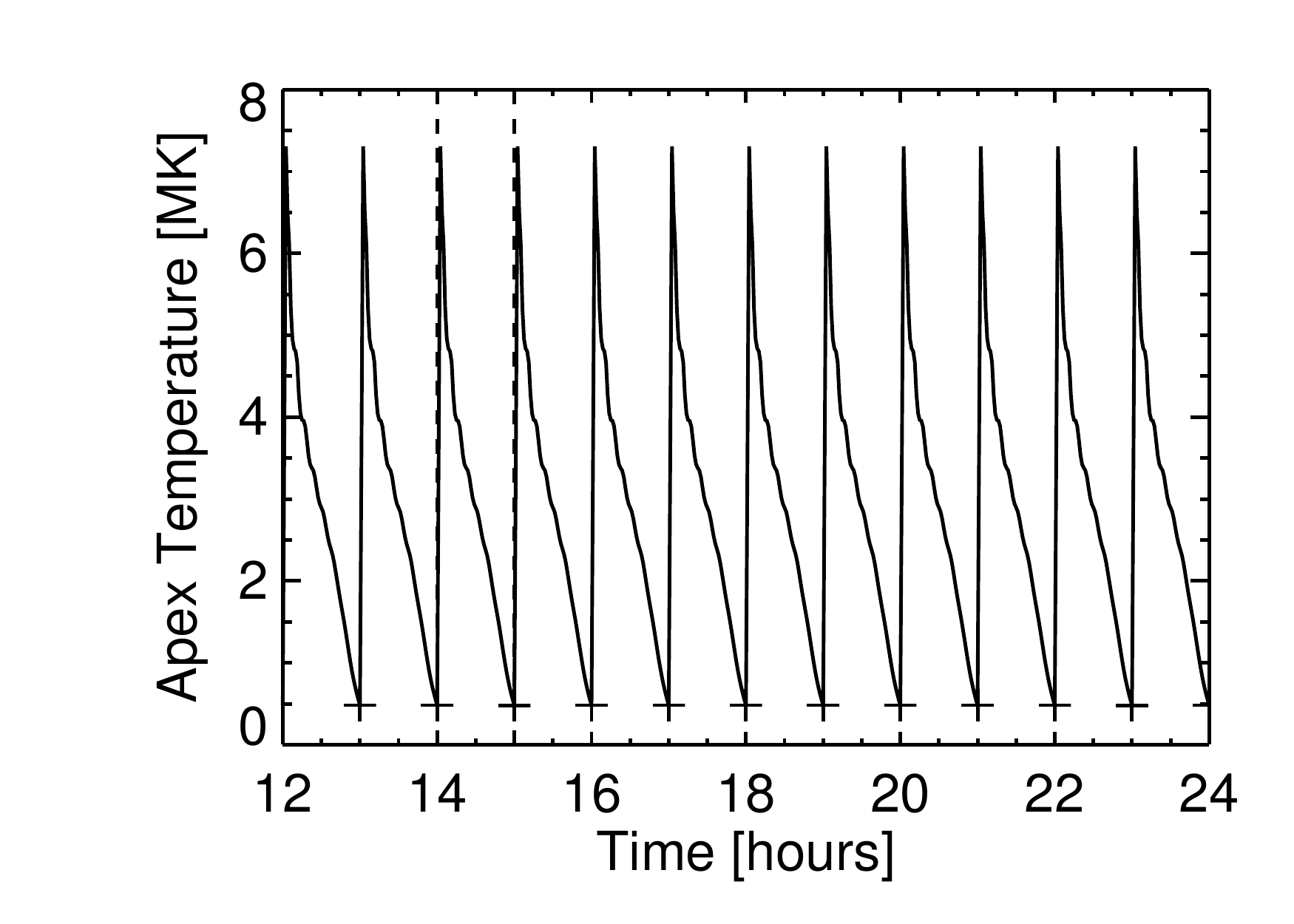}}}
\resizebox{.49\textwidth}{!}{\rotatebox{0}{\includegraphics{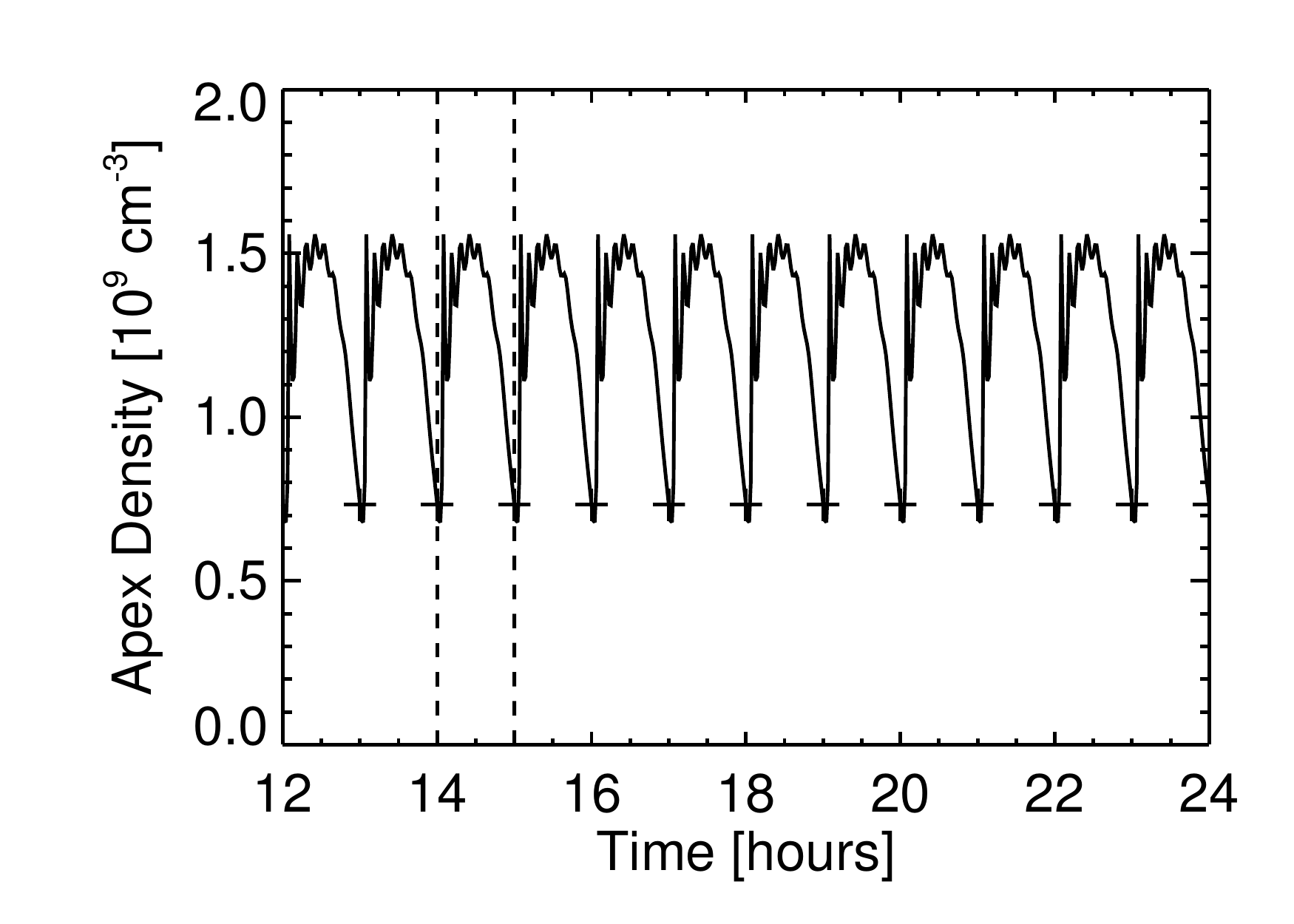}}}
\caption{The apex temperature and density for the example impulsive heating simulation.}
\label{fig:example_apex_imp}
\end{center}
\end{figure*}

The hydrodynamic solution for this impulsive heating function is shown in Figure~\ref{fig:example_map_imp}.  The average apex quantities are shown in Figure~\ref{fig:example_apex_imp}.  Note that this solution cycles much more rapidly than the footpoint solution,  but this is entirely artificial since we have chosen the solution to cycle at the cooling time, $\tau$, or 1 hour.  The range of apex temperatures are 0.48 to 7.3\,MK and the range of apex densities are 6.8$\times 10^8$ - 1.6$\times 10^9$ cm$^{-3}$.   These values for both the footpoint and impulsive heating solutions are given in Table~\ref{tab:density_temp_cycle}.

\begin{deluxetable}{ccc}
\tablecaption{Footpoint vs Impulsive Heating Results}
\tabletypesize{\scriptsize}
\tablewidth{0pt}
\tablehead{
\colhead{ } & \colhead{Footpoint} & \colhead{Impusive} }
\startdata
Temperature Range & 1.6 - 3.2 MK & 0.48 - 7.3 MK \\
Density Range & 8.4$\times 10^8$ - 1.9$\times 10^9$ cm$^{-3}$  &  6.8$\times 10^8$ - 1.6$\times 10^9$ cm$^{-3}$\\
Cycle Time/Cooling Time & 3.17 hours & 1 hour \\
\enddata
\label{tab:density_temp_cycle}
\end{deluxetable}

\subsection{Identifying Observables}

\begin{figure*}[t!]
\begin{center}
\resizebox{.49\textwidth}{!}{\includegraphics{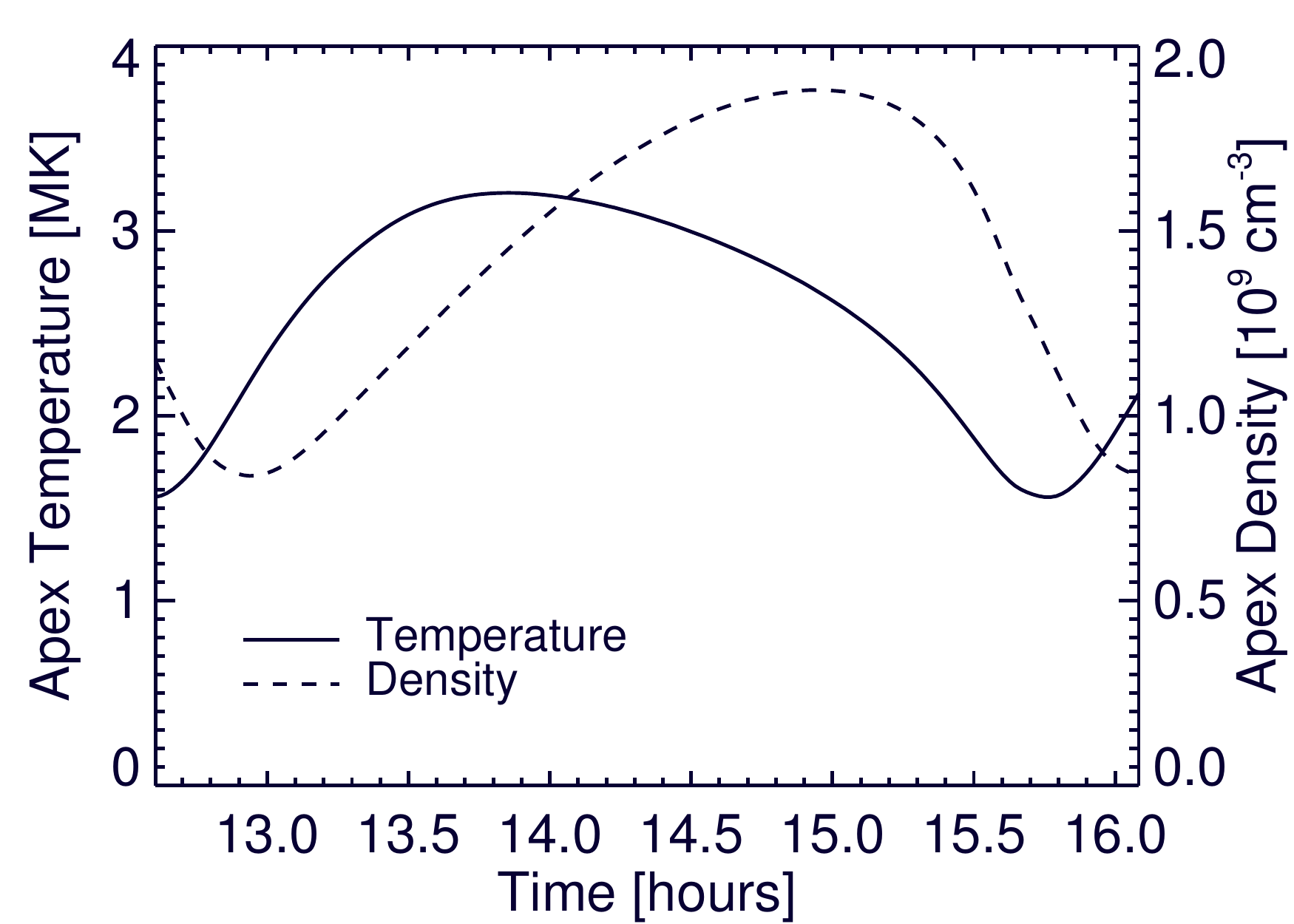}}
\resizebox{.49\textwidth}{!}{\includegraphics{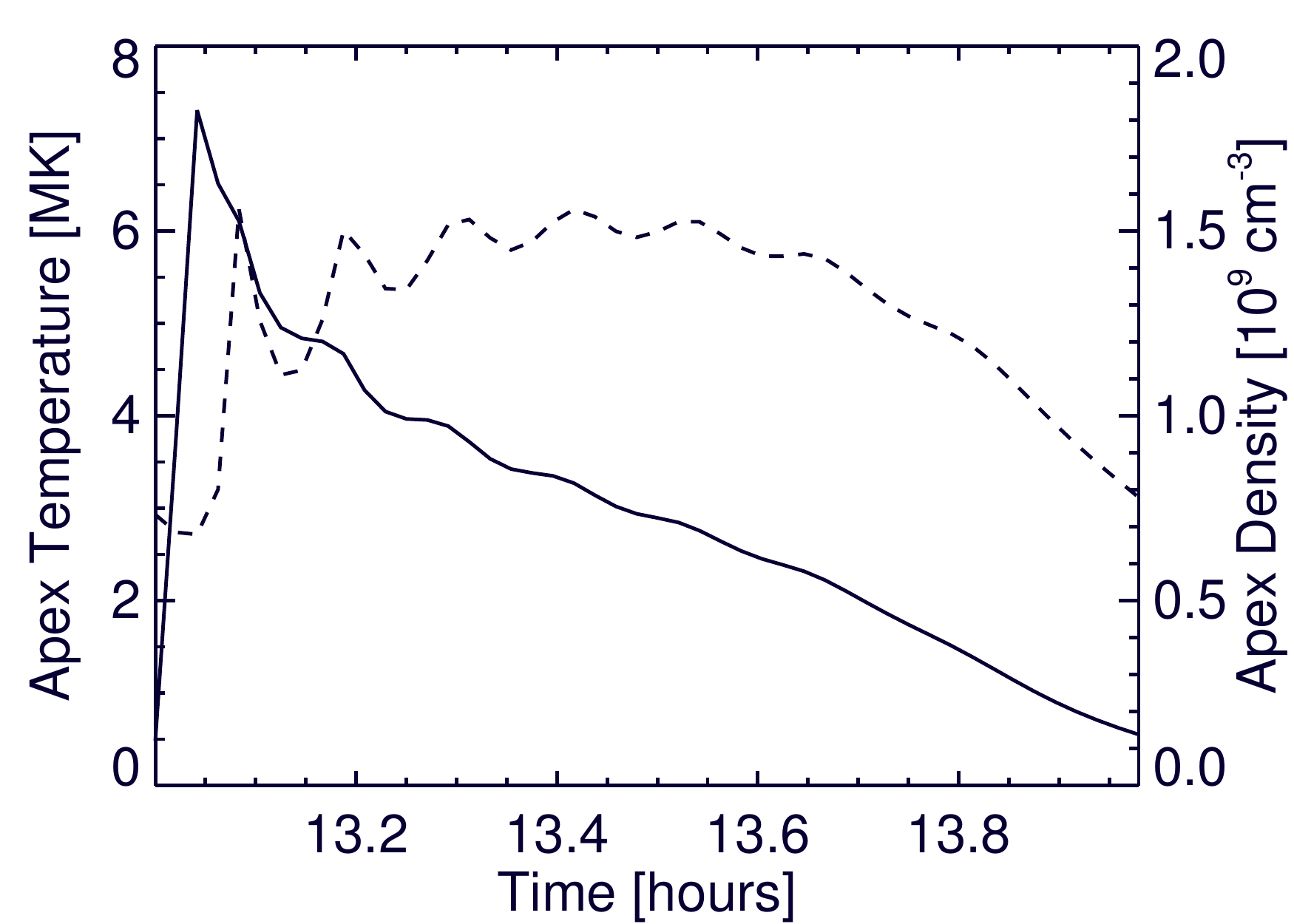}}
\caption{Example temperature and density evolution of the footpoint (left) and impulsive (right) heating simulation.  For the footpoint solution, we are showing slightly more than a single cycle, from the temperature minimum of one cycle to the density minimum.  For the impulsive heating solution, the temperature and density minima roughly coincide.}
\label{fig:norm}
\end{center}
\end{figure*}

Hydrodynamic solutions for low-frequency, impulsive heating and highly-stratified heating that result in thermal non-equilibrium share some important similarities, allowing them both to qualitatively match the observed properties of loops.   In both cases the plasma goes from high temperatures to cool temperatures.   However, despite the fact that the two example simulations have the same average power, the different location and temporal evolution of the heating produce temperature and density evolutions that are quite distinct.  For comparison,  the  apex temperature and density evolution for a single cycle of footpoint and impulsive heating simulations are shown in Figure~\ref{fig:norm}.

The first thing to note is that the cycle time of the footpoint heating solutions is much longer than the cooling time of an impulsively heated loop.  The cycle time is 3.17 hours, while the cooling time is 1 hour.   The maximum and minimum temperatures reached are also quite different.  The impulsive heating solution reaches temperatures greater than 7 MK and returns to ambient temperatures (roughly 0.5 MK), while the footpoint solution cycles between a maximum temperature of   3.2 MK and a minimum temperature of 1.6 MK.  Hence the impulsive heating solution goes through a wider range of temperatures and goes through them more quickly than does the footpoint solution.  This difference in the temperature evolution ought to manifest itself in the time delays between images or spectral intensities sensitive to different temperature plasma.  

Another difference between footpoint and impulsive heating solutions is the behavior of the density evolution.  For impulsive heating, the density peaks roughly when the cooling switches from being dominated by conduction to being dominated by radiation.  The density then first decreases slowly, then rapidly, throughout the rest of the evolution.  In the footpoint heating solution, the density increases slowly while the temperature decreases.  The peak of the density occurs just before the loop cools rapidly at the end of the cycle.   The evolution of the density should impact the relative intensity in the emission between various channels.   

\begin{figure*}[t!]
\begin{center}
\resizebox{.49\textwidth}{!}{\rotatebox{0}{\includegraphics{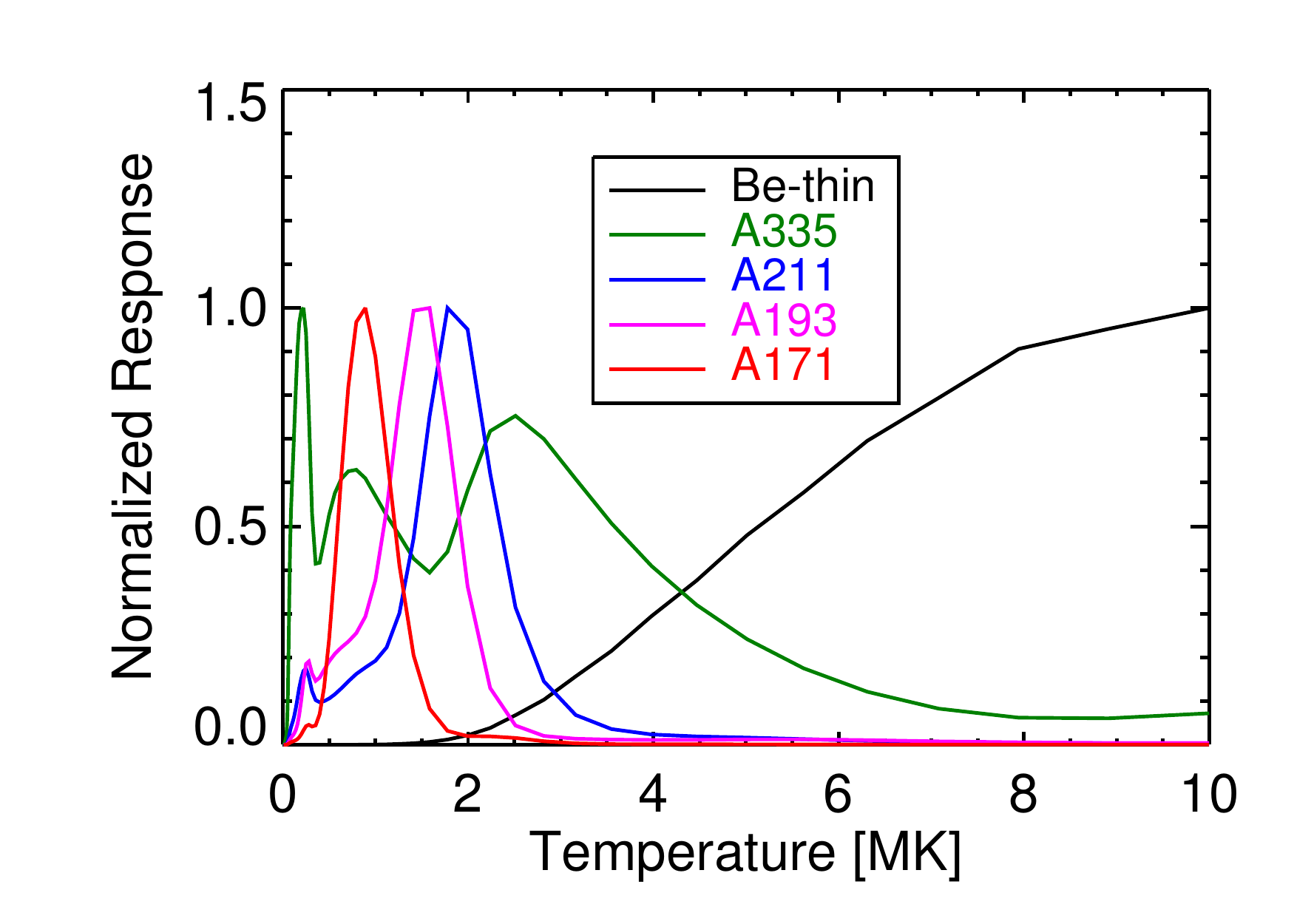}}}
\caption{AIA response functions for the 335, 211, 193, and 171\,\AA\ channels and the XRT Be-thin response function.}
\label{fig:resp}
\end{center}
\end{figure*}

Based on the above behavior, we suggest that delay times of the appearance of the loop between hot and cool channels (i.e., the time lag), as well as the relative intensity in the different channels, may be able to discriminate between the two heating scenarios.  To demonstrate this, we use the apex temperature and density values and calculate the emergent intensity of the loop in four EUV channels from {\it Solar Dynamics Observatory's} Atmospheric Imaging Assembly (AIA, \citealt{lemen2012}) and  one X-ray channel from {\it Hinode's} X-Ray Telescope (XRT; \citealt{golub2007}) using the equation
\begin{equation}
I_{\rm ch}(t) = \int n(t)^2 R_{\rm ch}(T(t)) ds
\end{equation}
where $R_{\rm ch}$ is the response function of the channel (see Figure~\ref{fig:resp}) and $ds$ is the depth of the loop.   Previous studies have found that loops are 1-2 Mm wide \citep{2005ApJ...634L.193A, 2008ApJ...680.1477A}.  We use 1\,Mm as the depth in this study.  Note that we are assuming this is the line-of-sight depth of the loop at the apex.  Because the cross sectional area in the loop expands, this implies the loop is much narrower at the footpoints.  

\begin{figure*}[t!]
\begin{center}
\resizebox{.32\textwidth}{!}{\rotatebox{0}{\includegraphics{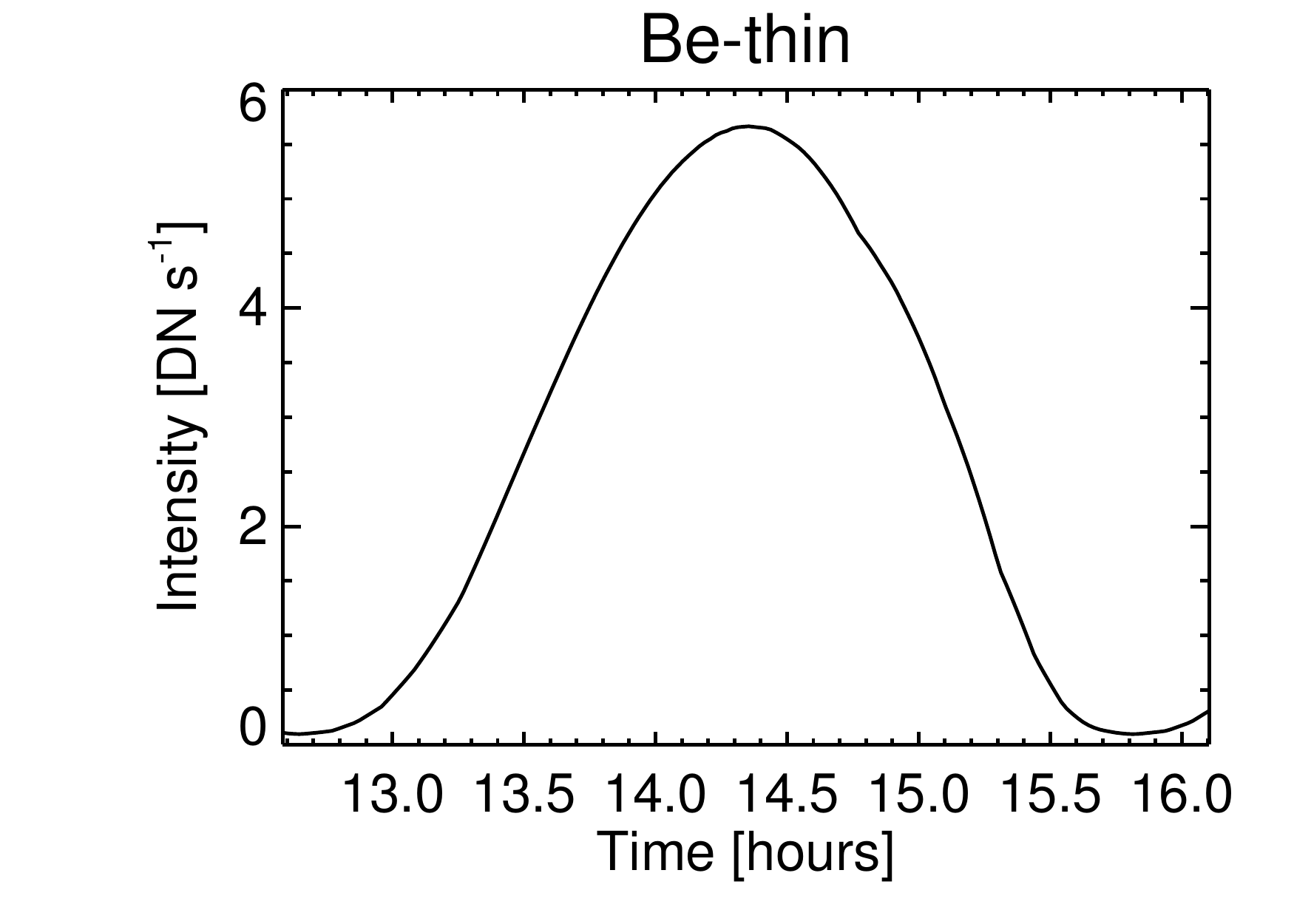}}}
\resizebox{.32\textwidth}{!}{\rotatebox{0}{\includegraphics{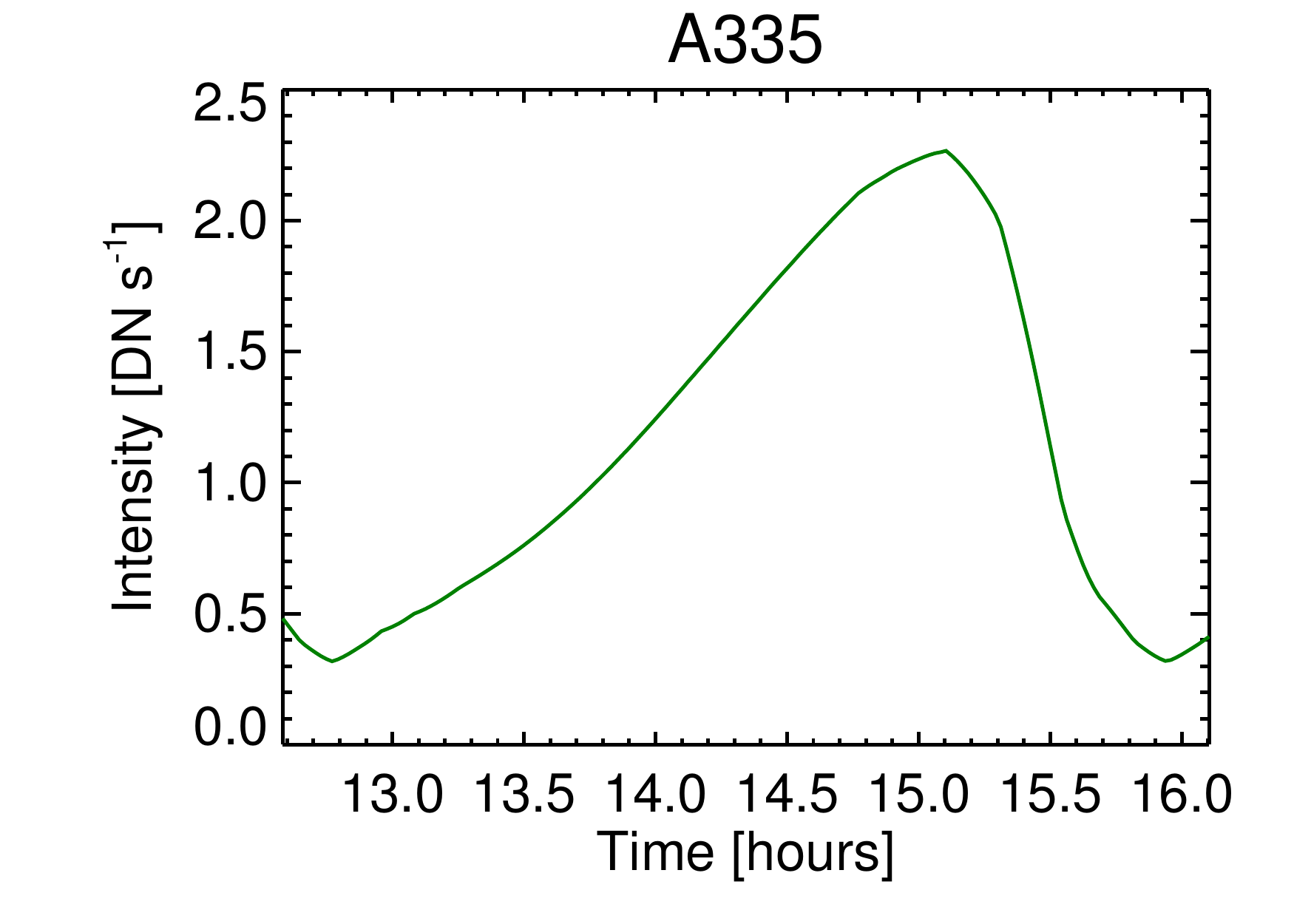}}}
\resizebox{.32\textwidth}{!}{\rotatebox{0}{\includegraphics{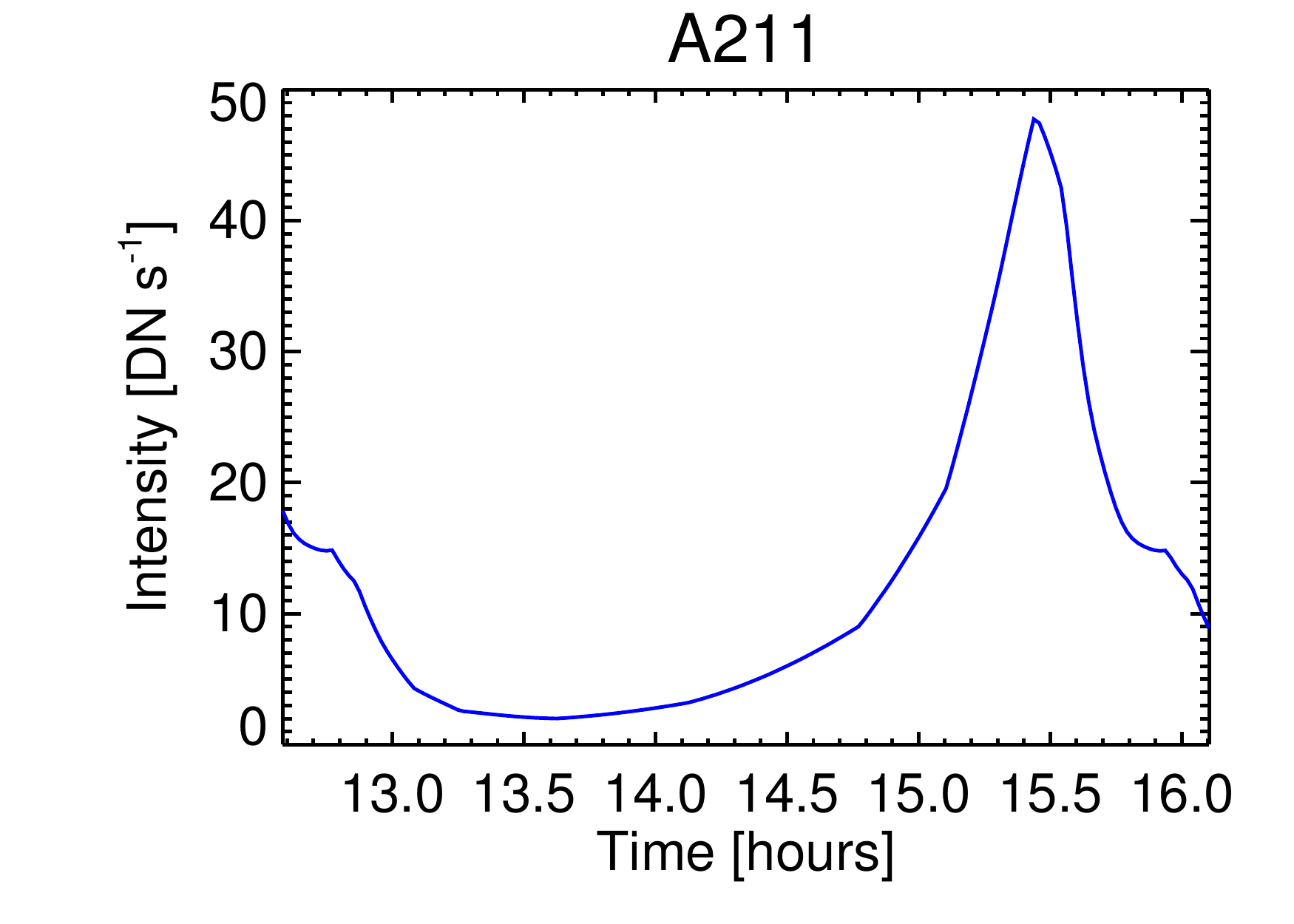}}}
\resizebox{.32\textwidth}{!}{\rotatebox{0}{\includegraphics{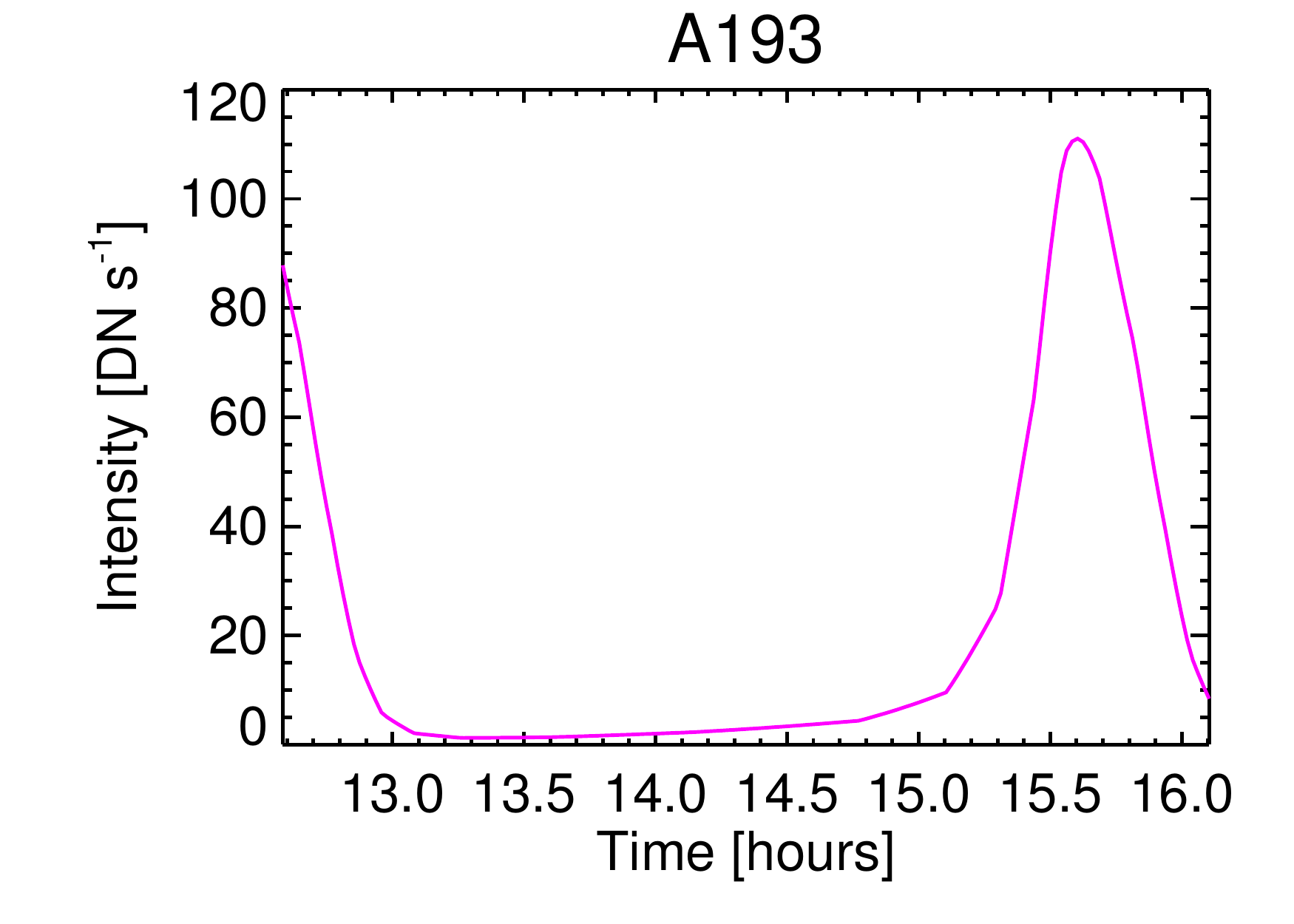}}}
\resizebox{.32\textwidth}{!}{\rotatebox{0}{\includegraphics{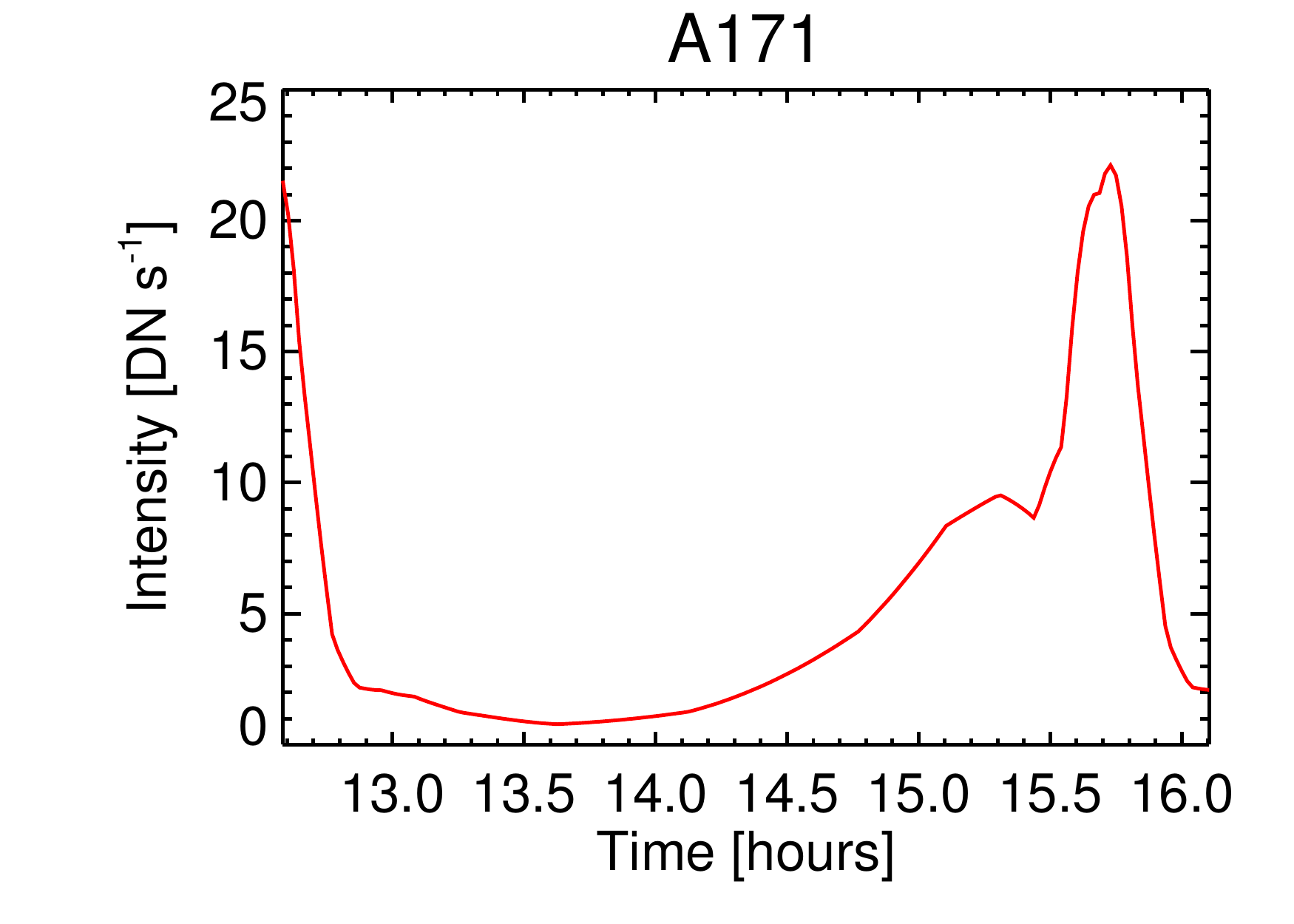}}}
\resizebox{.32\textwidth}{!}{\rotatebox{0}{\includegraphics{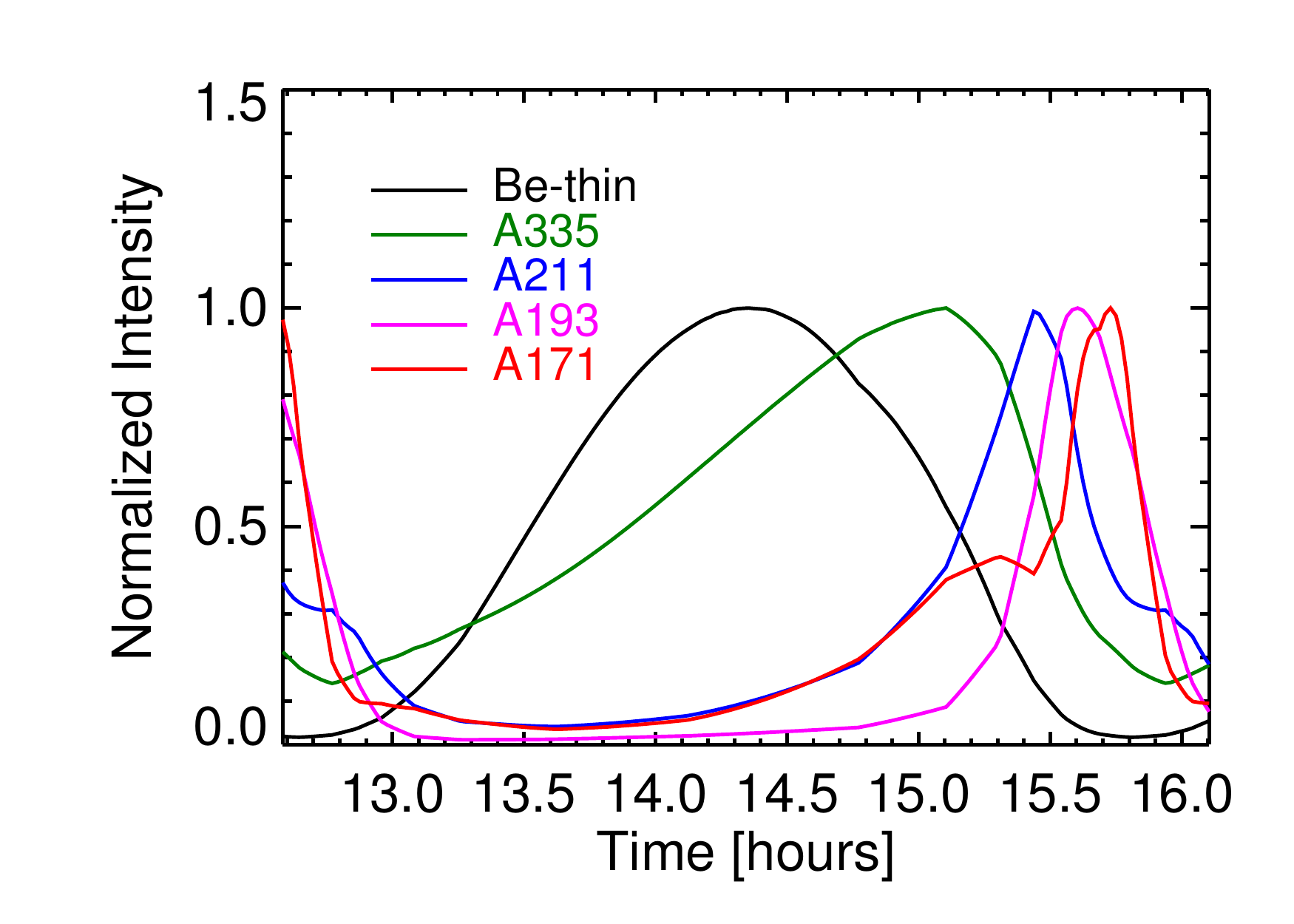}}}
\caption{The light curve of the apex of the loop in Hinode XRT Be-thin and four AIA channels for one cycle of the example footpoint heated simulation.  The bottom right panel shows the normalized light curves for all channels.}
\label{fig:example_lc}
\end{center}
\end{figure*}

\begin{figure*}[t!]
\begin{center}
\resizebox{.32\textwidth}{!}{\rotatebox{0}{\includegraphics{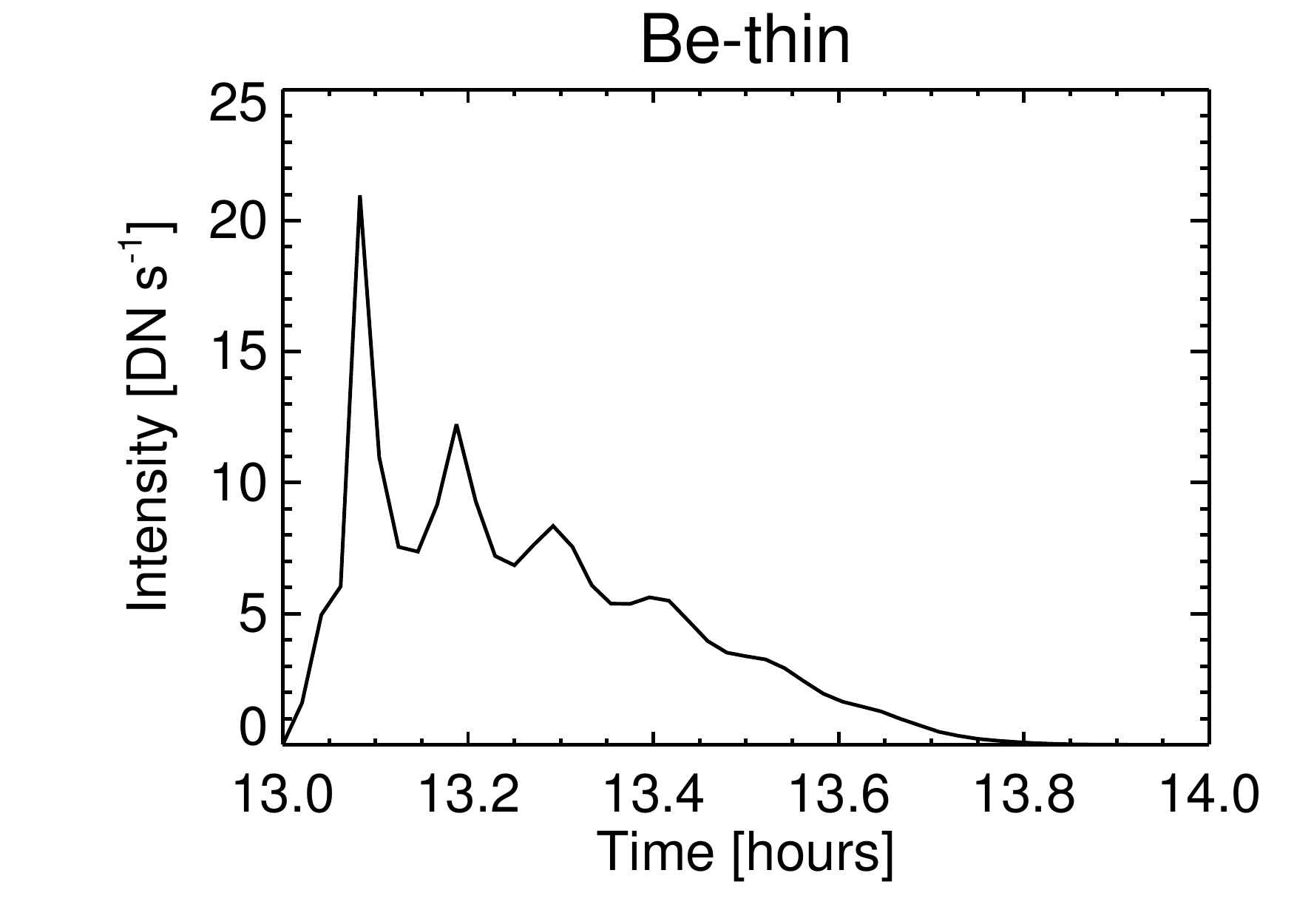}}}
\resizebox{.32\textwidth}{!}{\rotatebox{0}{\includegraphics{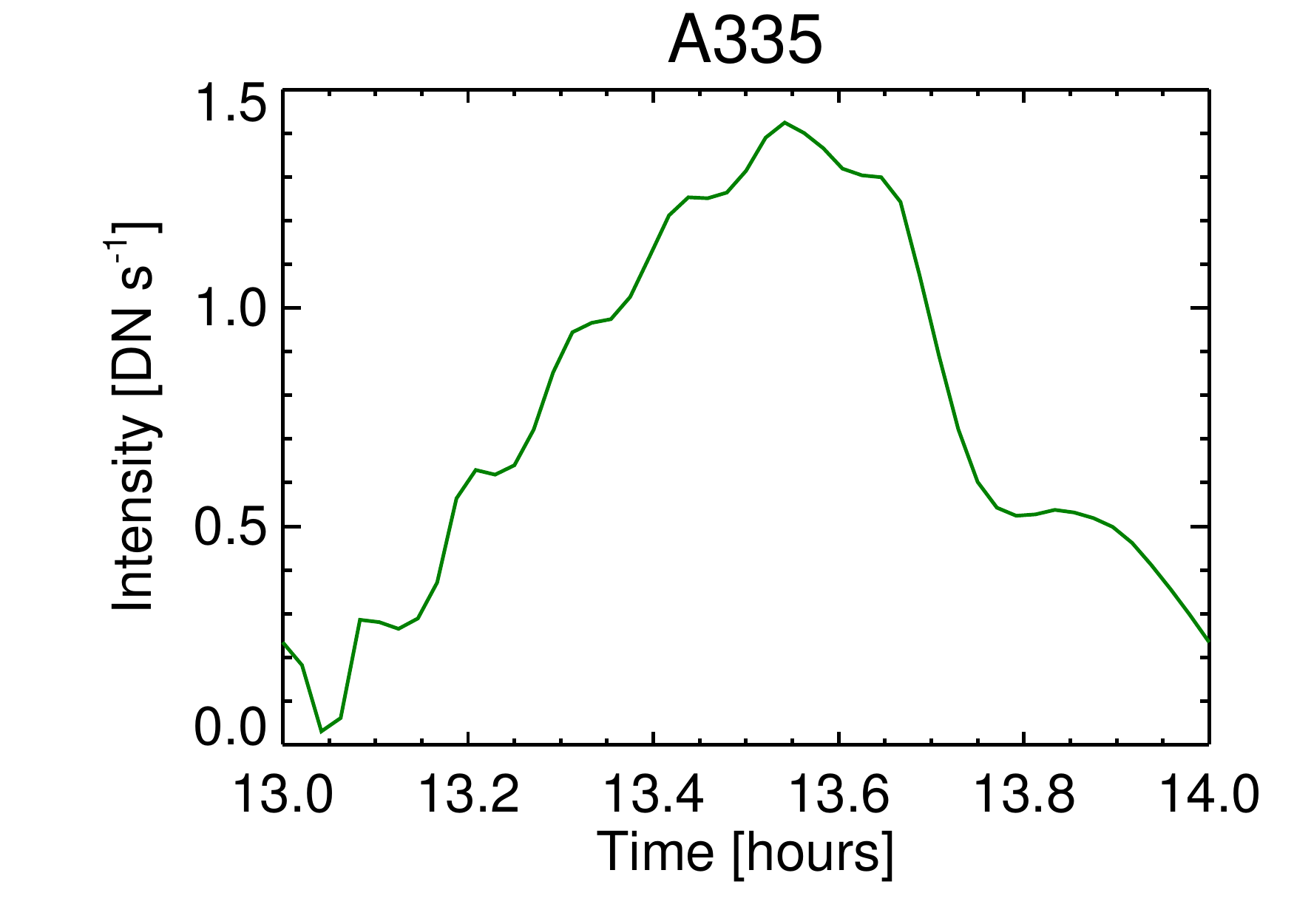}}}
\resizebox{.32\textwidth}{!}{\rotatebox{0}{\includegraphics{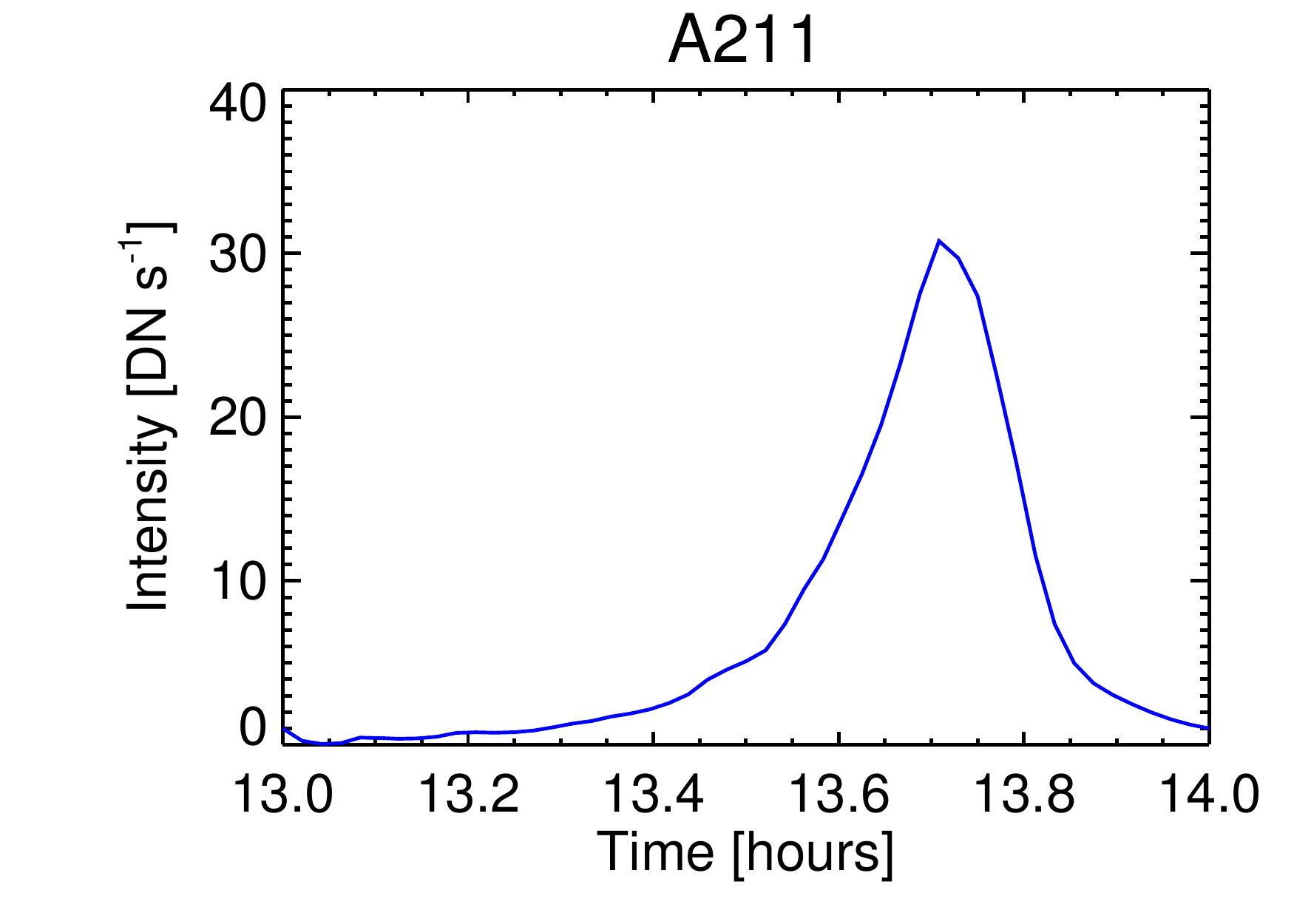}}}
\resizebox{.32\textwidth}{!}{\rotatebox{0}{\includegraphics{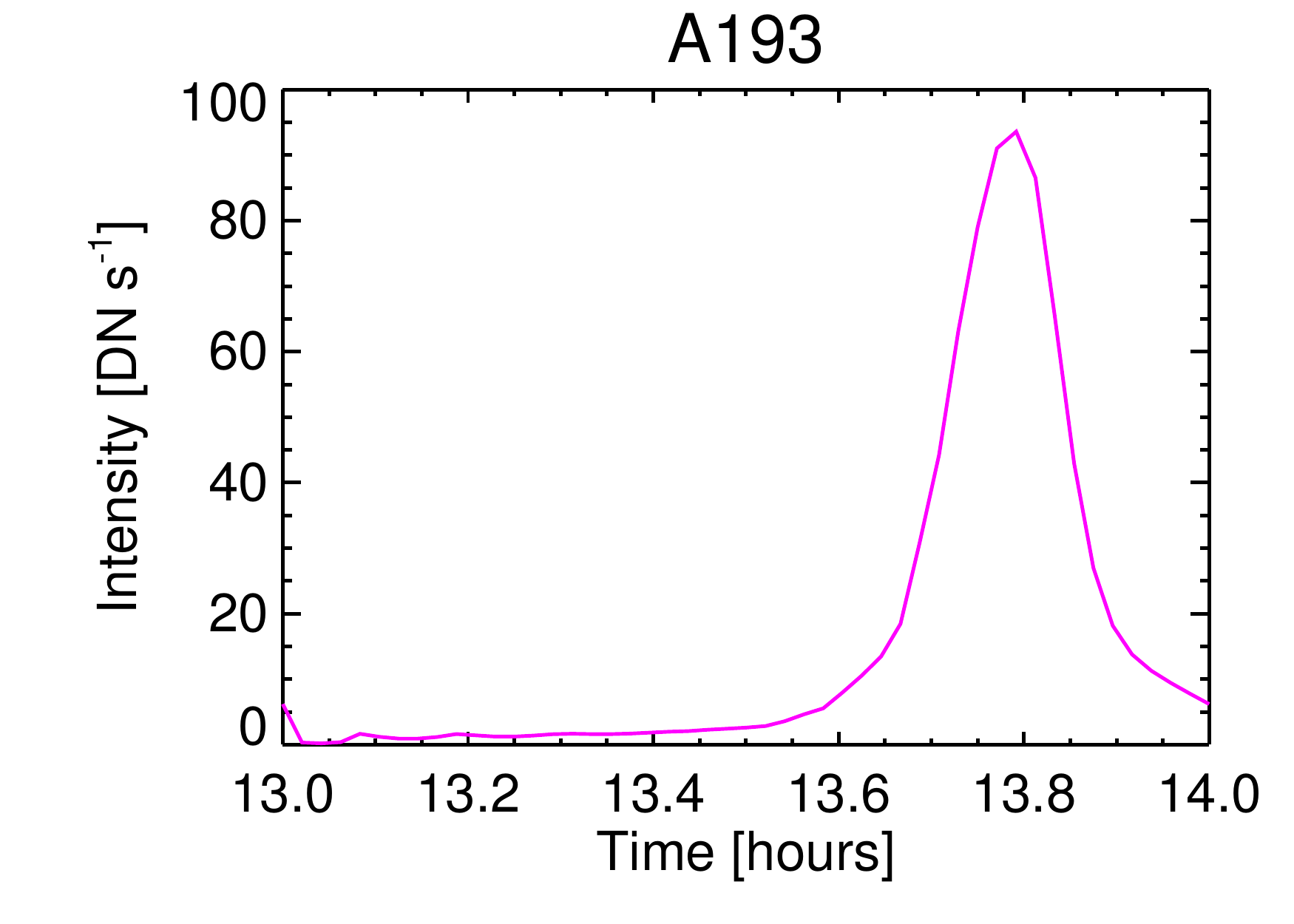}}}
\resizebox{.32\textwidth}{!}{\rotatebox{0}{\includegraphics{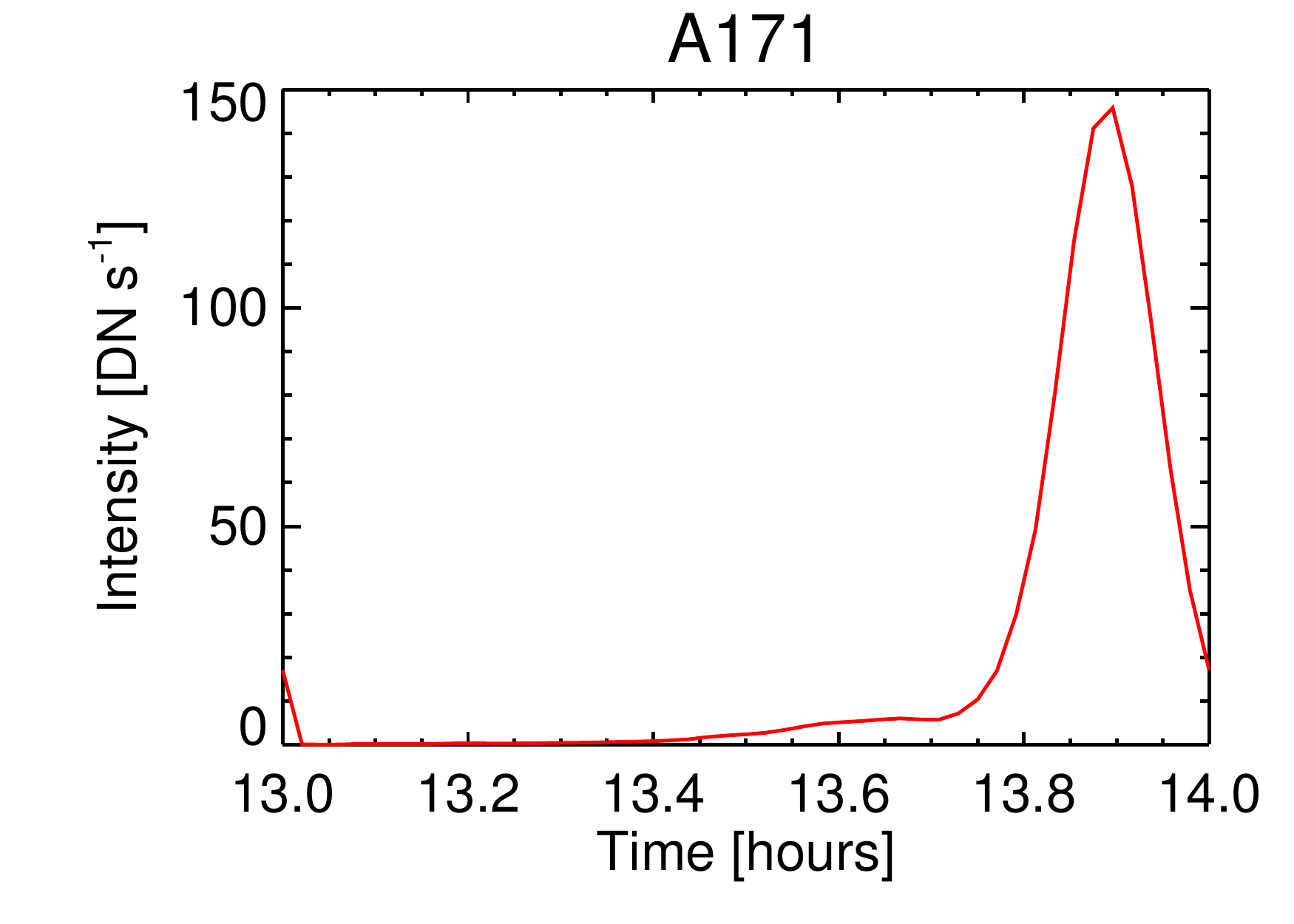}}}
\resizebox{.32\textwidth}{!}{\rotatebox{0}{\includegraphics{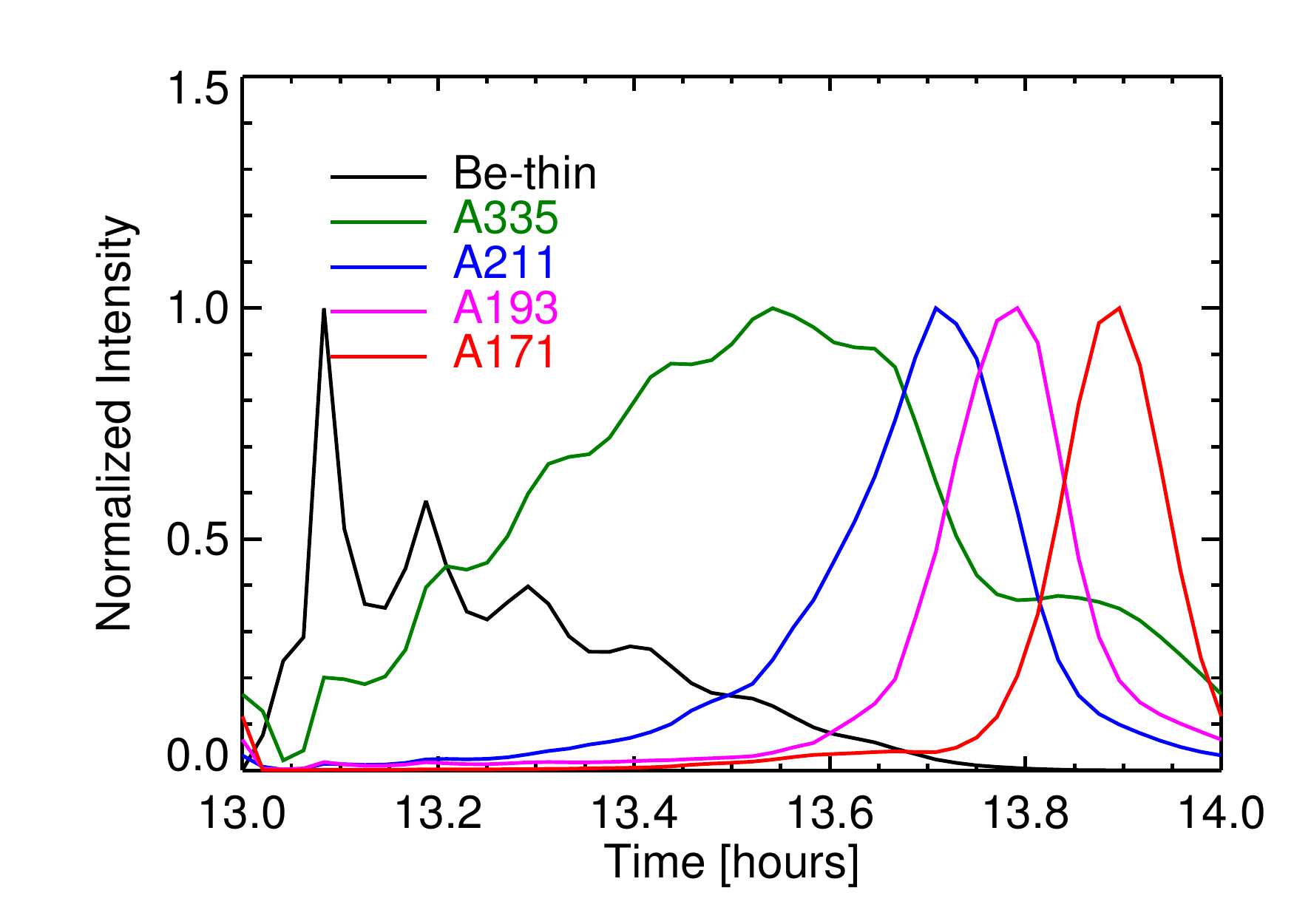}}}
\caption{The light curve of the apex of the loop in Hinode XRT Be-thin and four AIA channels for one cycle of the example impulsive heating simulation.  The bottom right panel shows the normalized light curves for all channels.}
\label{fig:example_lc_imp}
\end{center}
\end{figure*}

The calculated light curves for a single cycle are shown in Figures~\ref{fig:example_lc} and \ref{fig:example_lc_imp} for footpoint and impulsive heating simulations, respectively.    As expected, in both simulations, the hotter channels (i.e., XRT Be-thin and AIA 335\,\AA) precede the cooler channels in the light curves.   

The time delay between different channels has been measured several different ways, such as the difference in the time of the peak of the light curve in different channels or the difference in time when the light curve rises above half the maximum intensity.  More recently, \cite{viall2012} used the IDL C\_CORRELATE function to determine the time delay.  This function calculates the cross correlation value as a function of the time lag.  \cite{viall2012} took the time lag with the highest cross-correlation value.    When processing the data, \cite{viall2012} selected a window of time (2-12 hours) and allowed for time lags of up to $\pm$ 2 hours.  

In this paper, we are considering ideal solutions that have multiple cycles.  If we took a similar 12 hour light curve, it is just as likely, for instance, for the XRT Be-thin peak of one cycle to be correlated with the AIA 171\,\AA\ peak of the previous cycle as it is to be correlated with the peak of the next cycle.  Hence, we only consider the light cures of a single cycle of the solutions and allow time lags within that limit.  This implies that the time lags will always be less than the cycle time.  The time lags for these two example simulations are given in Table~\ref{tab:combs}.  We also calculate the ratio between the maximum intensity in each channel in the light curves.  The ratios of those two intensities are also given in Table~\ref{tab:combs}. 

Comparing the time lags from the footpoint and impulsive heating solutions, we find the time lags calculated from the footpoint solution tend to be much longer than the time lags calculated from the impulsive heating solution when comparing channels sensitive to significantly different temperatures.  For instance, the Be-thin-A171 channel pair have a predicted time lag of 4,230 s for the footpoint heated solution and 2,760 s for the impulsive heated solution.    The time delays are much closer when comparing two channels with similar temperatures; for instance, the A211-A171 channel pair predict almost identical time lags (780 s vs 630 s).  It is interesting to note that the footpoint-heated solution predicts a time lag of 0 in the A193-A171 channel pair, even though the A171 intensity peaks after the A193 intensity, as seen in the lower right panel of Figure \ref{fig:example_lc}.  This is because the solution never cools through the 193 or 171\,\AA\ passband; the minimum temperature the solution reaches is 1.6 MK.  The resulting light curve in A171 is broad, with the highest cross-correlation coefficient with the A193 light curve at  a time delay of 0 s.  The intensity ratios for the Be-thin to AIA channels tend to be larger in the footpoint-heated solutions, while the intensity ratios between the AIA channels tend to be larger in the impulsively-heated solution.   We completed identical analyses for every loop simulated in this study.  

\begin{deluxetable}{c|cc|cc}
\tablecaption{Footpoint vs Impulsive Heating Results for Example Solutions}
\tabletypesize{\scriptsize}
\tablewidth{0pt}
\tablehead{
\colhead{Channel } & \colhead{Footpoint}  & \colhead{Footpoint}& \colhead{Impulsive} & \colhead{Impulsive}\\
\colhead{Pair}         & \colhead{Time }    & \colhead{Int. Rat.}        & \colhead{Time}    & \colhead{Int. Rat.}\\
\colhead{[Ch. 1 - Ch. 2]}         & \colhead{Lag [s]}   & \colhead{[Ch. 2 / Ch. 1]}         & \colhead{Lag [s]}   & \colhead{[Ch. 2 / Ch. 1]}}
\startdata
        Be-thin-A335 &        990 &       0.40 &       1200 &       0.07 \\
        Be-thin-A211 &       3870 &       8.43 &       1950 &       1.47 \\
        Be-thin-A193 &       4830 &      19.60 &       2340 &       4.46 \\
        Be-thin-A171 &       4230 &       3.90 &       2760 &       6.91 \\
           A335-A211 &       1770 &      21.06 &        510 &      21.58 \\
           A335-A193 &       2430 &      48.99 &        840 &      65.71 \\
           A335-A171 &       2070 &       9.76 &       1230 &     101.73 \\
           A211-A193 &        690 &       2.33 &        240 &       3.04 \\
           A211-A171 &        780 &       0.46 &        630 &       4.71 \\
           A193-A171 &          0 &       0.20 &        390 &       1.55 \\

\enddata
\label{tab:combs}
\end{deluxetable}

\section{VARYING THE MAGNITUDE OF THE HEATING}

In this section, we vary the magnitude of the heating in both footpoint and impulsive heating simulations to determine the effect such a change has on the observables.  

\subsection{Footpoint heating}

\begin{figure*}[t!]
\begin{center}
\resizebox{.75\textwidth}{!}{\rotatebox{0}{\includegraphics{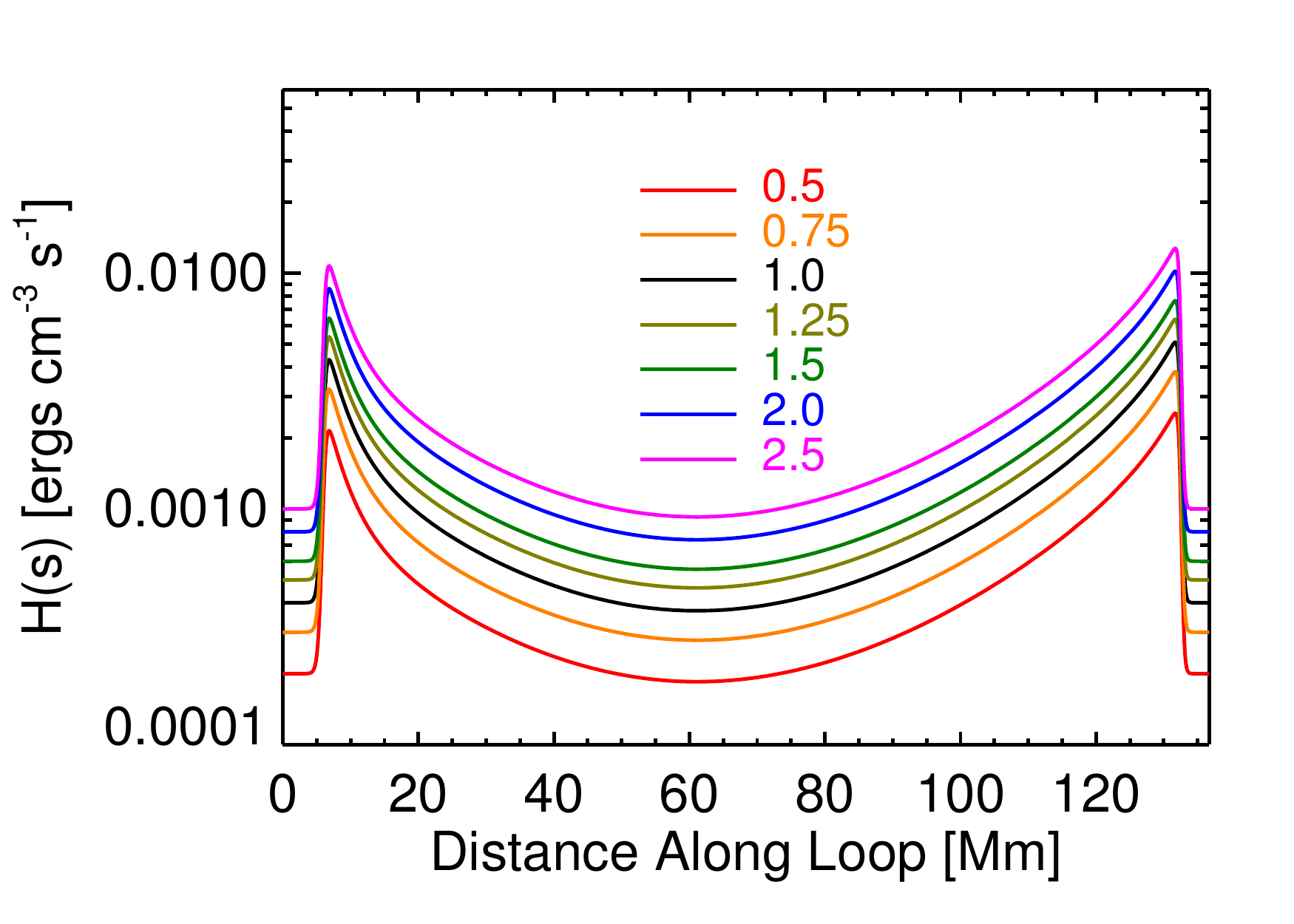}}}
\caption{The different steady heating functions used in this section.}
\label{fig:mag_heat}
\end{center}
\end{figure*}

The goal of this section is to investigate how changing the magnitude of the heating impacts the solutions of the hydrodynamic equations for footpoint heating.  We vary the magnitude of the heating by simply taking the original heating function, shown in Figure~\ref{fig:b_a_h}, and multiplying it by a factor, i.e.,
\begin{equation}
H_{\rm FP}(s) = M H_{\rm FP0}(s)
\end{equation}
where $H_{\rm FP0}(s)$ is the original heating function and $M$ is a multiplier that we vary between 0.5 and 2.5.  The heating functions we use in this section are shown in Figure~\ref{fig:mag_heat}.

\begin{figure*}[t!]
\begin{center}
\resizebox{1\textwidth}{!}{\rotatebox{90}{\includegraphics{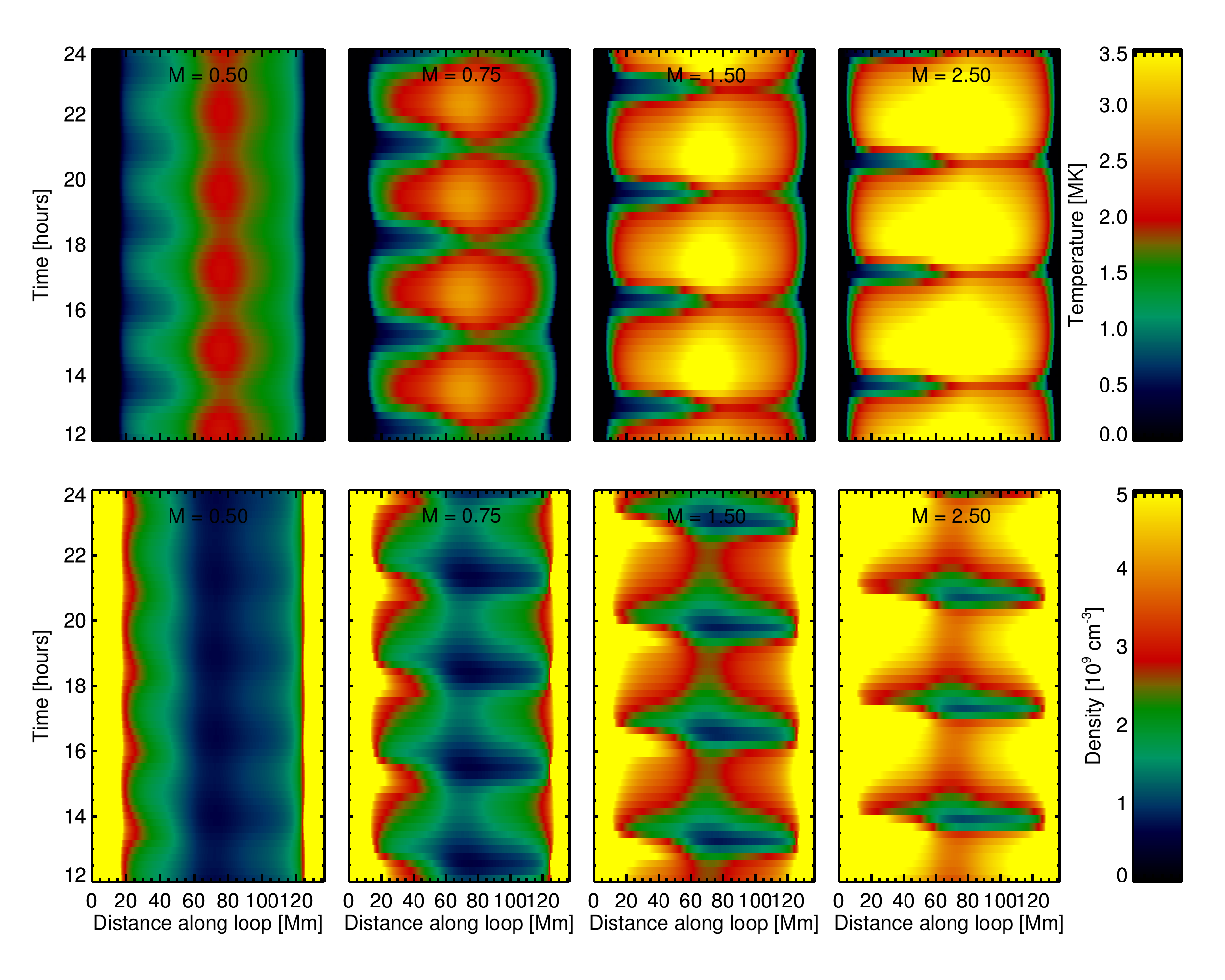}}}
\caption{A comparison of the temperature (top row) and density (bottom row) evolution for various heating magnitudes in footpoint heating simulations.  Note that this does not show the evolution for all heating magnitudes considered in this paper.}
\label{fig:temp_map_mag}
\end{center}
\end{figure*}

\begin{figure*}[t!]
\begin{center}
\resizebox{.49\textwidth}{!}{\rotatebox{0}{\includegraphics{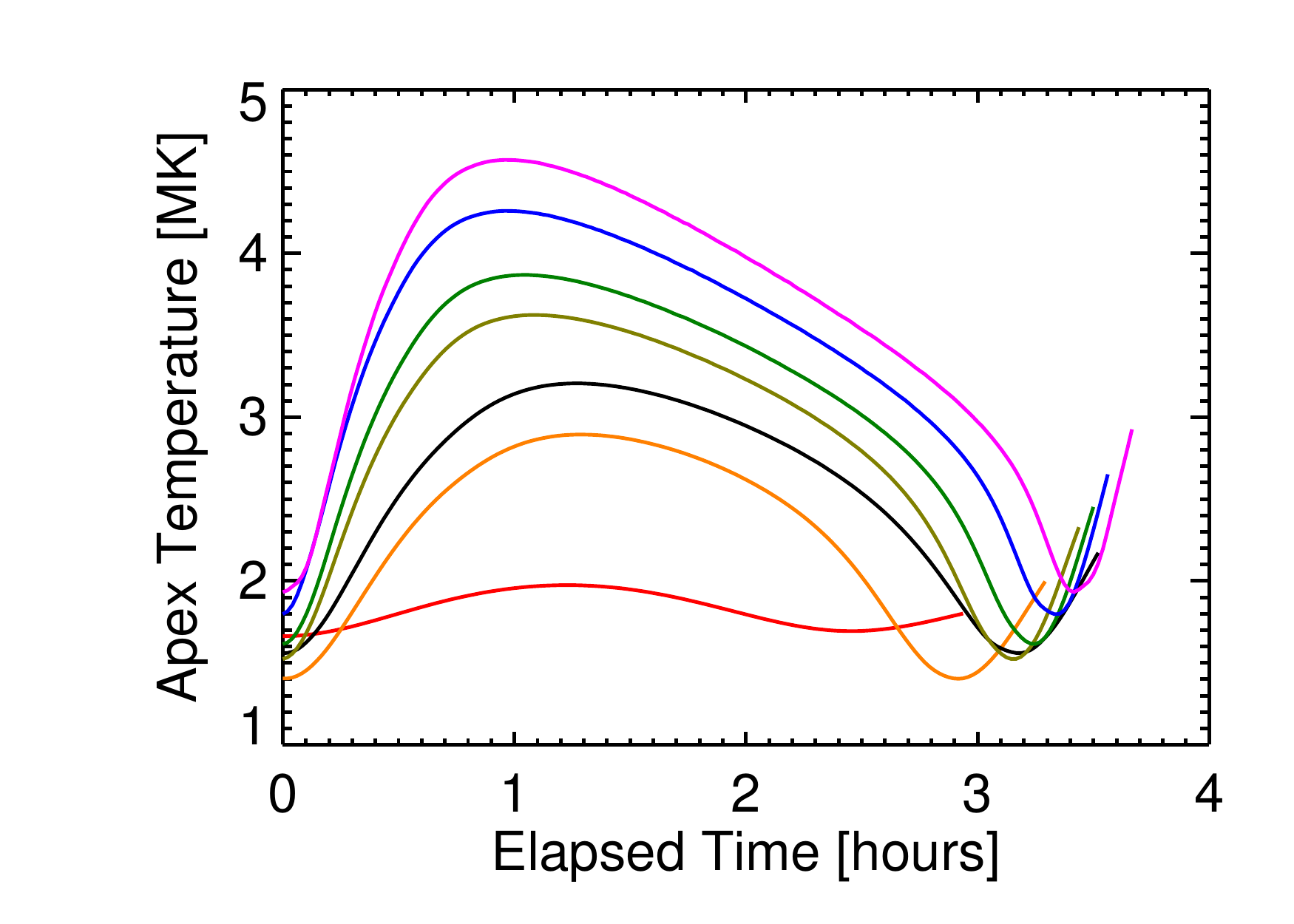}}}
\resizebox{.49\textwidth}{!}{\rotatebox{0}{\includegraphics{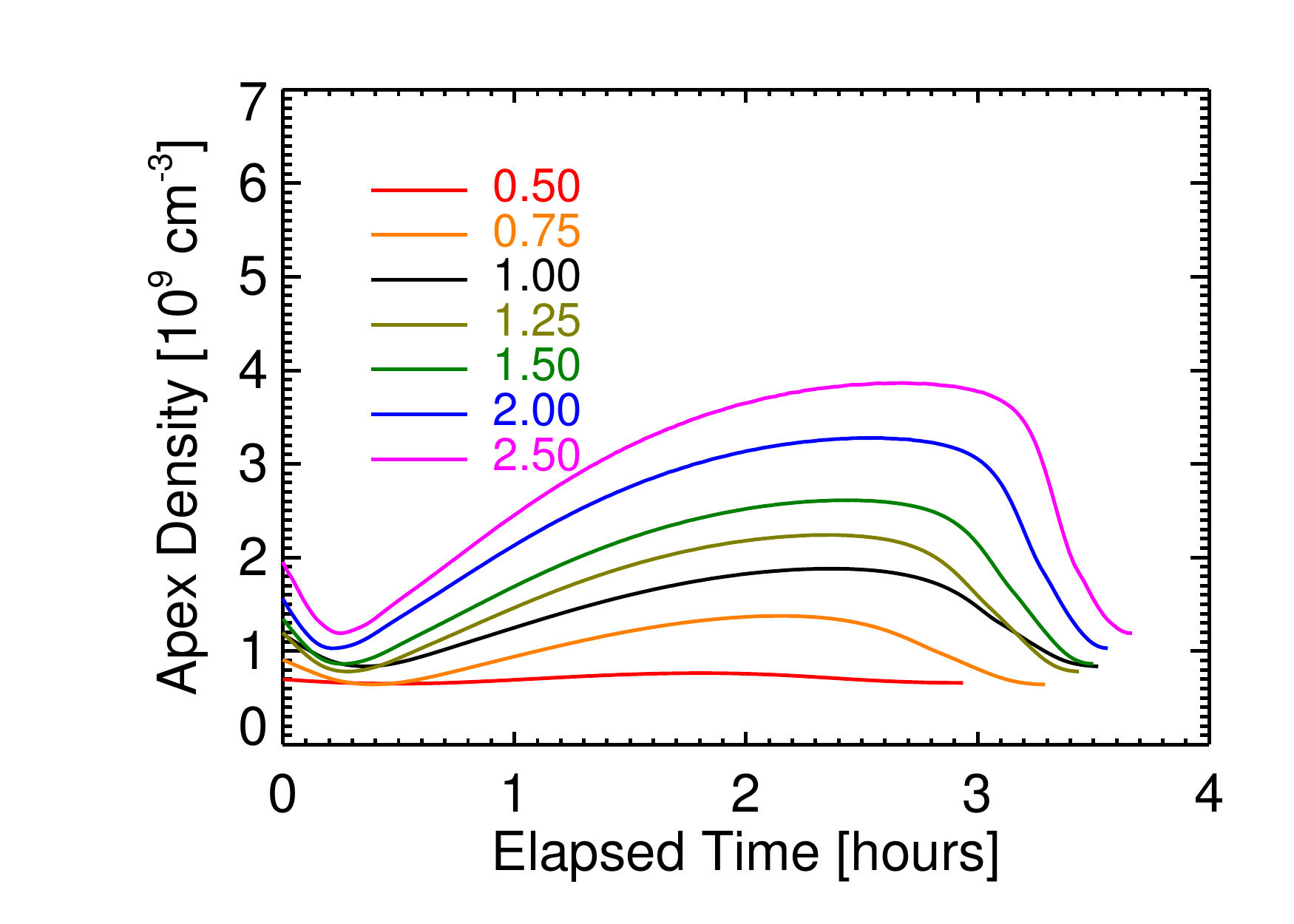}}}
\caption{Apex temperature and density values for different heating magnitudes in footpoint heating simulations.  We are showing slightly more than one cycle of the solution (from a minimum in apex temperature to the minimum in apex density).  We have shifted the solutions to start at the same time.  }
\label{fig:apex_mag_footpoint}
\end{center}
\end{figure*}

It is interesting to examine how changing the magnitude of the heating, but not the stratification, impacts the temperature and density of the solution.   These are shown in Figures~\ref{fig:temp_map_mag}.  The apex values of the temperature and density (calculated over the upper 10\% of the loop) are shown in Figure~\ref{fig:apex_mag_footpoint}.  For each simulation, we show slightly more the one cycle of data (from the minimum apex temperature to the minimum apex density) and we shift the solutions to all start at the same time.  

For all heating magnitudes, we find the solutions are in thermal non-equilibrium with incomplete condensations.    In general, the magnitudes of the temperature, density, and cycle period increase for higher heating magnitudes.   Comparing the evolution of the apex values in Figure~\ref{fig:apex_mag_footpoint}, we see that the maximum values for the temperature and density increase, but the minimum values for the temperature and density increase as well.  

\subsection{Impulsive Heating}

In this section, we show the solutions to the hydrodynamic equations for impulsive heating.  We vary the magnitude of heating in the same way as for the thermal non-equilibrium solution, i.e., by multiplying the original heating function by a factor, $M$, and calculating the resulting impulsive-heating parameters, i.e.,
\begin{equation}
H_{\rm imp0} / \tau = \frac{ \int_0^{\rm{s_{max}}} M H_{\rm FP0}(s) A(s) ds }{\delta  \int_0^{\rm{s_{max}}} A(s) ds}.
\end{equation}
We find that $\tau = 1$ hr is an acceptable time for all solutions to return to $\sim 0.5$ MK.    The resulting $H_{\rm imp0}$ for each value of $M$ are given in Table~\ref{tab:imp_params}.   The duration of the heating, $2\delta = 174$\,s, is the same for all simulations.

\begin{deluxetable}{cc}
\tablecaption{ Impulsive Heating Parameters}
\tabletypesize{\scriptsize}
\tablewidth{0pt}
\tablehead{
\colhead{ $M$} & \colhead{$H_{\rm imp0}$ [ergs cm$^{-3}$ s$^{-1}$]}  }
\startdata
0.5 & 1.58 $\times 10^{-2}$ \\
0.75 & 2.38$\times 10^{-2}$  \\
1.0 & 3.17$\times 10^{-2}$  \\
1.25 & 3.96$\times 10^{-2}$  \\
1.50 & 4.75$\times 10^{-2}$ \\
2.00 & 6.34$\times 10^{-2}$ \\
2.50 & 7.92$\times 10^{-2}$ \\
\enddata
\label{tab:imp_params}
\end{deluxetable}

\begin{figure*}[t!]
\begin{center}
\resizebox{1\textwidth}{!}{\rotatebox{90}{\includegraphics{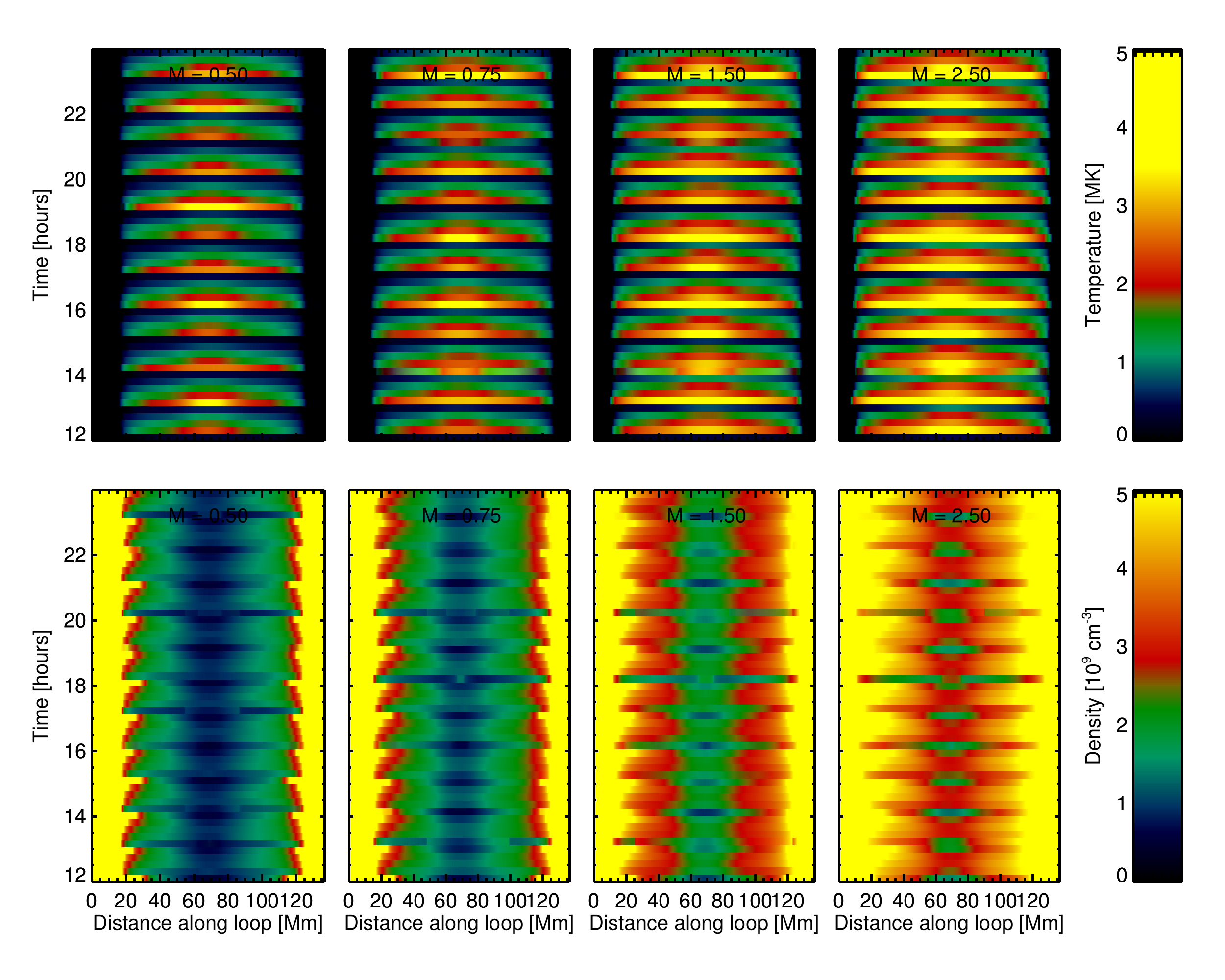}}}
\caption{A comparison of the temperature (top row) and density (bottom row) evolution for various heating magnitudes in impulsive heating simulations.  Note that this does not show the evolution for all heating magnitudes considered in this paper.}
\label{fig:temp_map_mag_imp}
\end{center}
\end{figure*}

\begin{figure*}[t!]
\begin{center}
\resizebox{.49\textwidth}{!}{\rotatebox{0}{\includegraphics{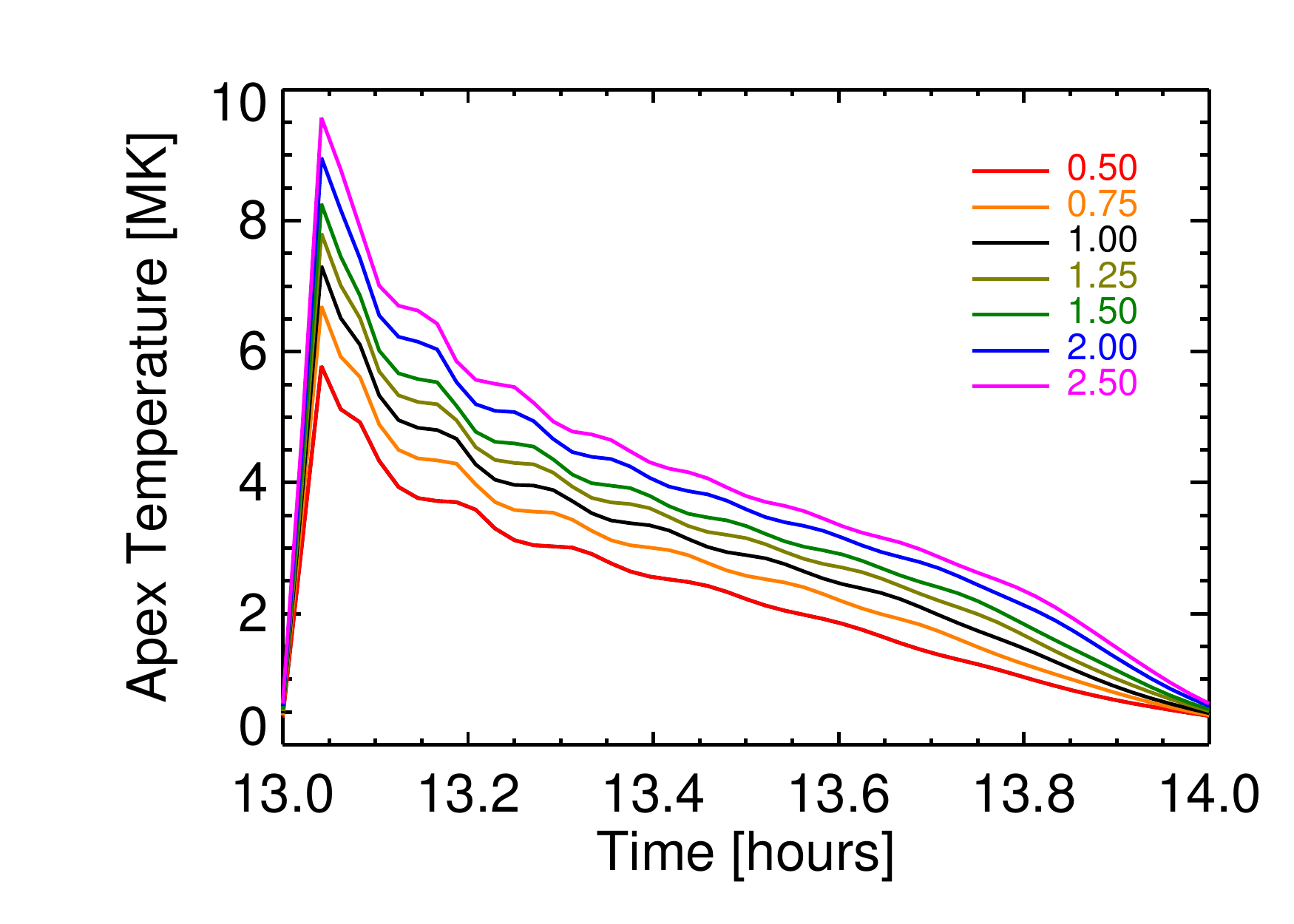}}}
\resizebox{.49\textwidth}{!}{\rotatebox{0}{\includegraphics{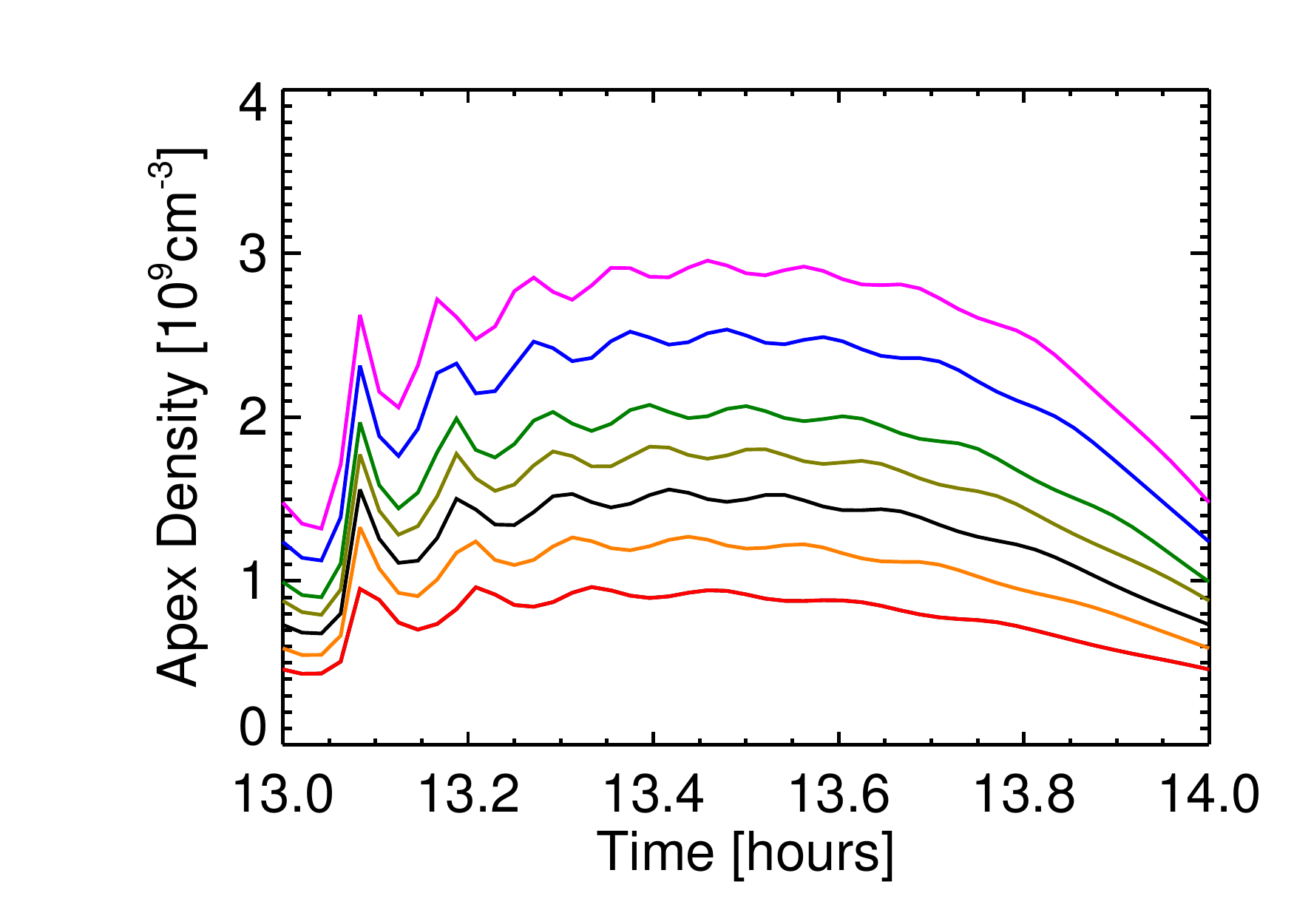}}}
\caption{Apex temperature and density values for different heating magnitude in impulsive heating simulations for a single cycle of heating.}
\label{fig:apex_mag_imp}
\end{center}
\end{figure*}

The resulting temperature and density evolutions of the simulations are shown in Figures~\ref{fig:temp_map_mag_imp} for each $M$.  The apex values are shown in Figure~\ref{fig:apex_mag_imp}.   The (artificial) cycle times are the same for all simulations (by design).  The maximum temperature  and density of the simulations increase as the  magnitudes of the impulsive heating event increase.

\subsection{Observables}

\begin{figure*}[t!]
\begin{center}
\resizebox{.32\textwidth}{!}{\rotatebox{0}{\includegraphics{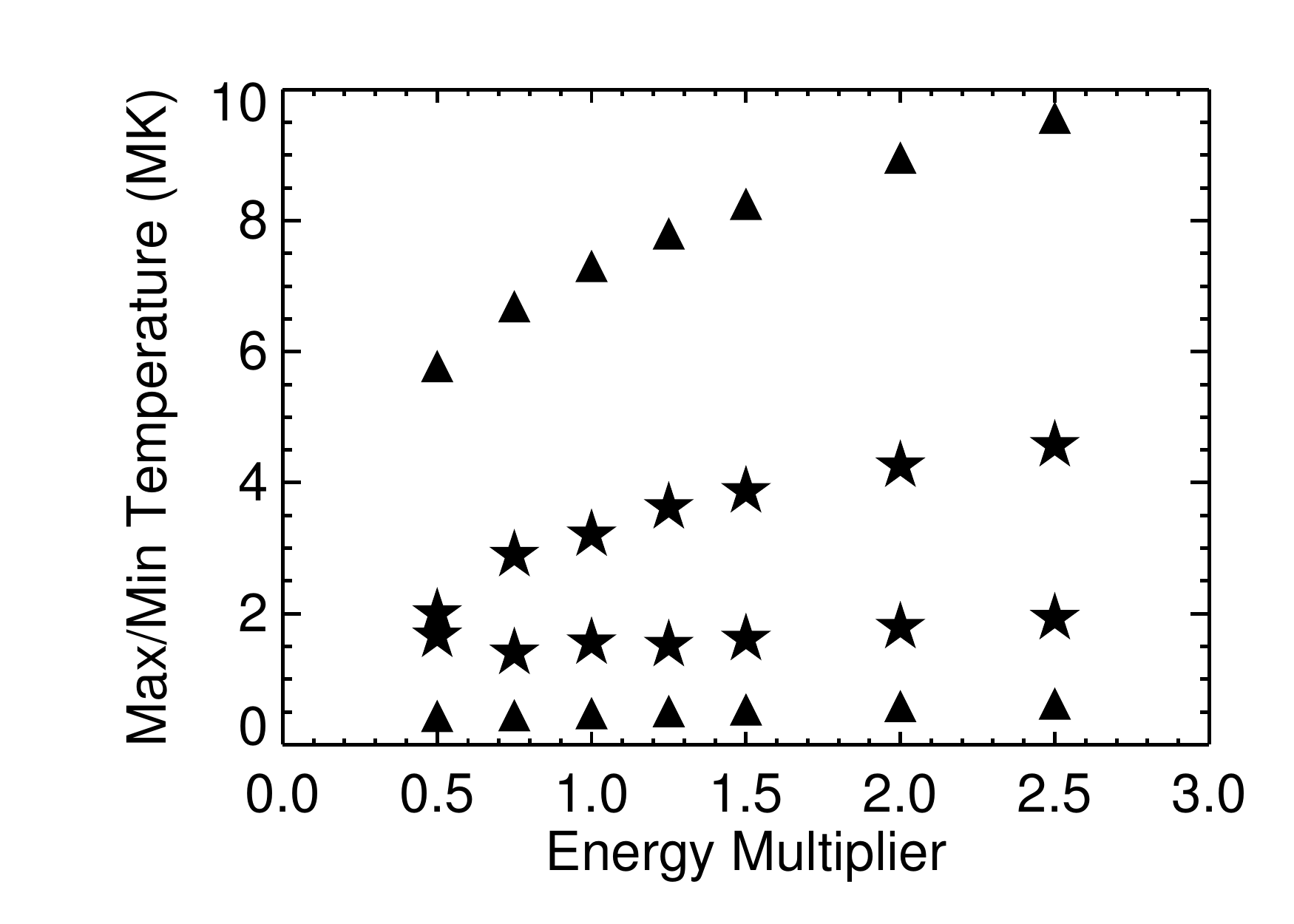}}}
\resizebox{.32\textwidth}{!}{\rotatebox{0}{\includegraphics{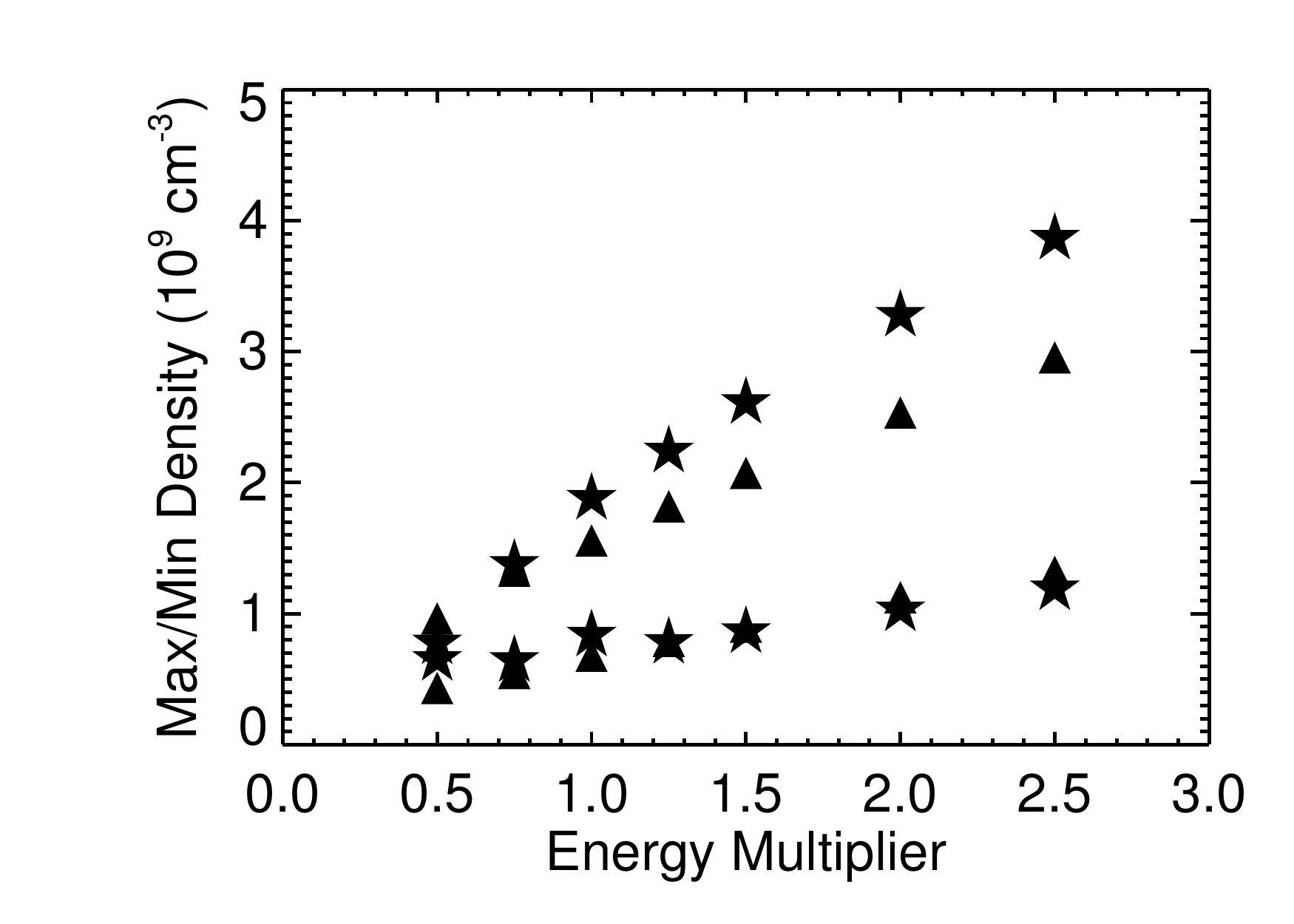}}}
\resizebox{.32\textwidth}{!}{\rotatebox{0}{\includegraphics{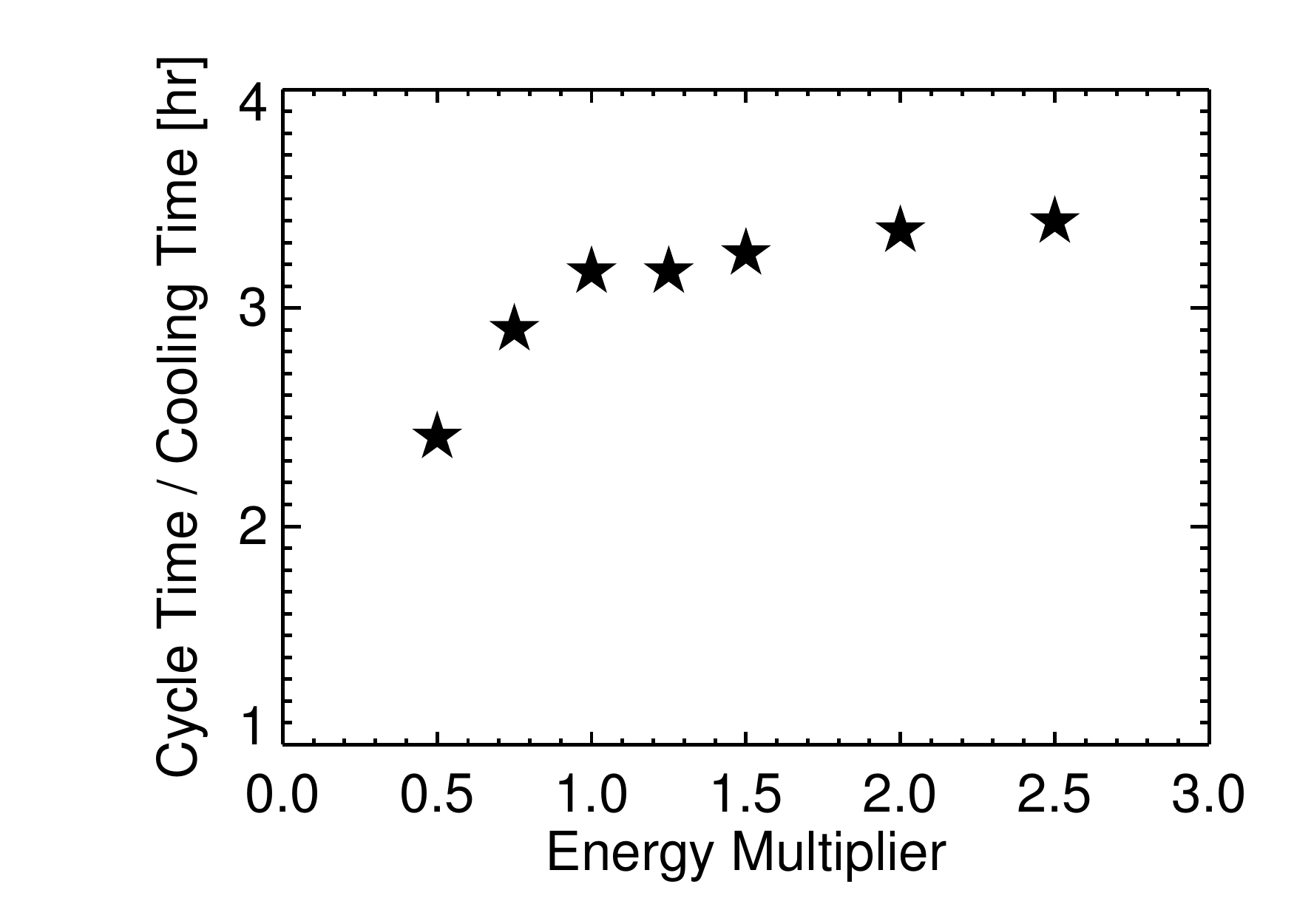}}}
\caption{A comparison of the minimum and maximum temperatures (left) and density (center) for the footpoint (star) and impulsive (triangle) simulations as a function of the heating magnitude multiplier.  The cycle time of the footpoint heated solutions are shown in the right panel.}
\label{fig:max_min_dens_temp}
\end{center}
\end{figure*}

In the above two subsections, we calculated a series of solutions to the hydrodynamic equations with different magnitudes of heating.  For all cases, the maximum temperature achieved in each simulation increased with heating magnitude; a comparison is shown in the left panel of Figure~\ref{fig:max_min_dens_temp}.  The minimum and maximum temperatures of the footpoint-heated solutions are shown with stars, and those of the impulsively-heated solutions are shown with triangles.  For the footpoint-heated solution, the minimum temperature increases slightly as well.  For the impulsively-heated solution, the minimum temperature is roughly 0.5\,MK, which we used to define the reheat time.  A similar pattern can be seen in the minimum and maximum apex density, shown in the middle panel of Figure~\ref{fig:max_min_dens_temp}.  

At least for this series of simulations, only the impulsive heating shows very hot ($>$ 4 MK) plasma. The observation of very hot plasma may be a discriminator between the two mechanisms.  Another distinction is the cycle time shown in the right panel of Figure~\ref{fig:max_min_dens_temp}.    All impulsively-heated solutions completed a full cycle (i.e., the apex temperature dropped to $\sim$ 0.5 MK) in 1 hour.  For footpoint-heated solutions, the cycle time varied between 2.4 and 3.4 hours.  We can conclude from this series of simulations that cycles from  the footpoint solutions are much longer than the cooling time of a plasma heated by simple impulsive heating.

\begin{figure*}[t!]
\begin{center}
\resizebox{.32\textwidth}{!}{\rotatebox{0}{\includegraphics{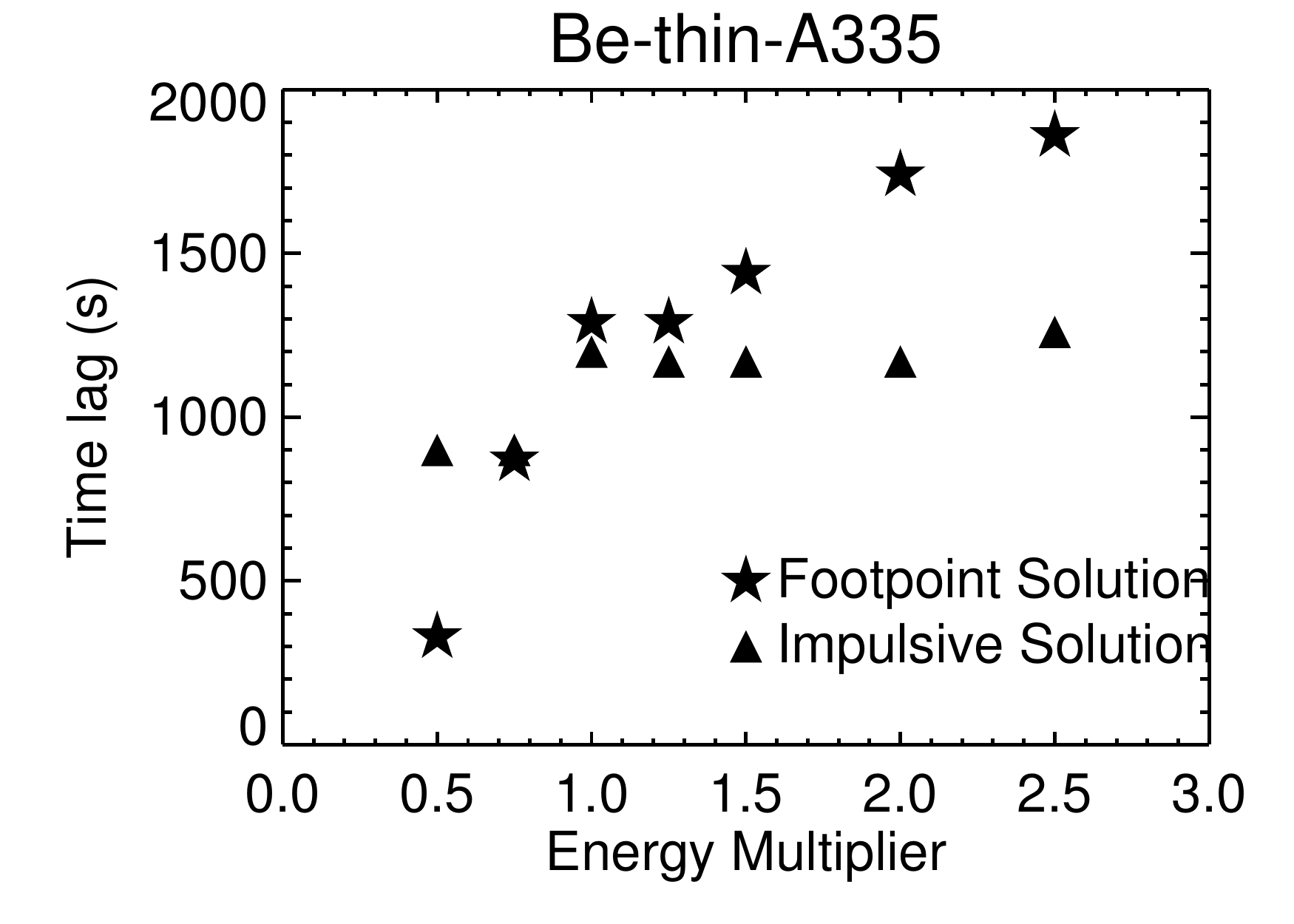}}}
\resizebox{.32\textwidth}{!}{\rotatebox{0}{\includegraphics{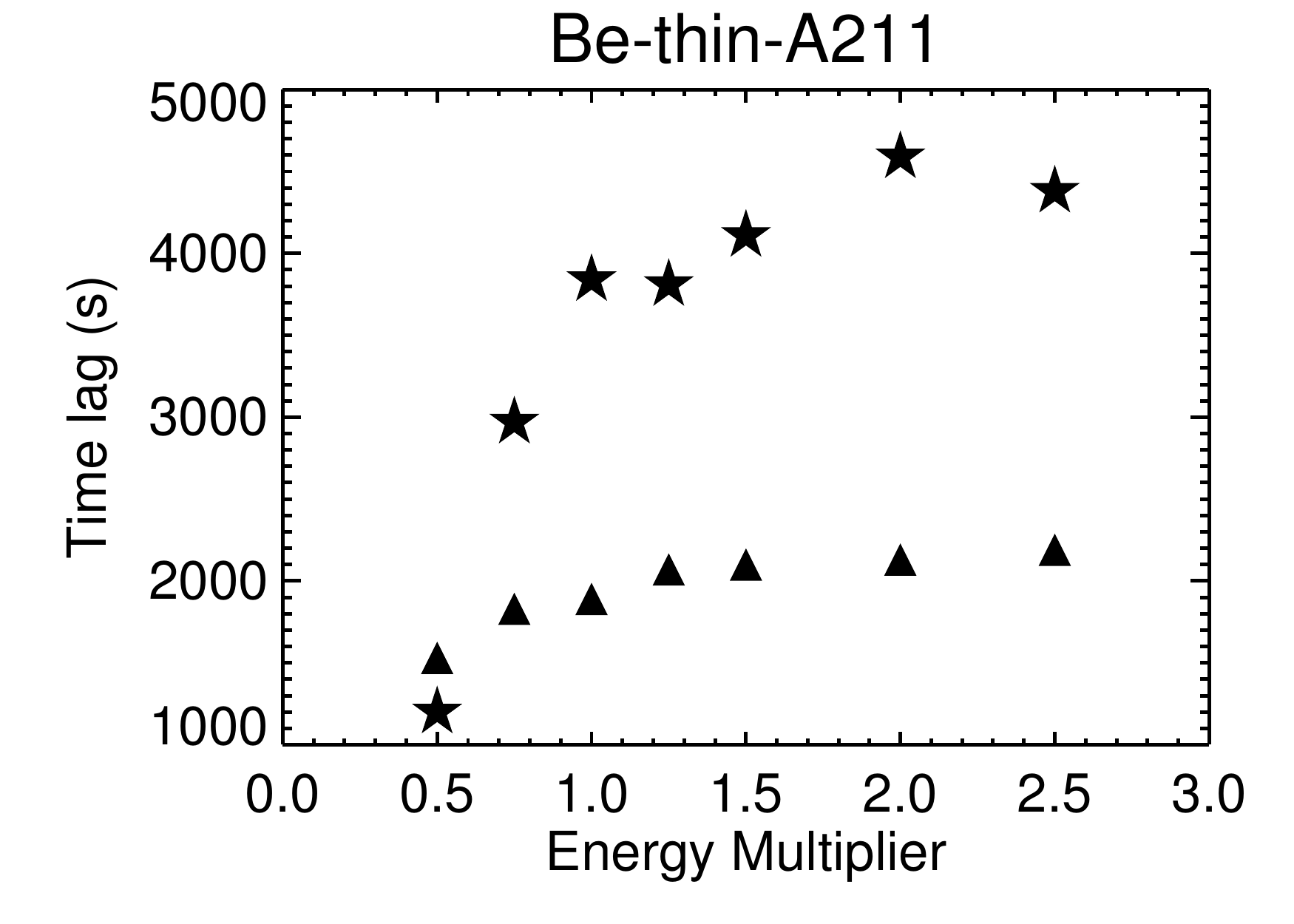}}}
\resizebox{.32\textwidth}{!}{\rotatebox{0}{\includegraphics{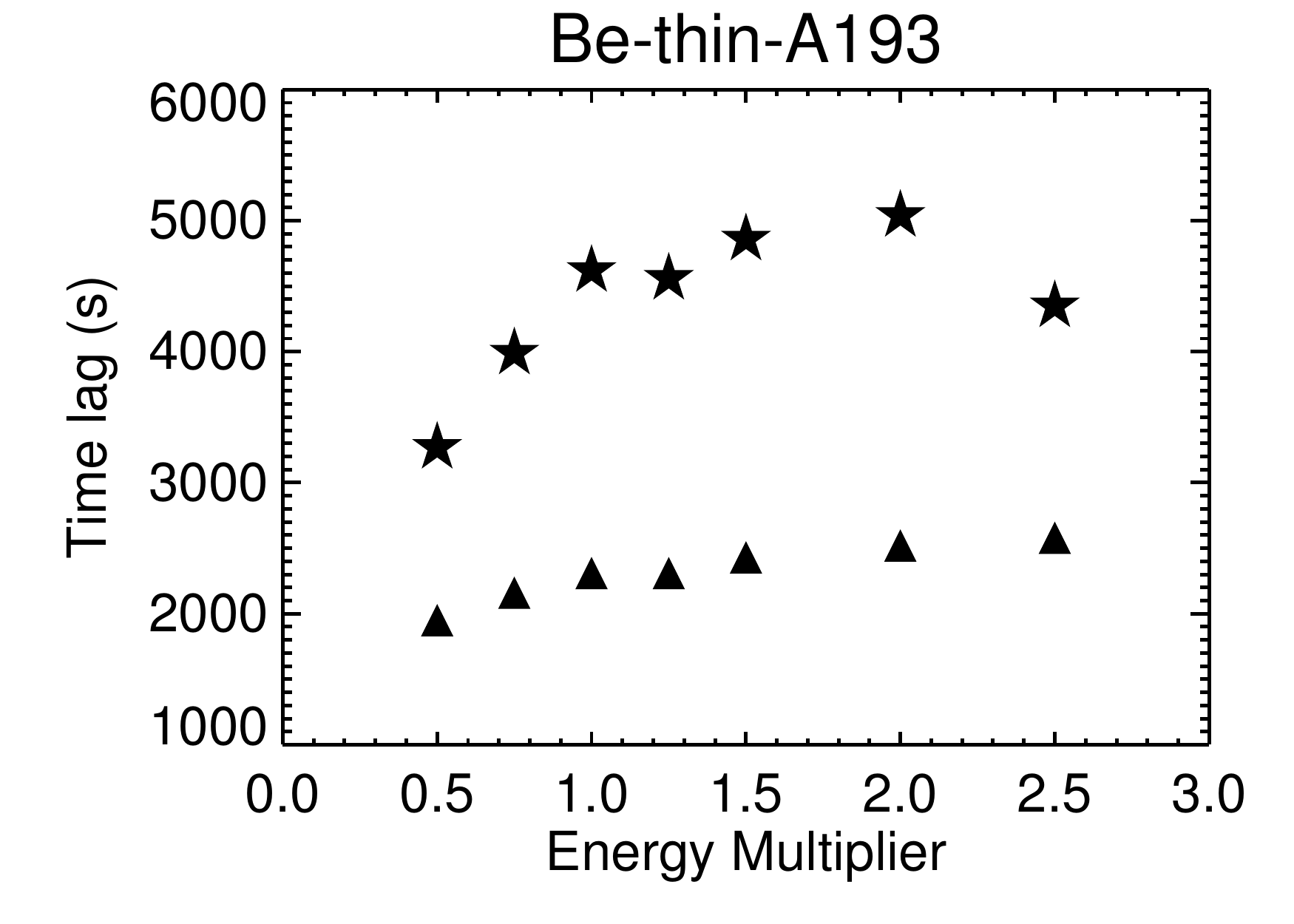}}}
\resizebox{.32\textwidth}{!}{\rotatebox{0}{\includegraphics{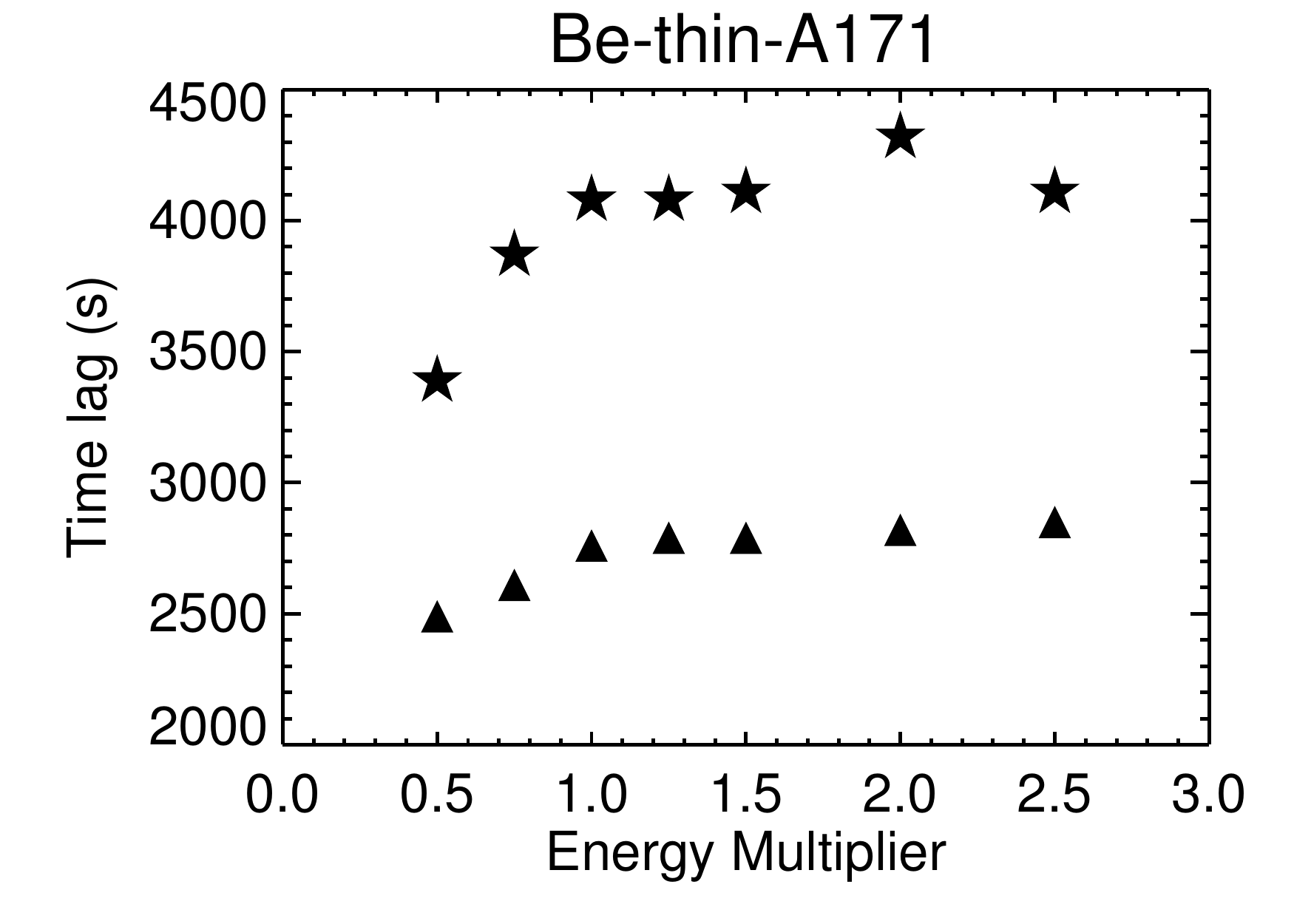}}}
\resizebox{.32\textwidth}{!}{\rotatebox{0}{\includegraphics{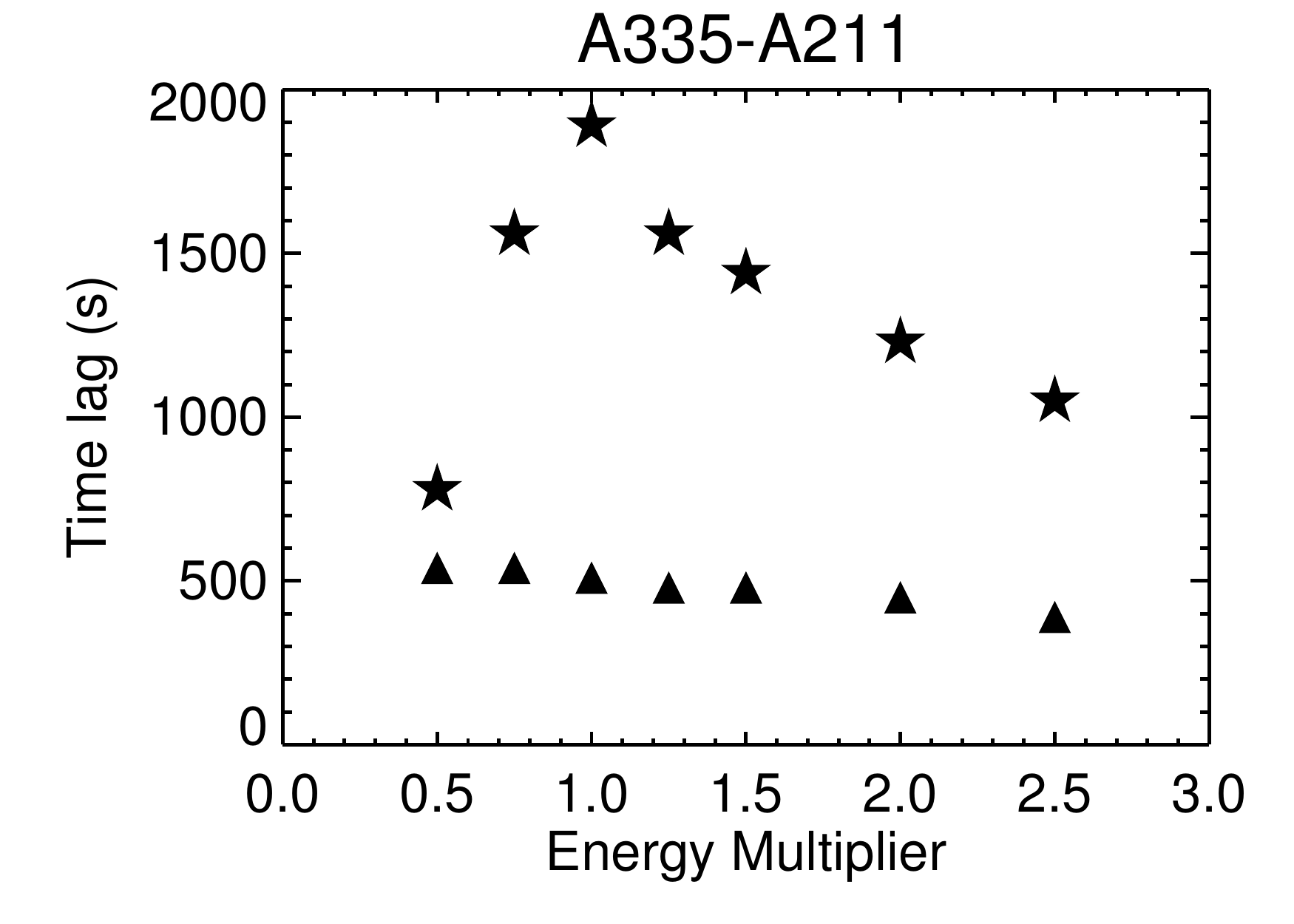}}}
\resizebox{.32\textwidth}{!}{\rotatebox{0}{\includegraphics{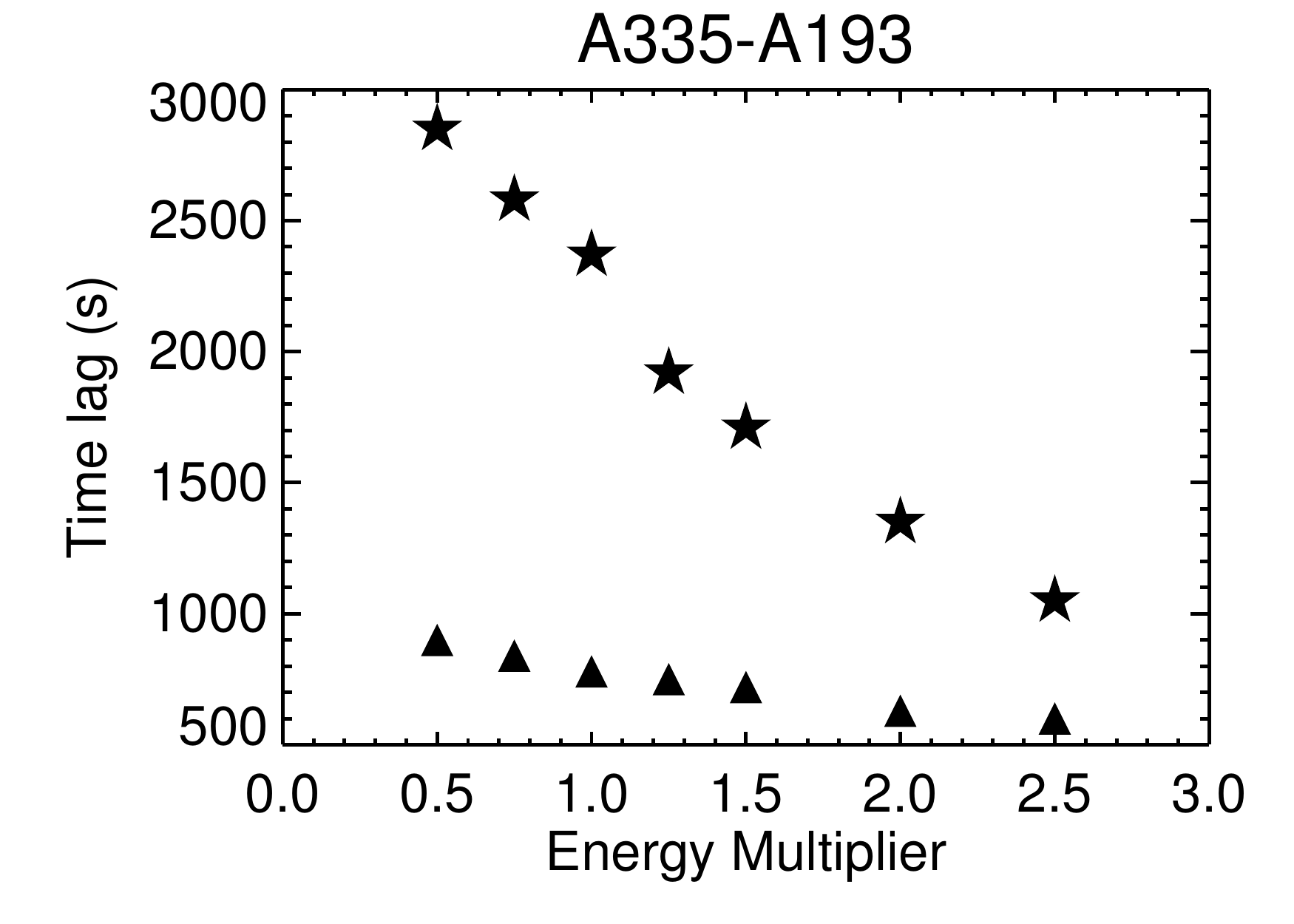}}}
\resizebox{.32\textwidth}{!}{\rotatebox{0}{\includegraphics{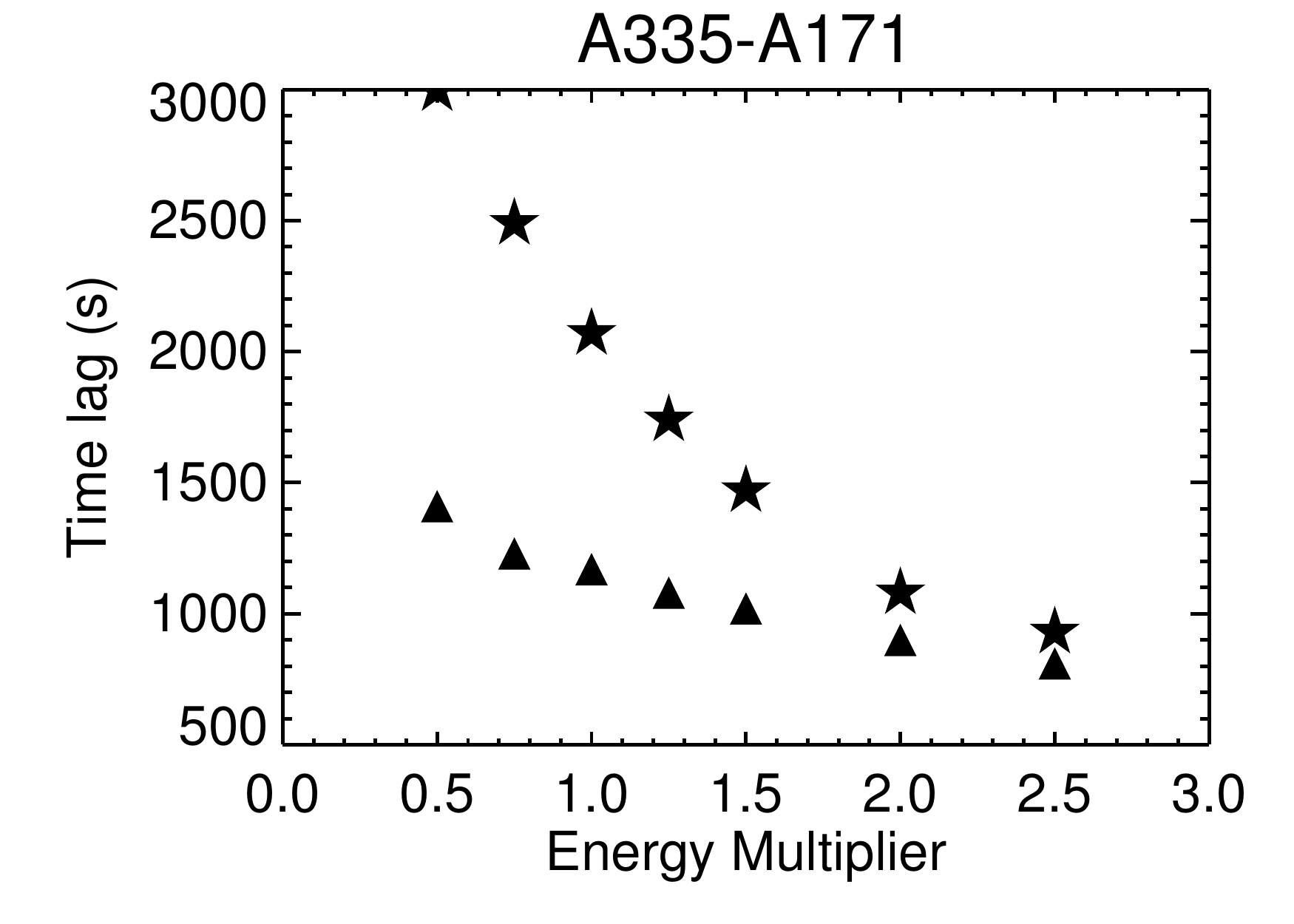}}}
\resizebox{.32\textwidth}{!}{\rotatebox{0}{\includegraphics{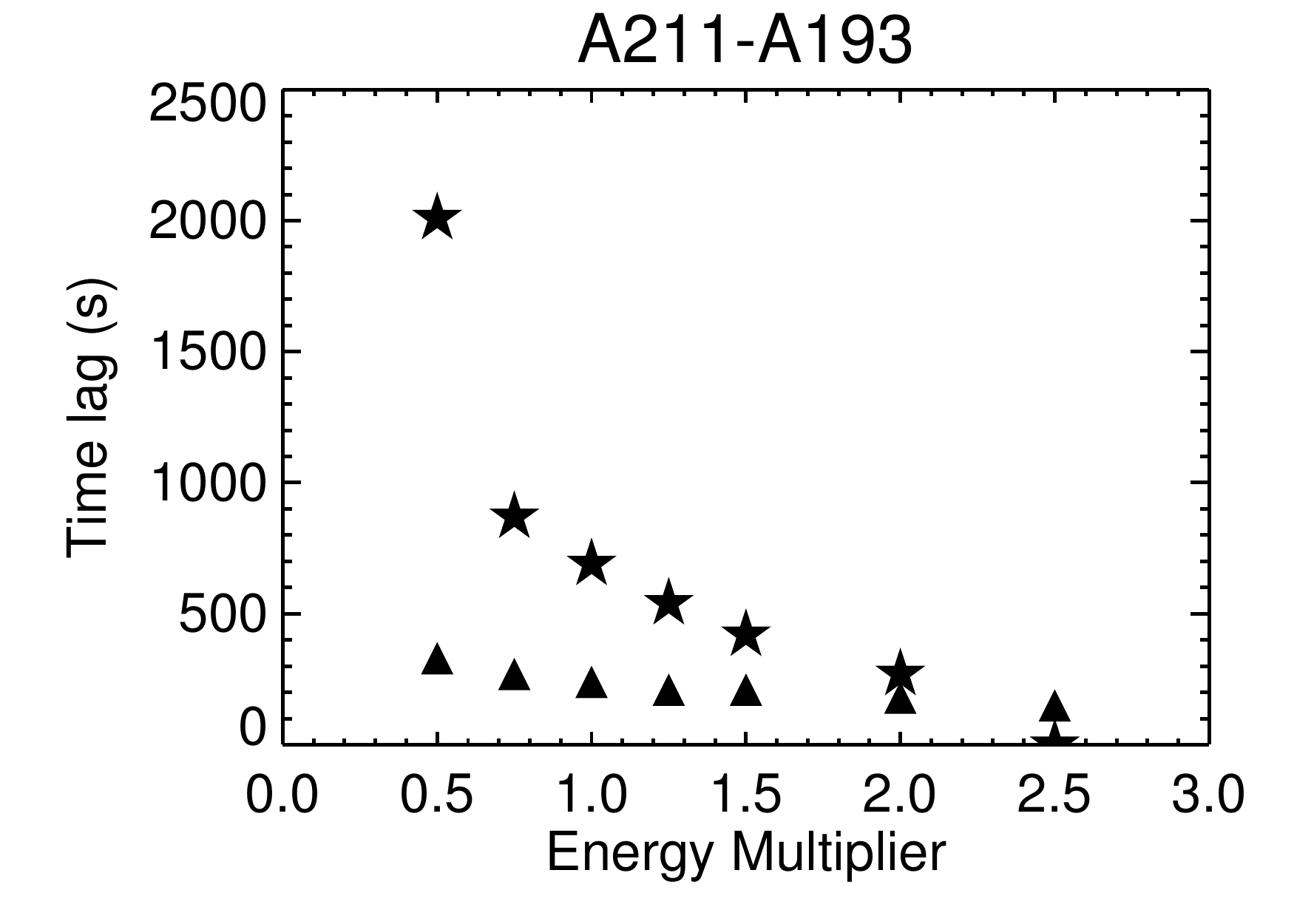}}}
\resizebox{.32\textwidth}{!}{\rotatebox{0}{\includegraphics{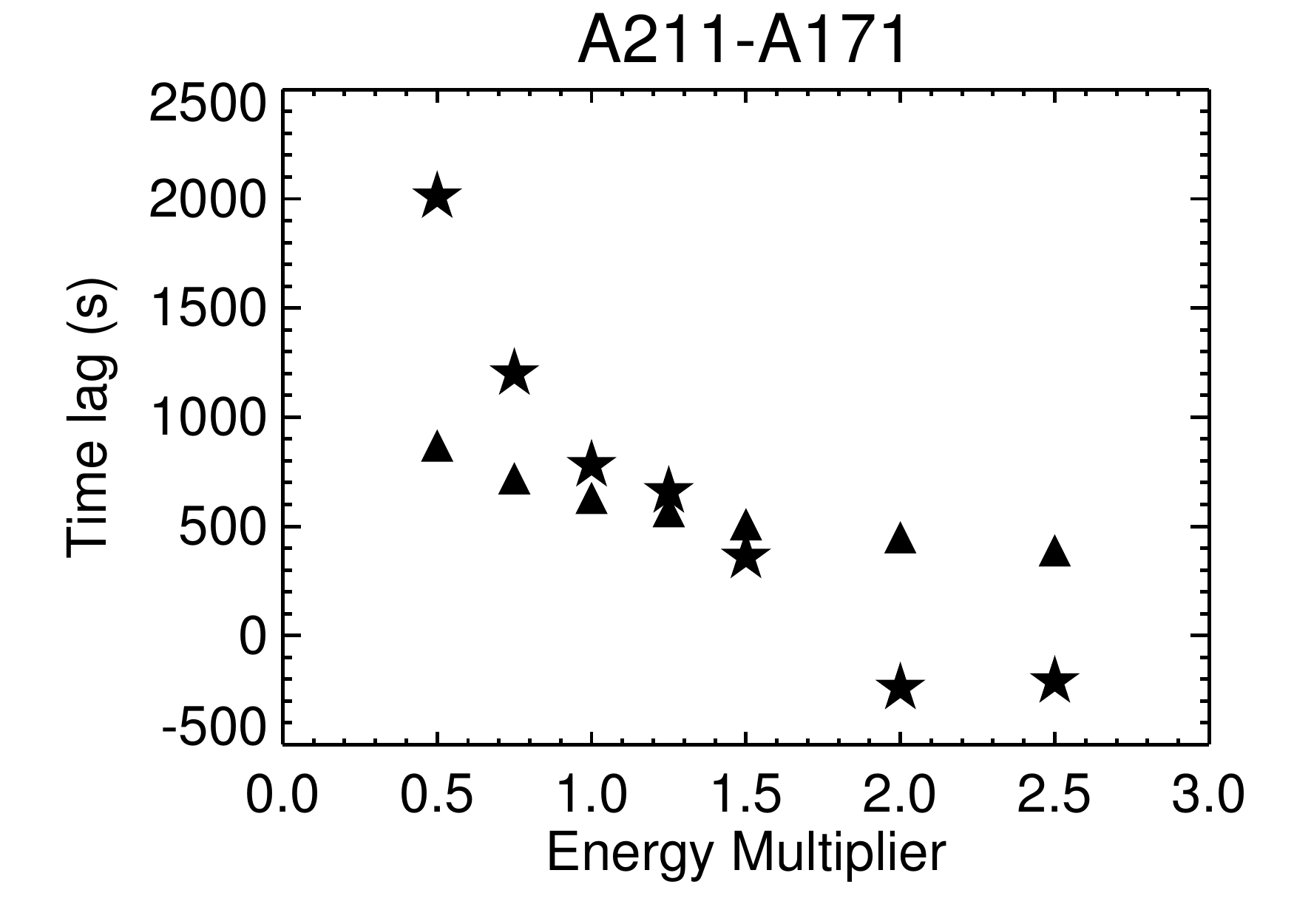}}}
\resizebox{.32\textwidth}{!}{\rotatebox{0}{\includegraphics{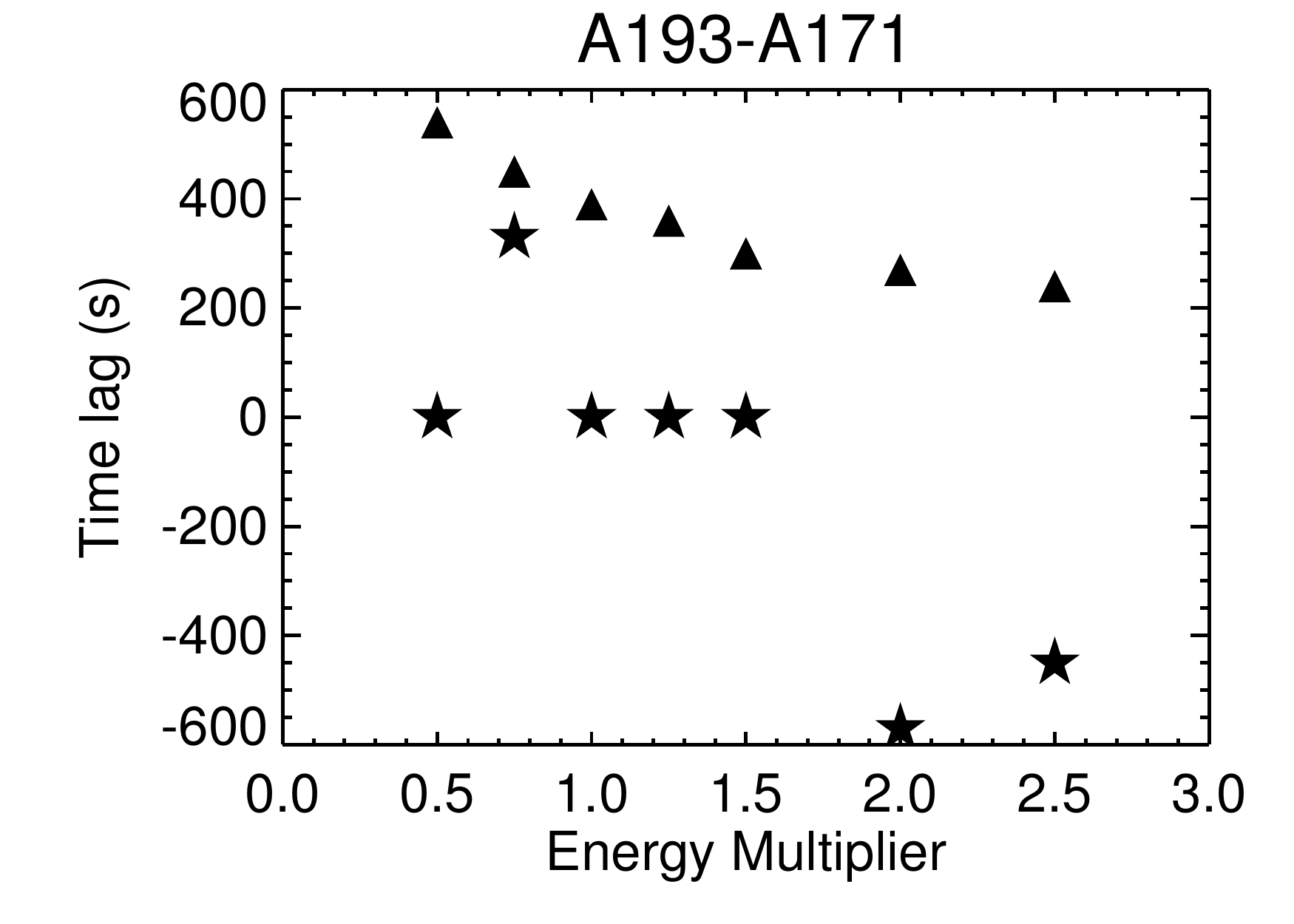}}}
\caption{Time lags for the different channel combinations and different heating magnitudes.  The footpoint solutions are shown with stars, the impulsive heating solutions are shown with triangles.}
\label{fig:time_lag_mag}
\end{center}
\end{figure*}

\begin{figure*}[t!]
\begin{center}
\resizebox{.32\textwidth}{!}{\rotatebox{0}{\includegraphics{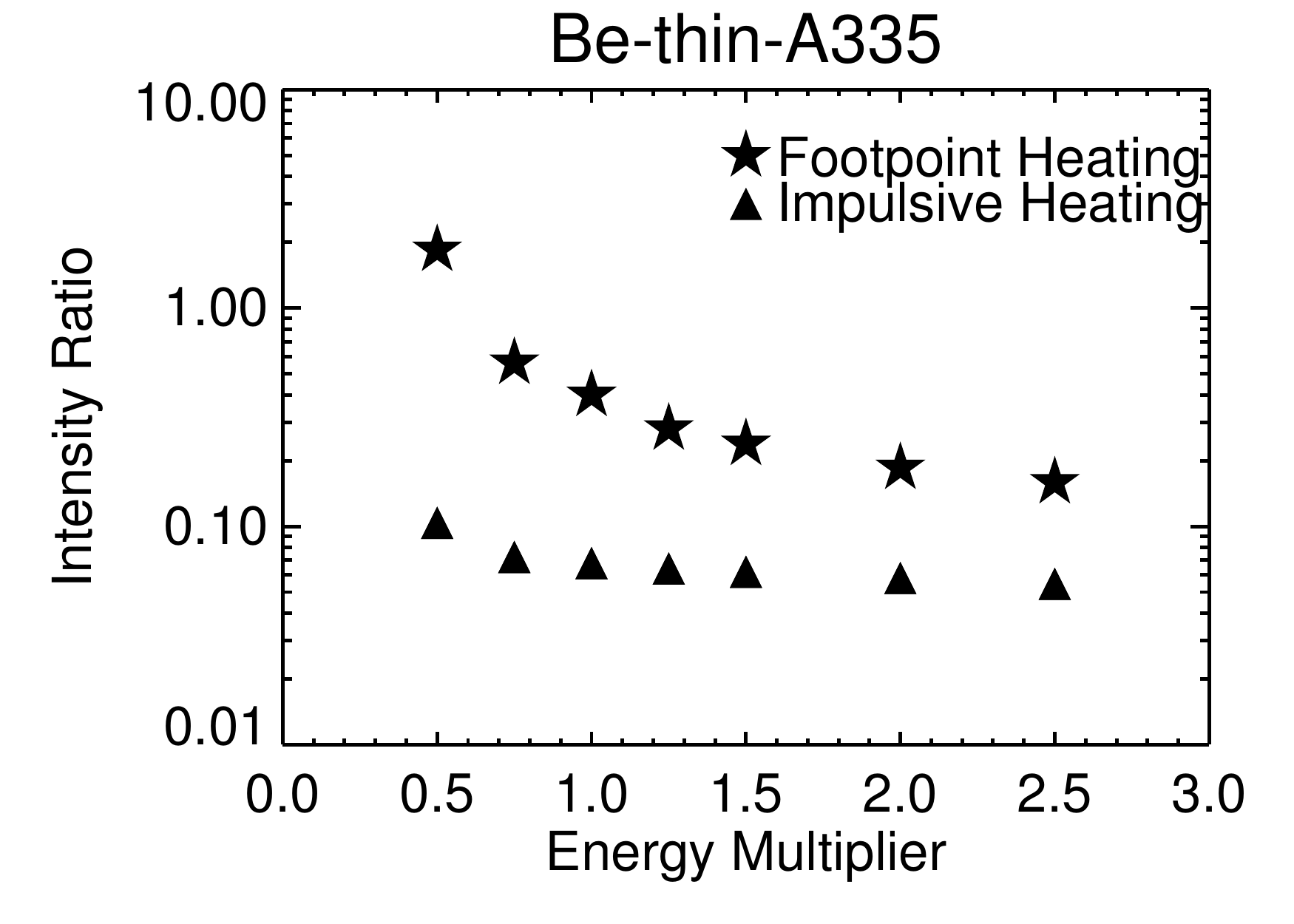}}}
\resizebox{.32\textwidth}{!}{\rotatebox{0}{\includegraphics{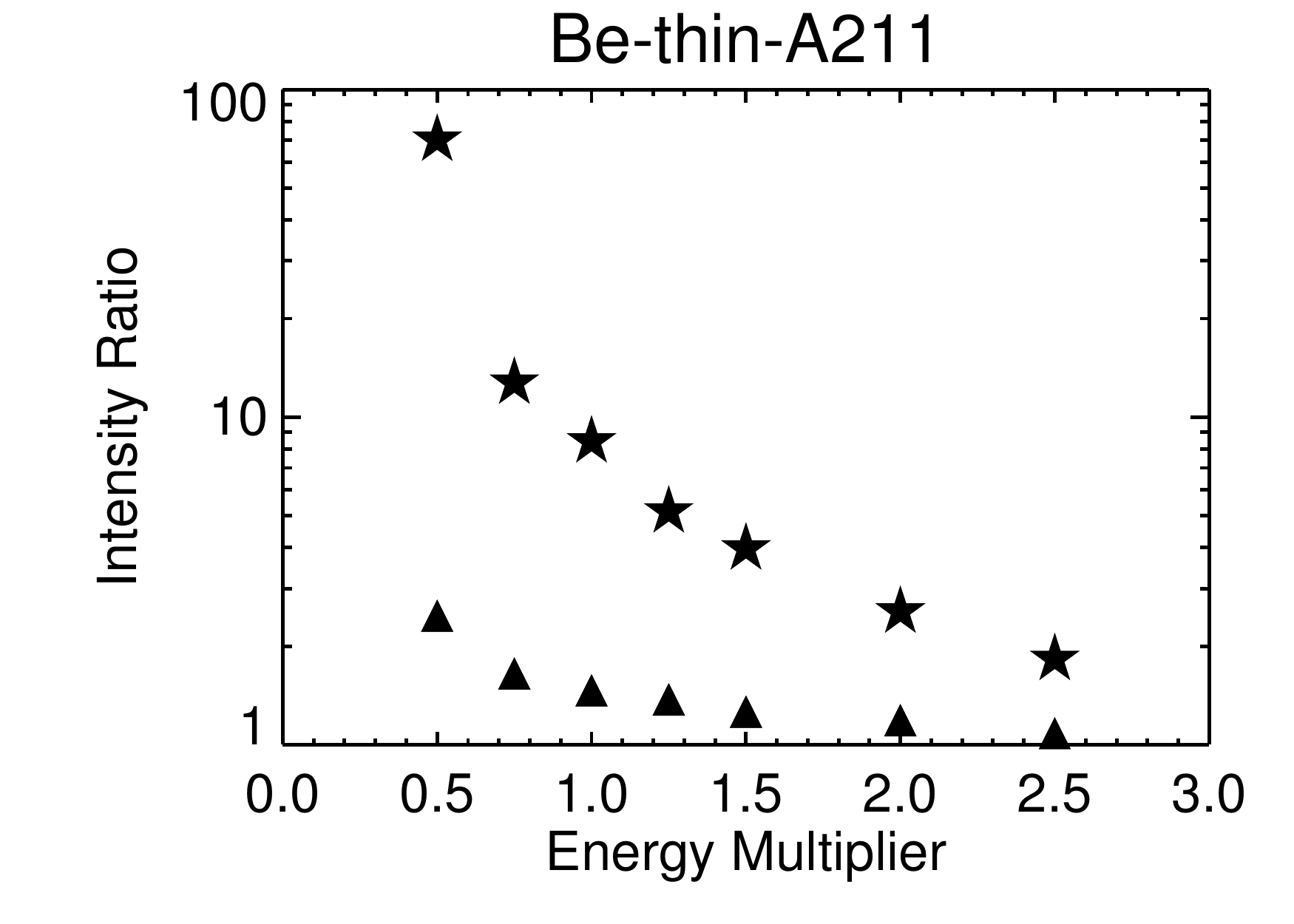}}}
\resizebox{.32\textwidth}{!}{\rotatebox{0}{\includegraphics{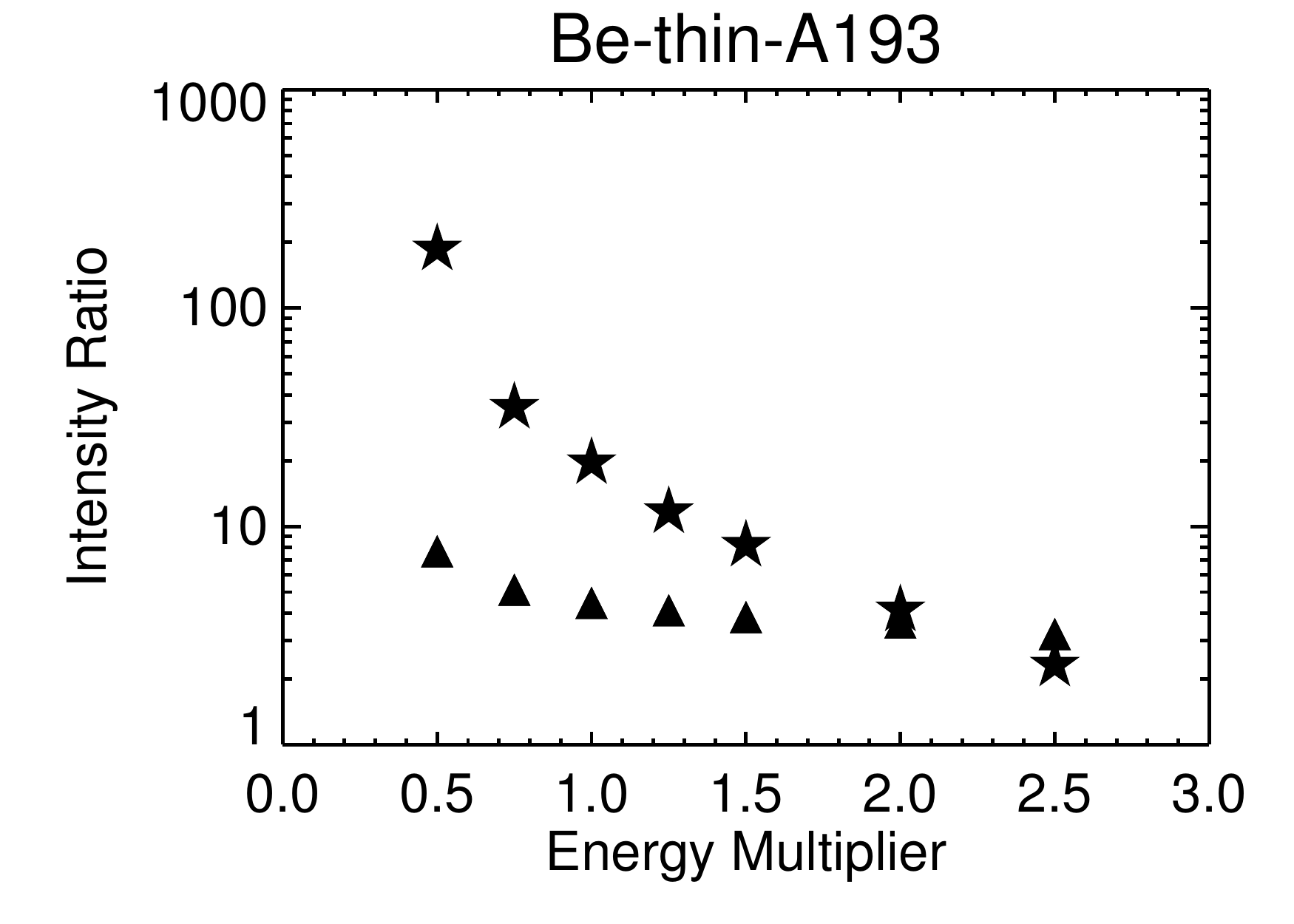}}}
\resizebox{.32\textwidth}{!}{\rotatebox{0}{\includegraphics{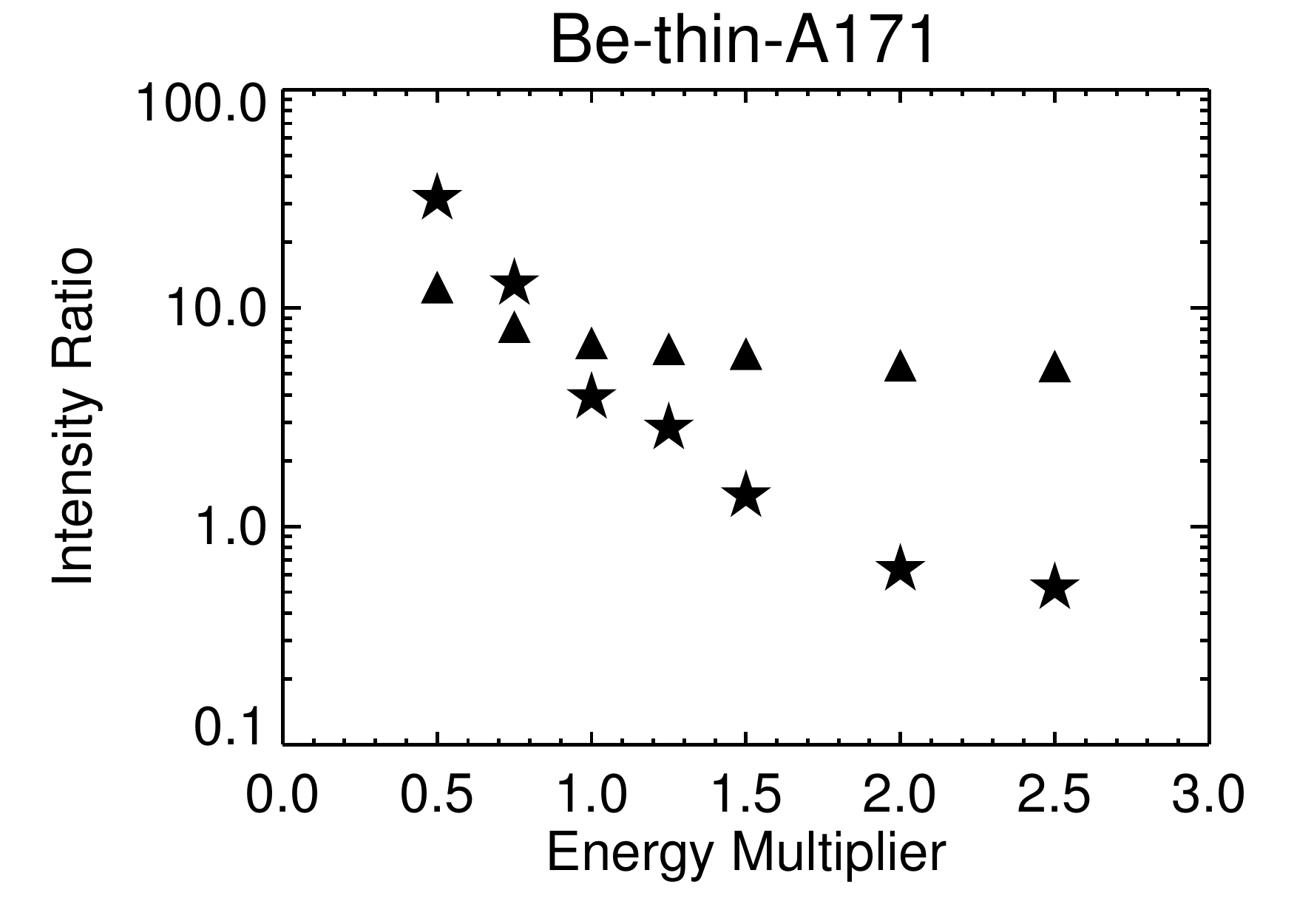}}}
\resizebox{.32\textwidth}{!}{\rotatebox{0}{\includegraphics{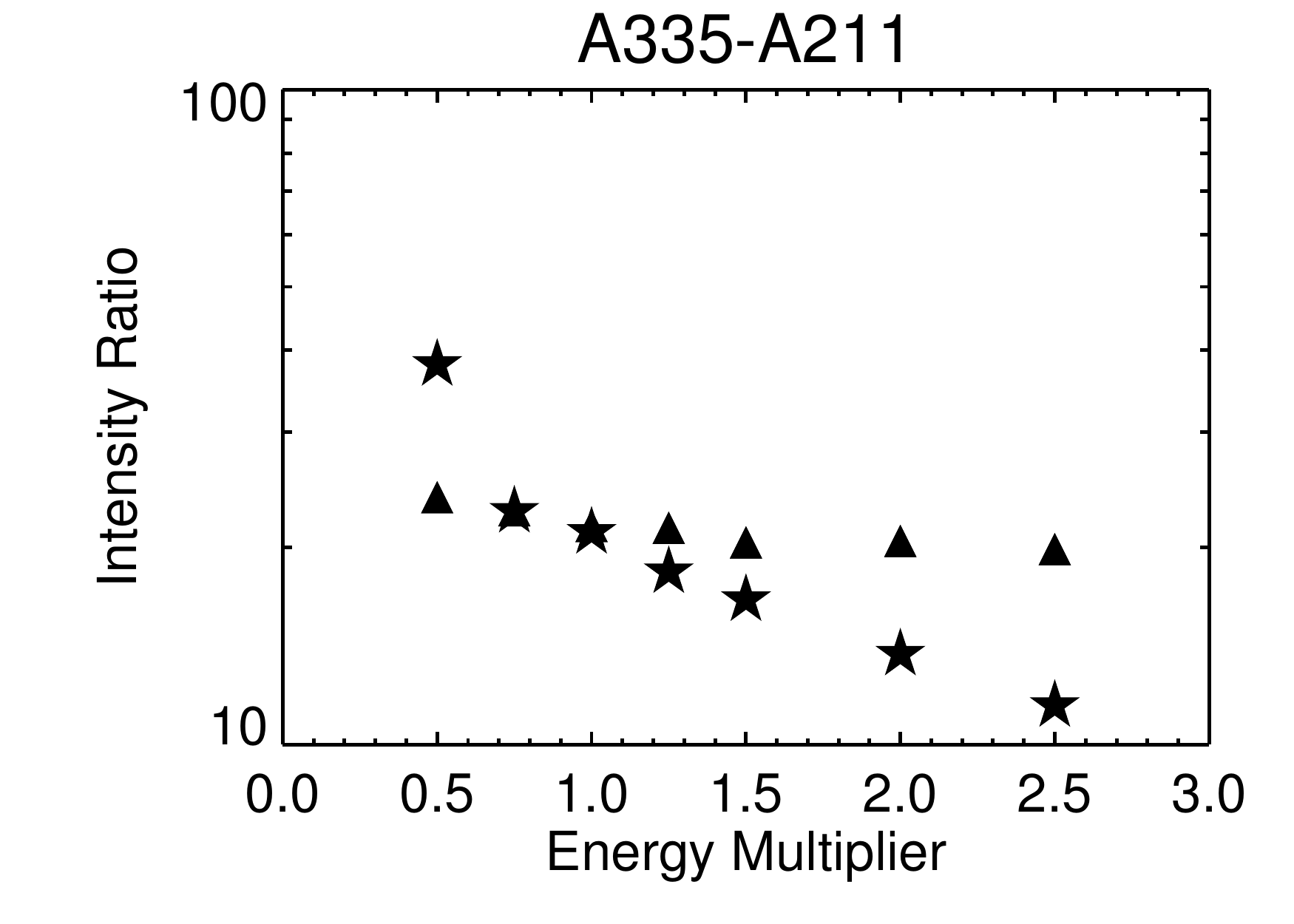}}}
\resizebox{.32\textwidth}{!}{\rotatebox{0}{\includegraphics{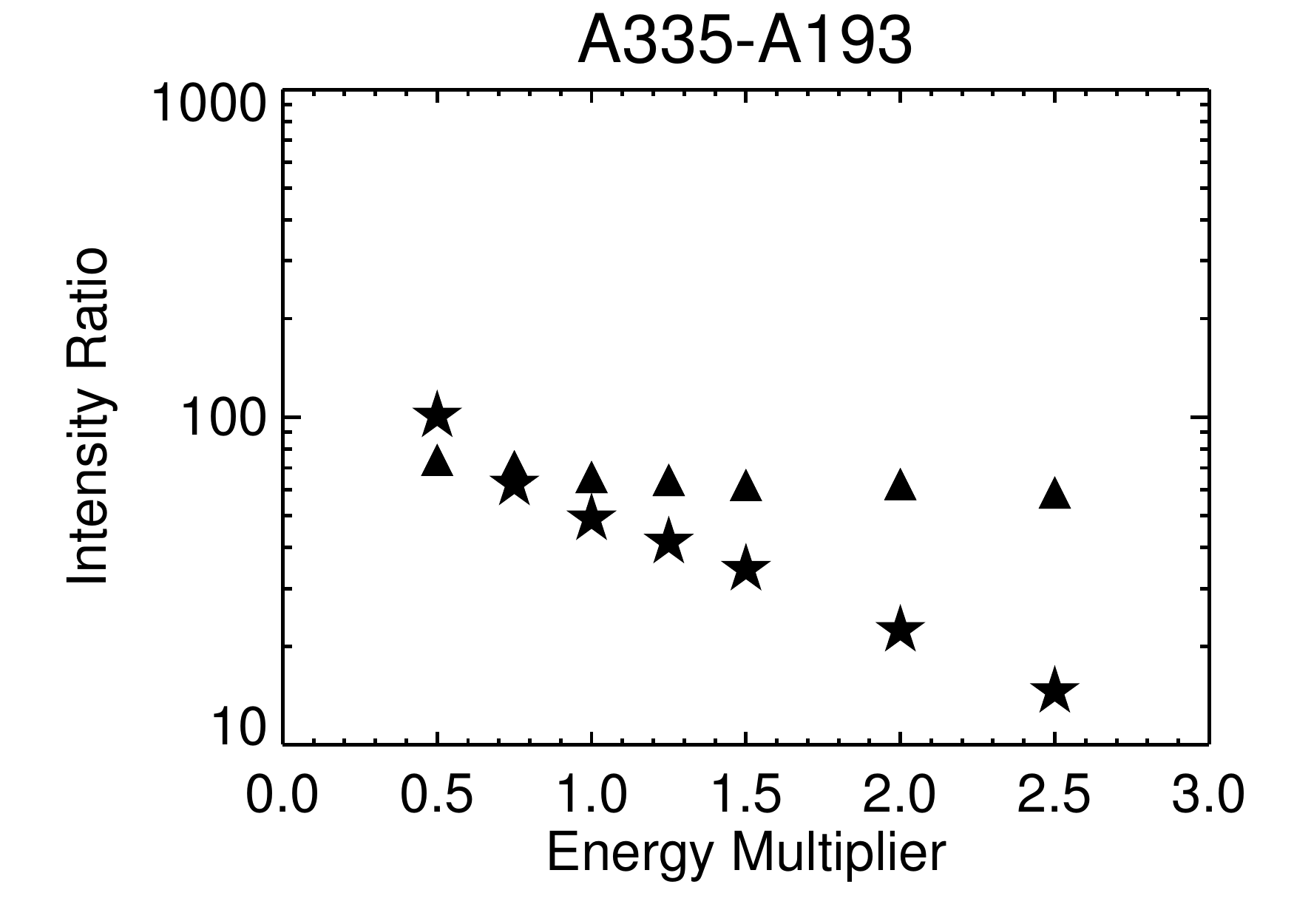}}}
\resizebox{.32\textwidth}{!}{\rotatebox{0}{\includegraphics{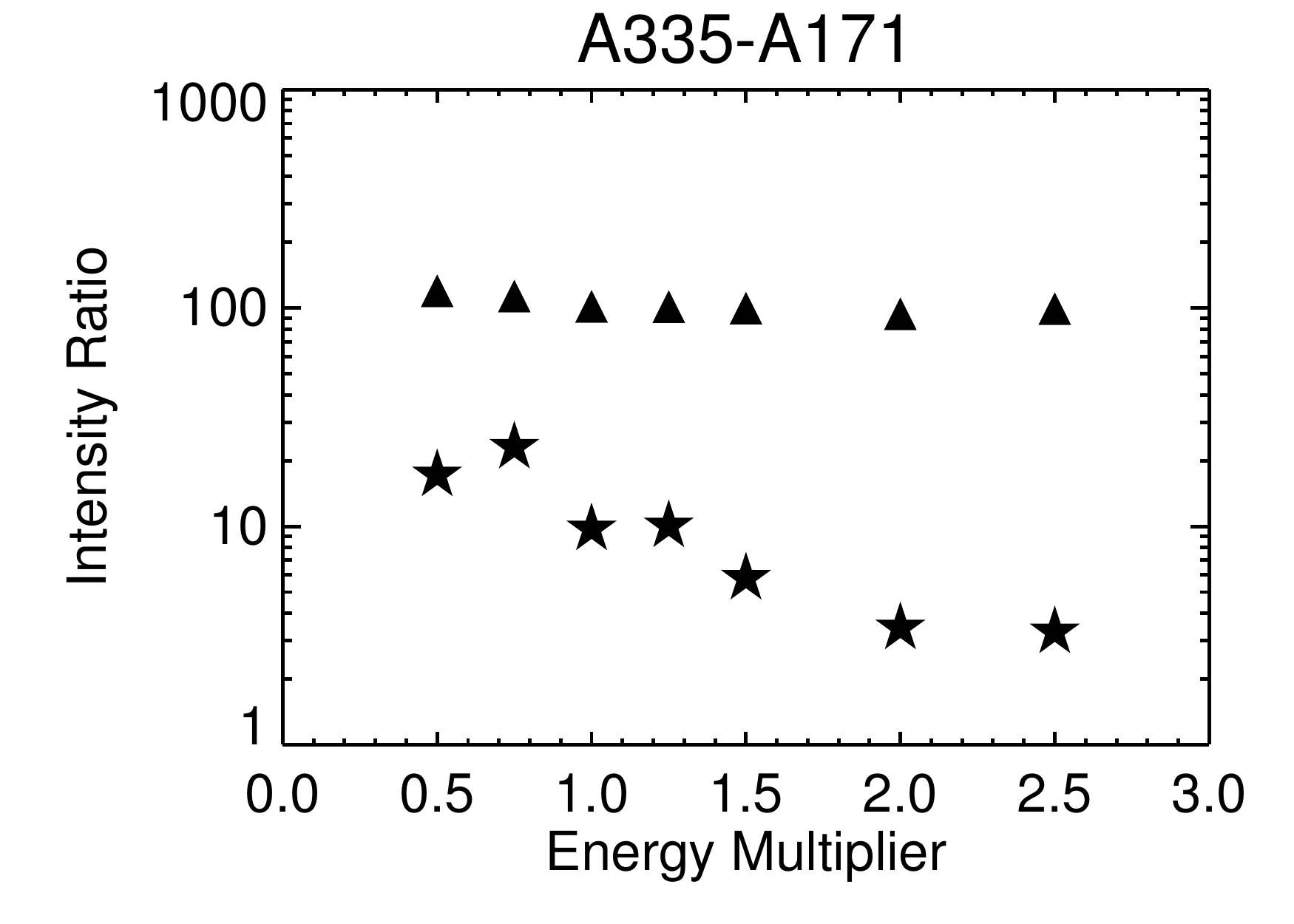}}}
\resizebox{.32\textwidth}{!}{\rotatebox{0}{\includegraphics{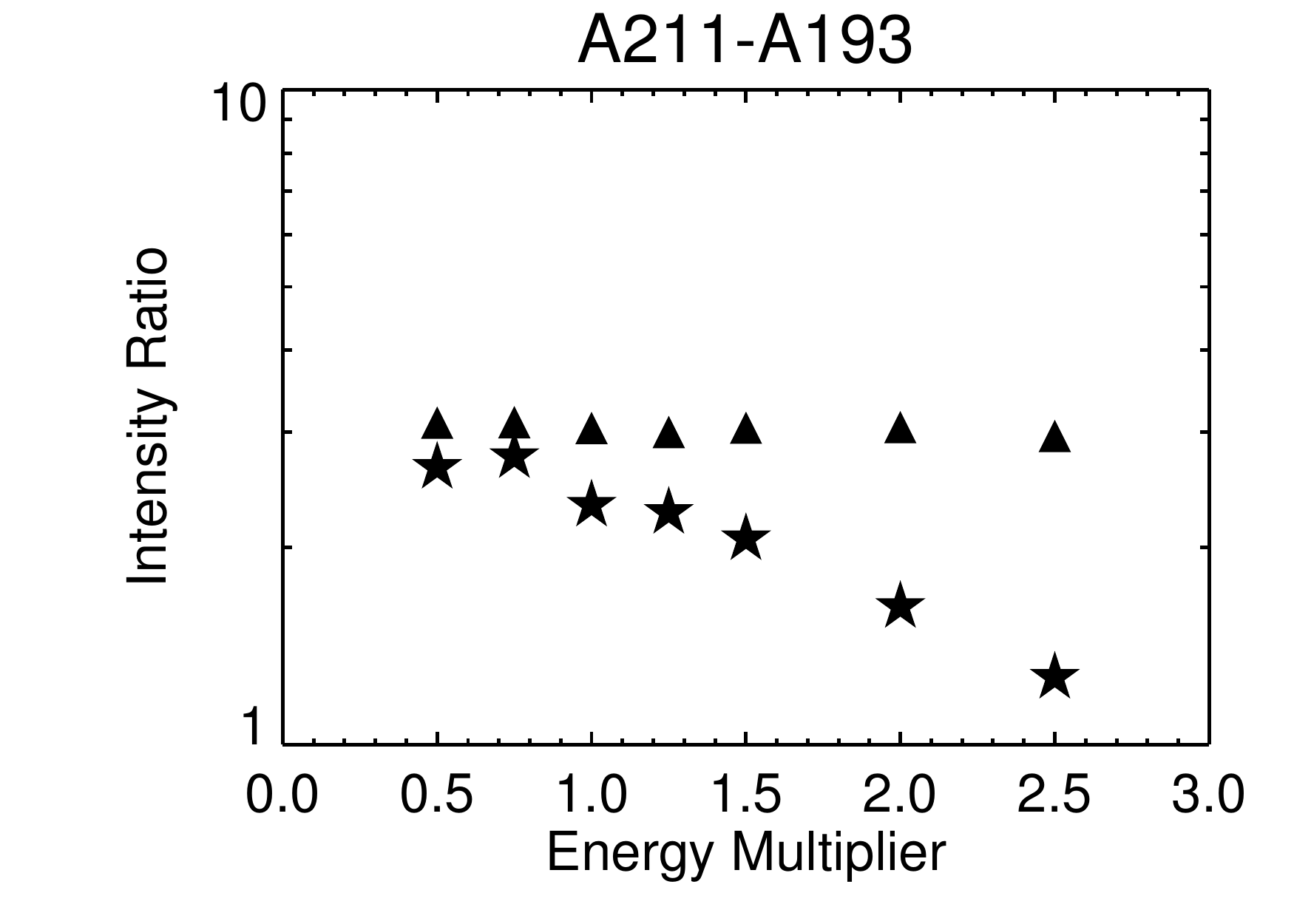}}}
\resizebox{.32\textwidth}{!}{\rotatebox{0}{\includegraphics{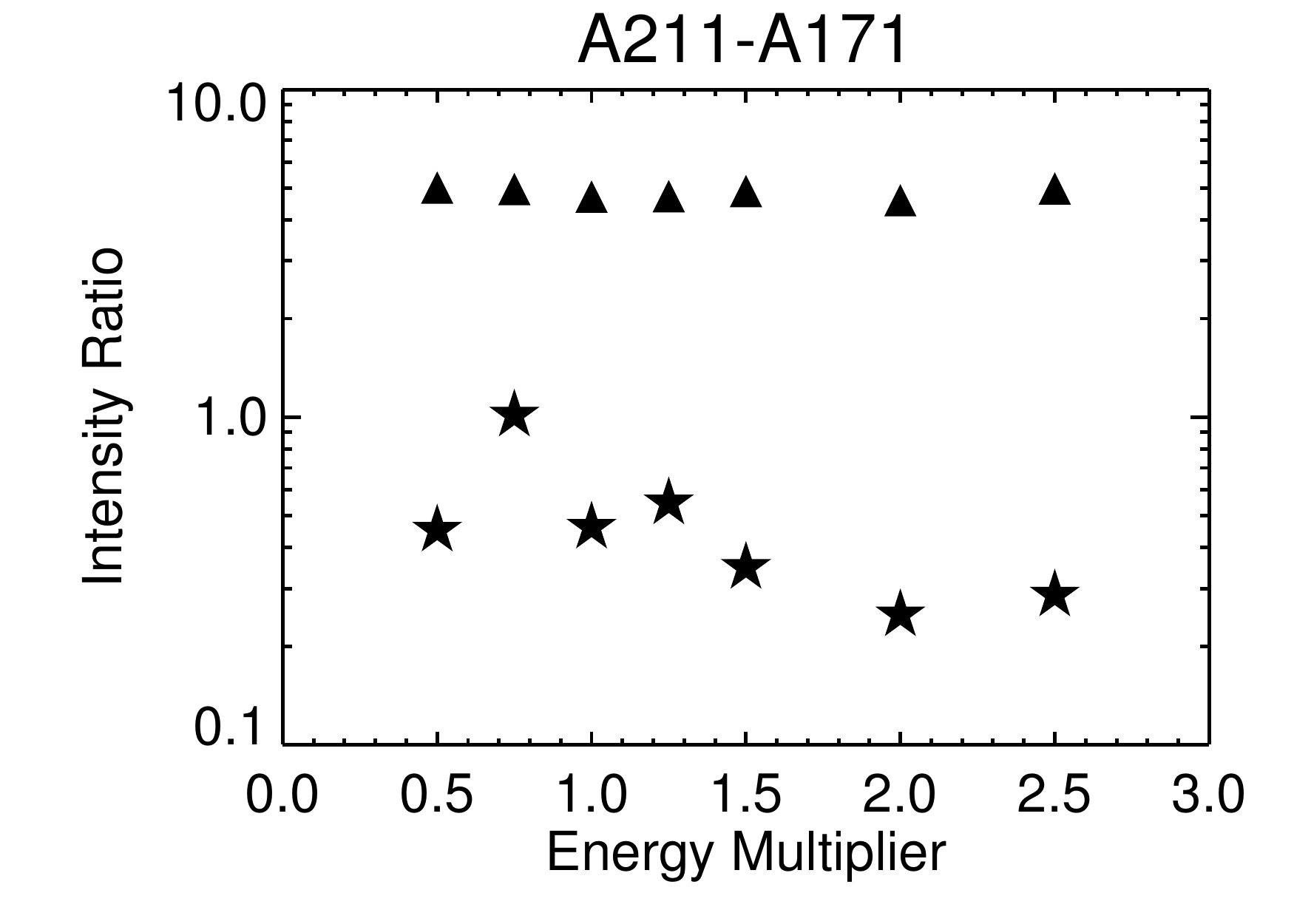}}}
\resizebox{.32\textwidth}{!}{\rotatebox{0}{\includegraphics{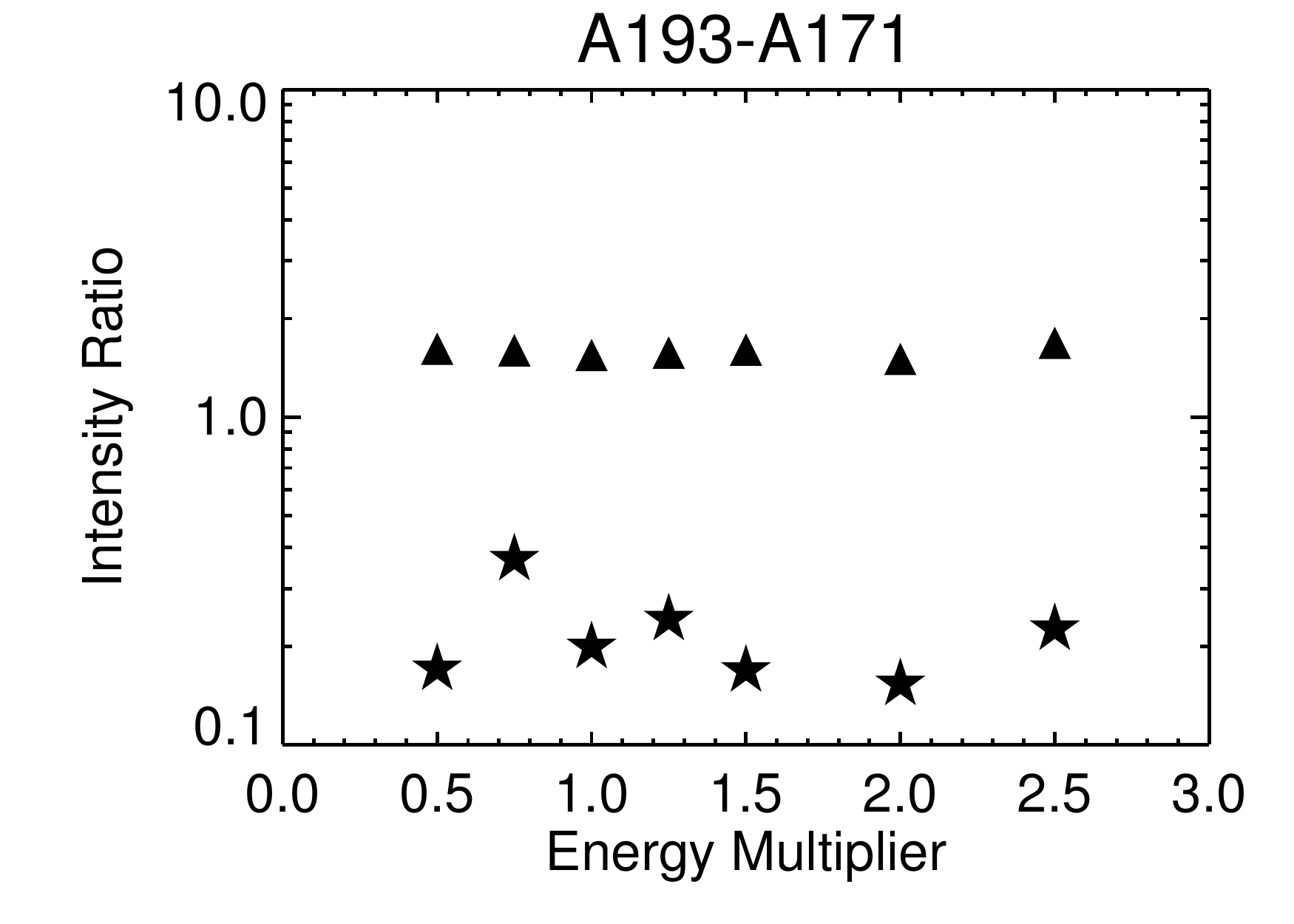}}}
\caption{The ratio of the peak intensity (Channel 2 / Channel 1) for  the different channel combinations and different heating magnitudes.  The footpoint solutions are shown with stars, the impulsive heating solutions are shown with triangles.}
\label{fig:int_rat_mag}
\end{center}
\end{figure*}

Next, we take the apex temperature and density and fold it through the XRT Be-thin and four AIA channel response functions, shown in Figure~\ref{fig:resp}.  We calculate the time lag in the light curves between each channel pair and the ratio of the peak intensity in the light curves for each pair.  We show these results in Figures~\ref{fig:time_lag_mag} and \ref{fig:int_rat_mag}.  The time lags and intensity ratios are shown as a function of the energy multiplier, $M$.  The footpoint-heated solutions are shown with stars and the impulsively-heated solutions are shown with triangles.  

Like the example simulation discussed in detail in Section 3, there are channel pairs where footpoint heating predicts much longer time lags than impulsive heating.  This is particularly clear when looking at the Be-thin-A193 and Be-thin-A171 channel pairs.  For instance, for the Be-thin-A193 channel pair, the range of time lags predicted by impulsive heating is 1,950- 2,580 s, while the range of time lags predicted by footpoint heating is 3,270-4,350 s.
For cooler channel pairs, there is more overlap in the predicted time lags.  For instance, in the A211-A171 channel pair,   at some energies footpoint heating predicts higher time lags, while at others impulsive heating predicts higher time lags.  At all energies, the time lags are roughly in the same range.  Footpoint heating can predict time lags of 0 or negative time lags, as shown in the A211-A171 and A193-A171 channel pairs.  As explained in Section 3, this occurs when the solution does not cool completely through the temperature response of the passband.

When comparing the intensity ratios (Figure~\ref{fig:int_rat_mag}), we find that for impulsively-heated solutions, the ratio of intensity in the cooler, narrow channels (AIA 171, 193, and 211\,\AA) is almost independent of the heating magnitude.  For footpoint-heated solutions, the ratio  varies more significantly.  

\section{VARYING THE STRATIFICATION OF THE HEATING}

In this section, we vary the stratification of the footpoint heating, while keeping the magnitude constant at the value of the original  footpoint simulation, discussion in Section 3.1.  We cannot vary the impulsive heating solution in a analogous way, hence we compare the results to a single impulsive heating simulation with the same total power, i.e., the one shown in Section 3.2.  

\subsection{Footpoint heating}

\begin{figure*}[t!]
\begin{center}
\resizebox{.75\textwidth}{!}{\rotatebox{0}{\includegraphics{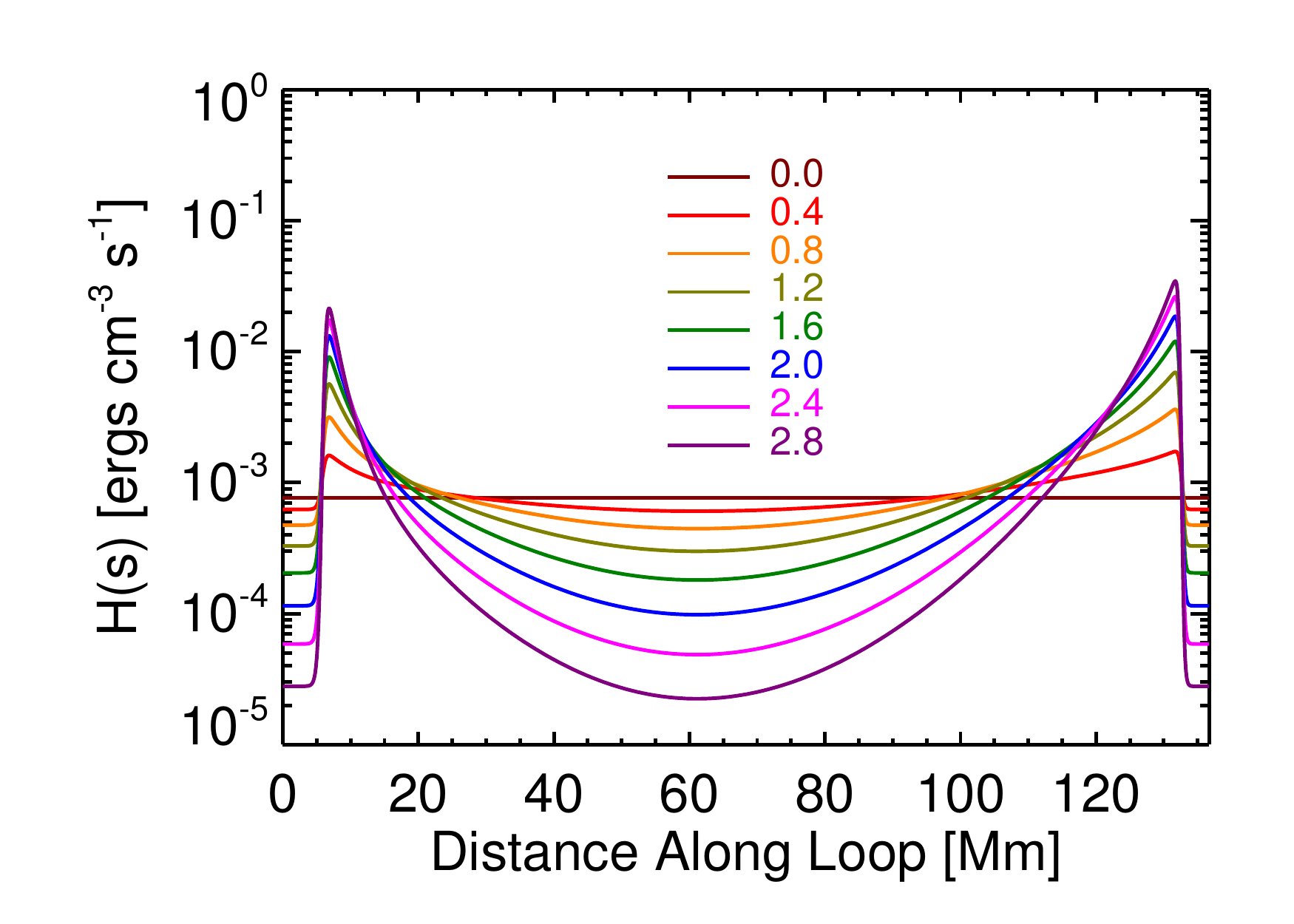}}}
\caption{The footpoint heating functions for different values of $\alpha$.  The integrated energy is constant, while the stratification is varied.}
\label{fig:exp_heat}
\end{center}
\end{figure*}

To vary the heating stratification, while keeping the total energy deposited along the loop the same, we raise the original 
heating function,  $H_{\rm FP0}(s)$, to an exponent, $\alpha$, i.e., 
\begin{equation}
H{\rm FP}(s) = H_{\rm FP0}(s)^{\alpha} \left( \frac{\int A(s) H_{\rm FP}0(s) ds}{\int A(s) H_{\rm FP0}(s)^\alpha ds}\right)
\end{equation}
where $A(s)$ is the area along the loop.  We vary $\alpha$ from 0 - 2.8.   Note that for $\alpha = 0$, the heating along the loop is uniform.   The heating functions for different $\alpha$ are shown in Figure~\ref{fig:exp_heat}.

\begin{figure*}[t!]
\begin{center}
\resizebox{\textwidth}{!}{\rotatebox{90}{\includegraphics{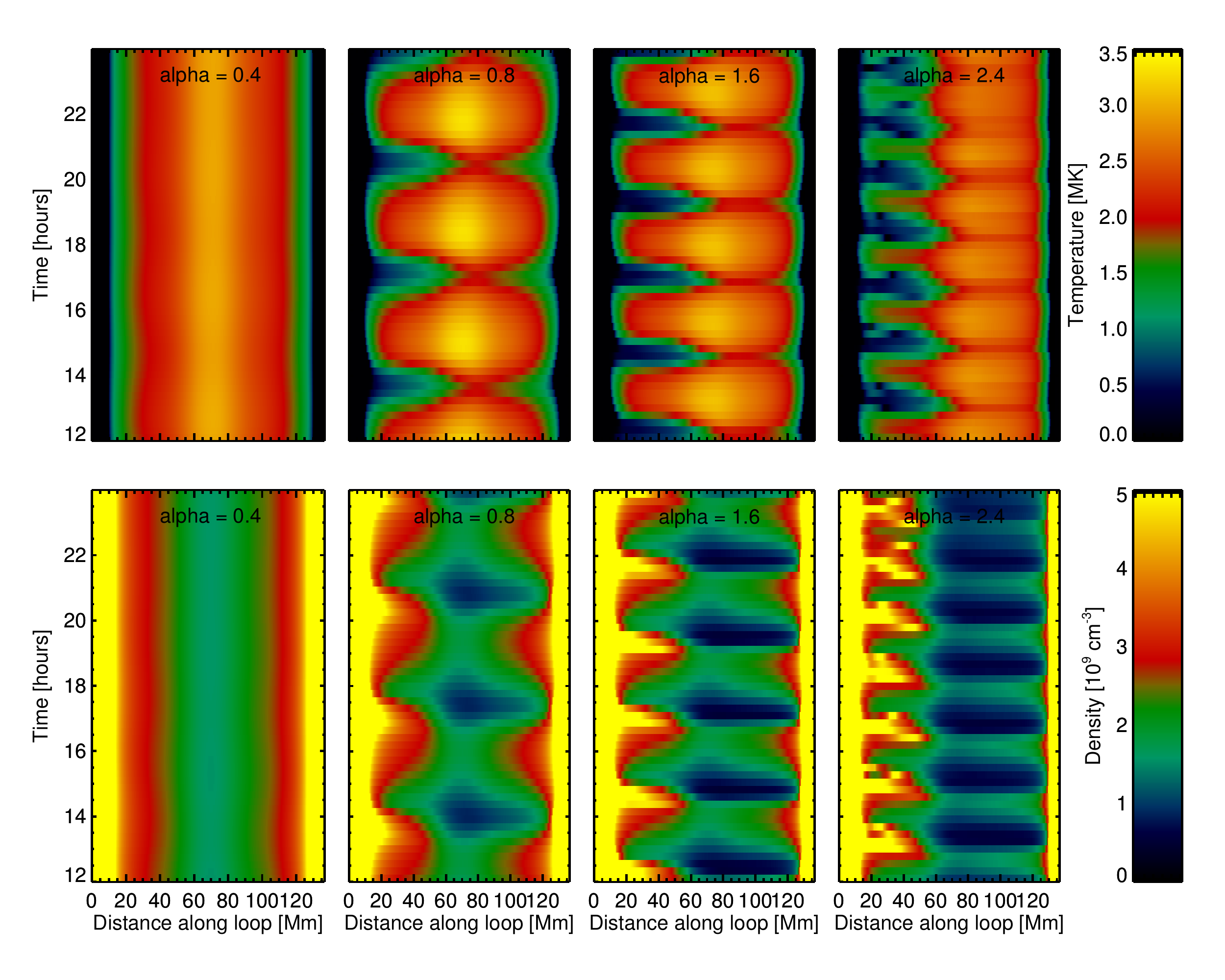}}}
\caption{A comparison of the temperature (top row) and density (bottom row) evolution for various heating stratifications in footpoint heating simulations.  Note that this does not show the evolution for all heating stratifications considered in this paper.}
\label{fig:temp_map_exp}
\end{center}
\end{figure*}

\begin{figure*}[t!]
\begin{center}
\resizebox{.49\textwidth}{!}{\rotatebox{0}{\includegraphics{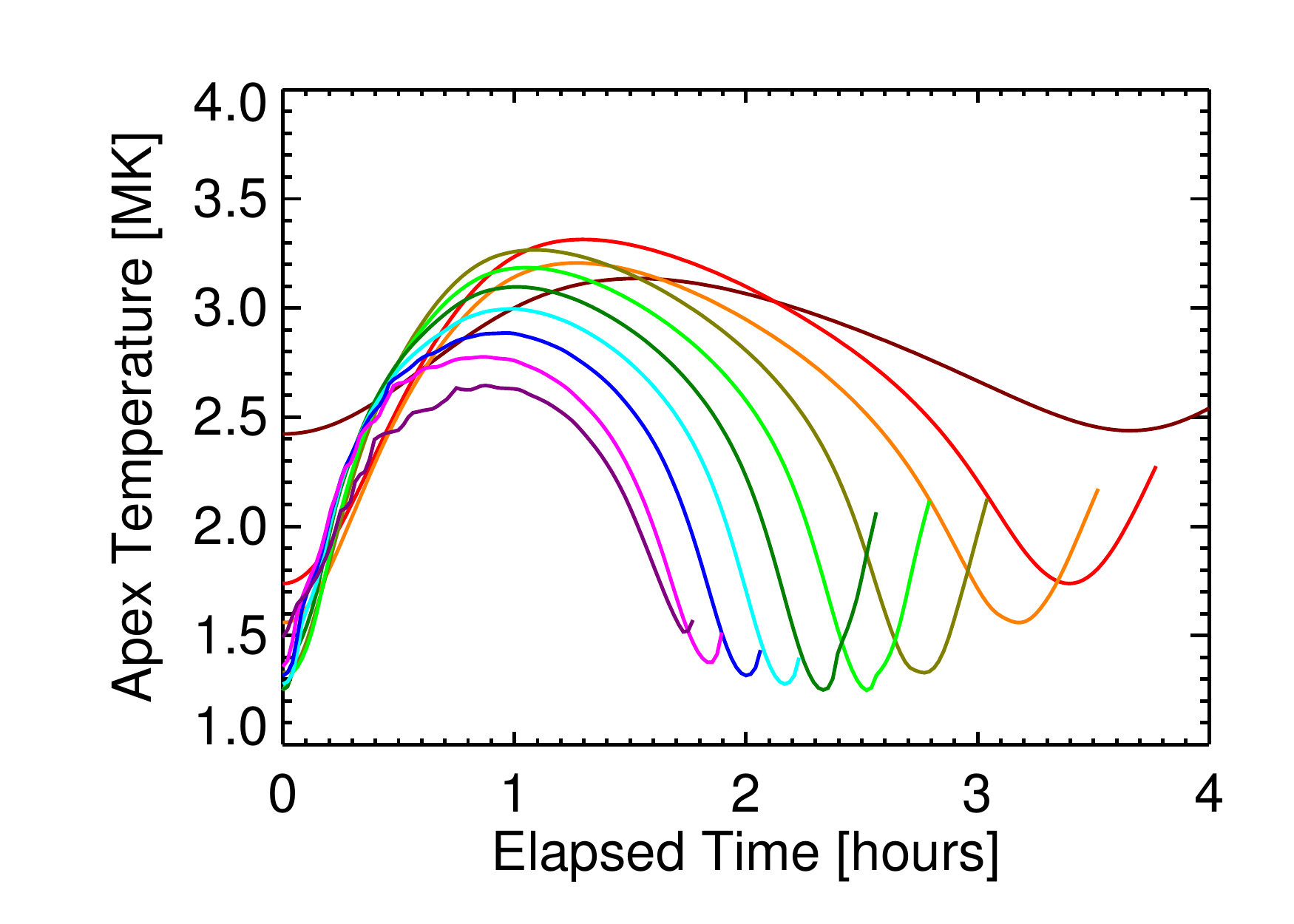}}}
\resizebox{.49\textwidth}{!}{\rotatebox{0}{\includegraphics{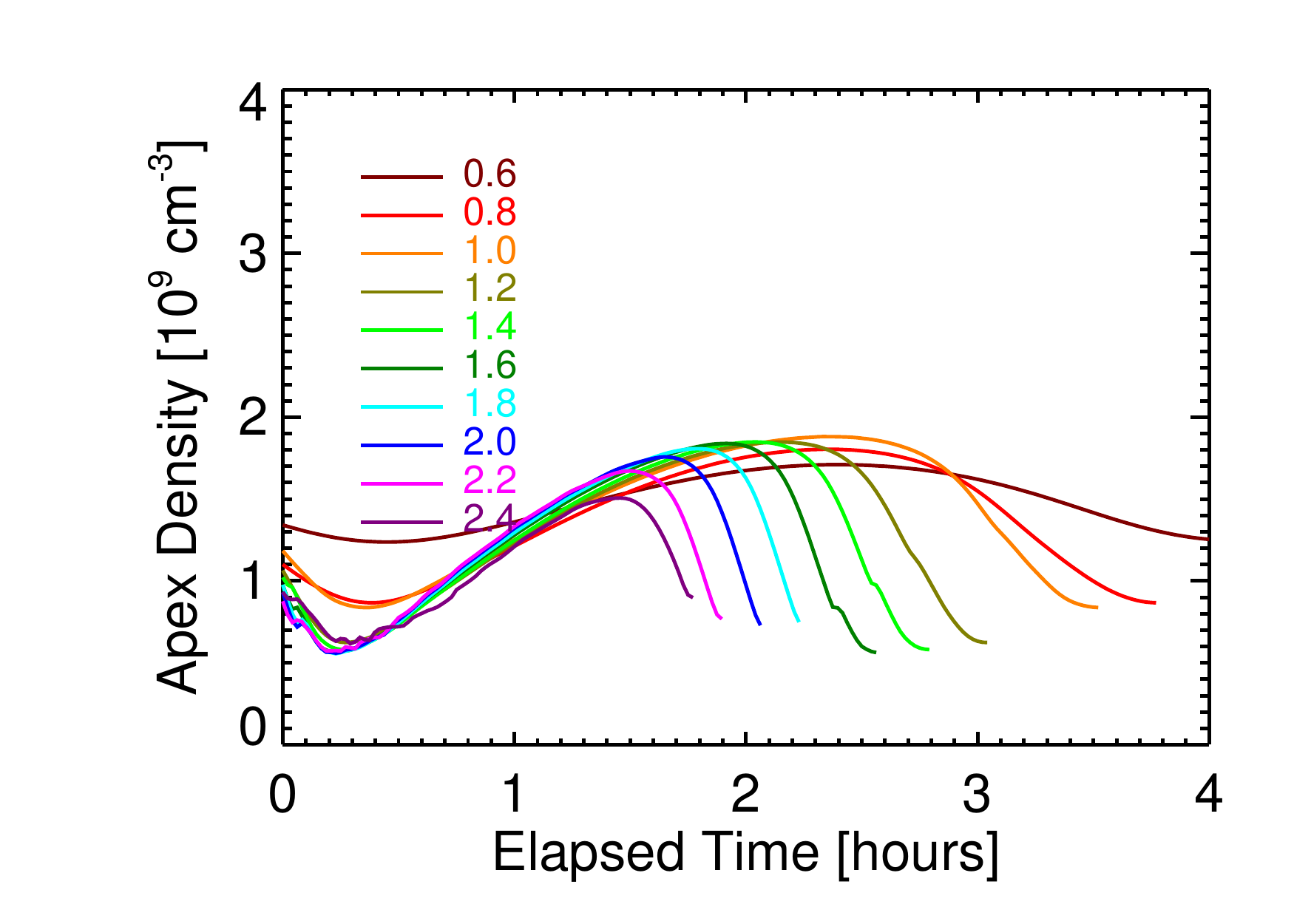}}}
\caption{Apex temperature and density values for different heating stratifications in footpoint heating simulations.}
\label{fig:apex_exp}
\end{center}
\end{figure*}

The resulting temperature and density evolutions are shown in Figures~\ref{fig:temp_map_exp} for different $\alpha$.  The apex values are shown in Figure~\ref{fig:apex_exp}.  For $\alpha = 0$, the heating is uniform, so the temperature and density quickly go to a steady-state solution soon after the start of the simulation.  As $\alpha$ increases, the cycle time of the solution decreases.  At large values of $\alpha$, the solution approaches a steady state again.  This can be understood by the inherent asymmetry of the heating and geometry.  As these large $\alpha$, the asymmetry grows and a sizable syphon flow develops between the two footpoints.  

\subsection{Observables}

\begin{figure*}[t!]
\begin{center}
\resizebox{.32\textwidth}{!}{\rotatebox{0}{\includegraphics{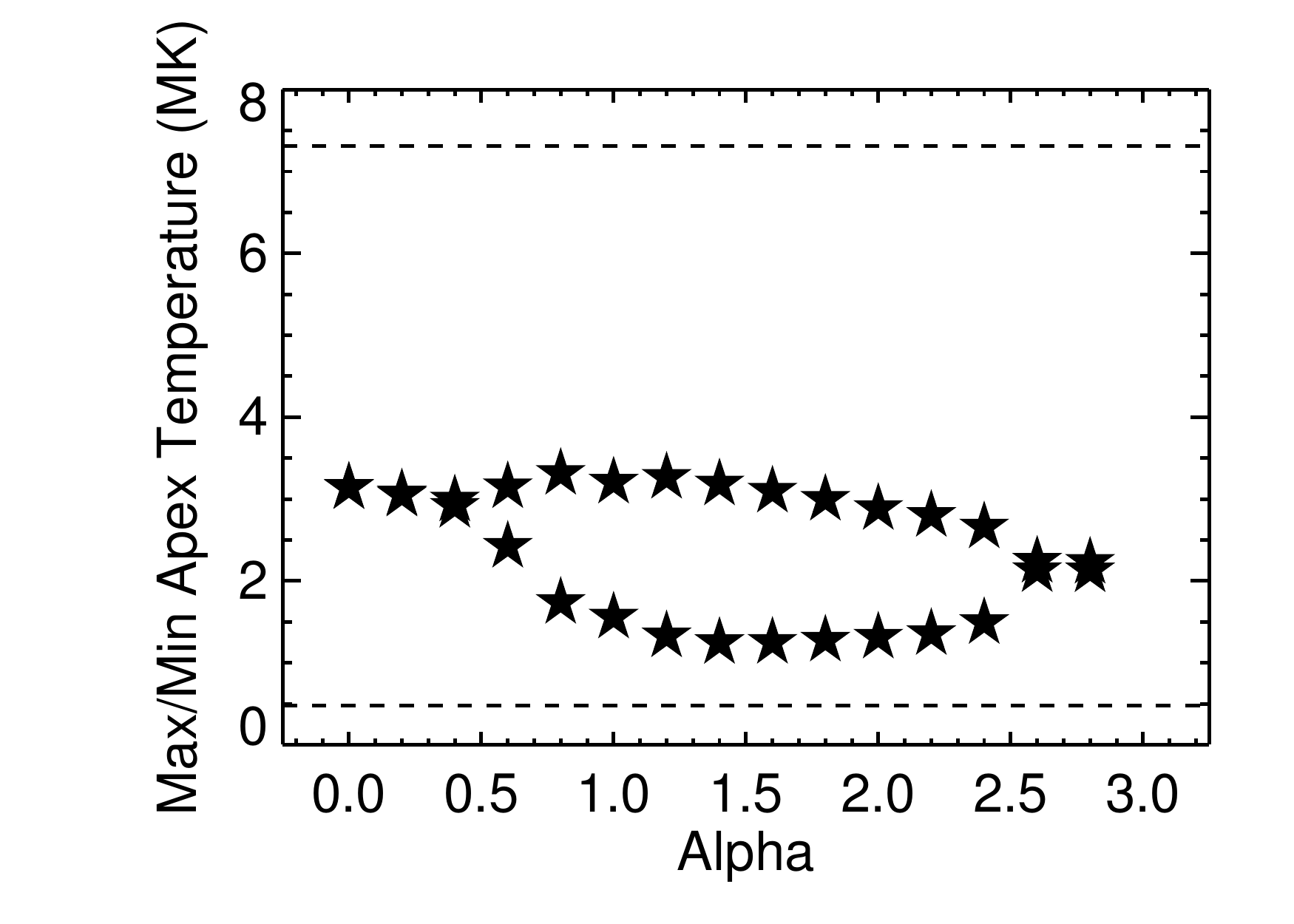}}}
\resizebox{.32\textwidth}{!}{\rotatebox{0}{\includegraphics{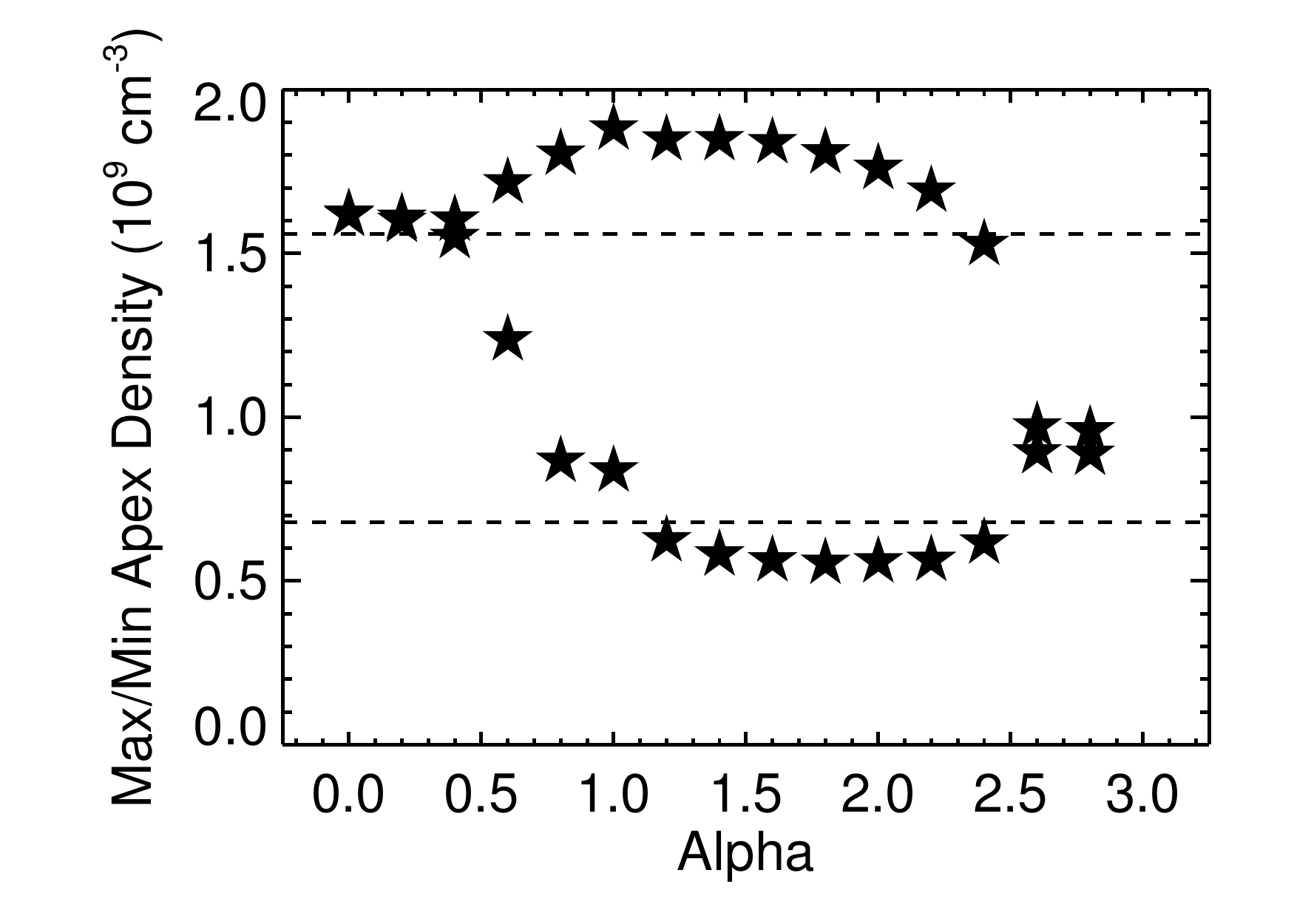}}}
\resizebox{.32\textwidth}{!}{\rotatebox{0}{\includegraphics{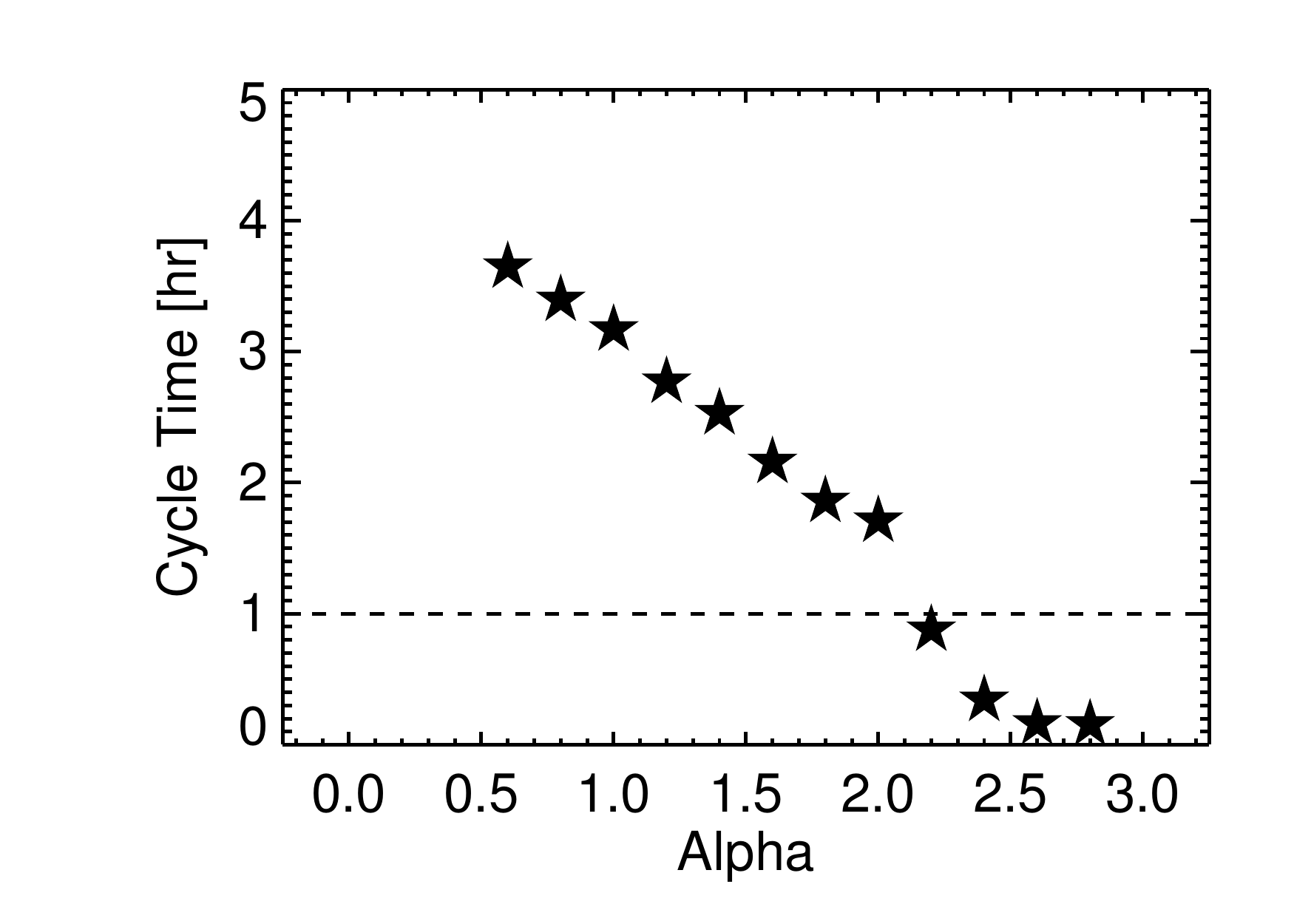}}}
\caption{A comparison of the minimum and maximum temperatures (left) and density (center) for the footpoint (star) solutions as a function of the exponent.  The values for the single impulsive heating solution with the same heating magnitude is shown as dashed lines. The right panel shows the cycle time of the footpoint heated solutions with stars, while the cooling time of the impulsively heating solution is shown as a dashed line.}
\label{fig:max_min_dens_temp_exp}
\end{center}
\end{figure*}

The maximum and minimum apex temperature and density and the cycle times are shown as stars for the footpoint heated solutions in Figure~\ref{fig:max_min_dens_temp_exp}.  The comparable values for the single impulsive heating solution are shown as dashed lines.  As discussed above and further demonstrated in this plot, TNE solutions only exist for a small range of $\alpha$.  For small $\alpha$ (near-uniform heating, $\alpha < 0.5$) there is no difference in the minimum and maximum temperature or density.   We do not show a cycle time for these values of $\alpha$.   Additionally, for very large $\alpha$, the solutions cycle very quickly between two almost identical apex temperature and density values.  Only for the median range of $\alpha$ establishes a TNE solution.  The maximum and minimum temperature range is much narrower than expected for the impulsively-heated solution, and similar to the range predicted by changing the heating magnitude, implying that at least for this simulation, we cannot generate high-temperature plasma by changing the heating stratification.  The cycle times of the footpoint simulations are generally longer than the cooling time  of the impulsively heated loop (shown as a dashed line in the right panel of Figure~\ref{fig:max_min_dens_temp_exp}), though we can achieve rapid cycling through a small range of temperatures for highly stratified heating.

\begin{figure*}[t!]
\begin{center}
\resizebox{.32\textwidth}{!}{\rotatebox{0}{\includegraphics{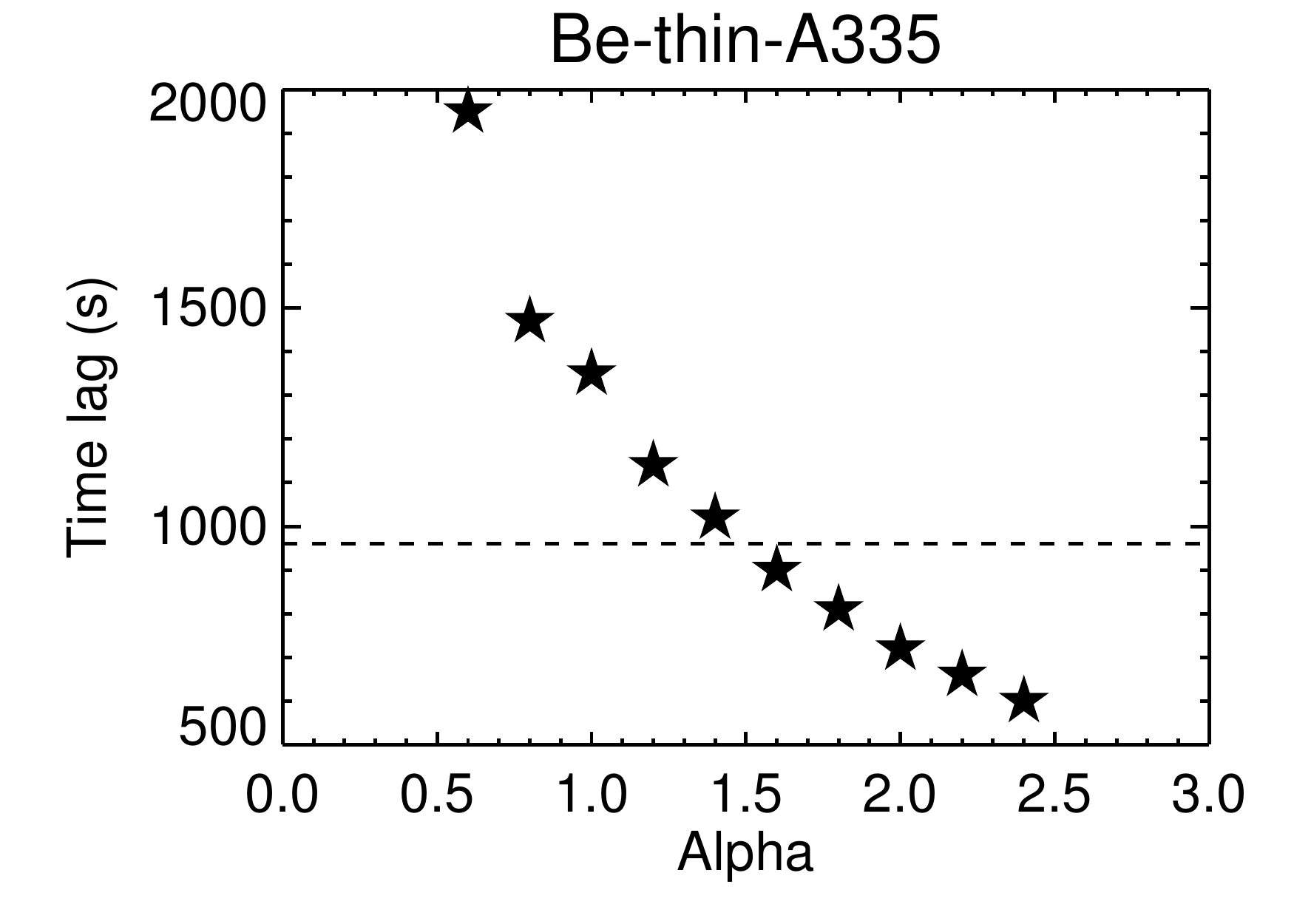}}}
\resizebox{.32\textwidth}{!}{\rotatebox{0}{\includegraphics{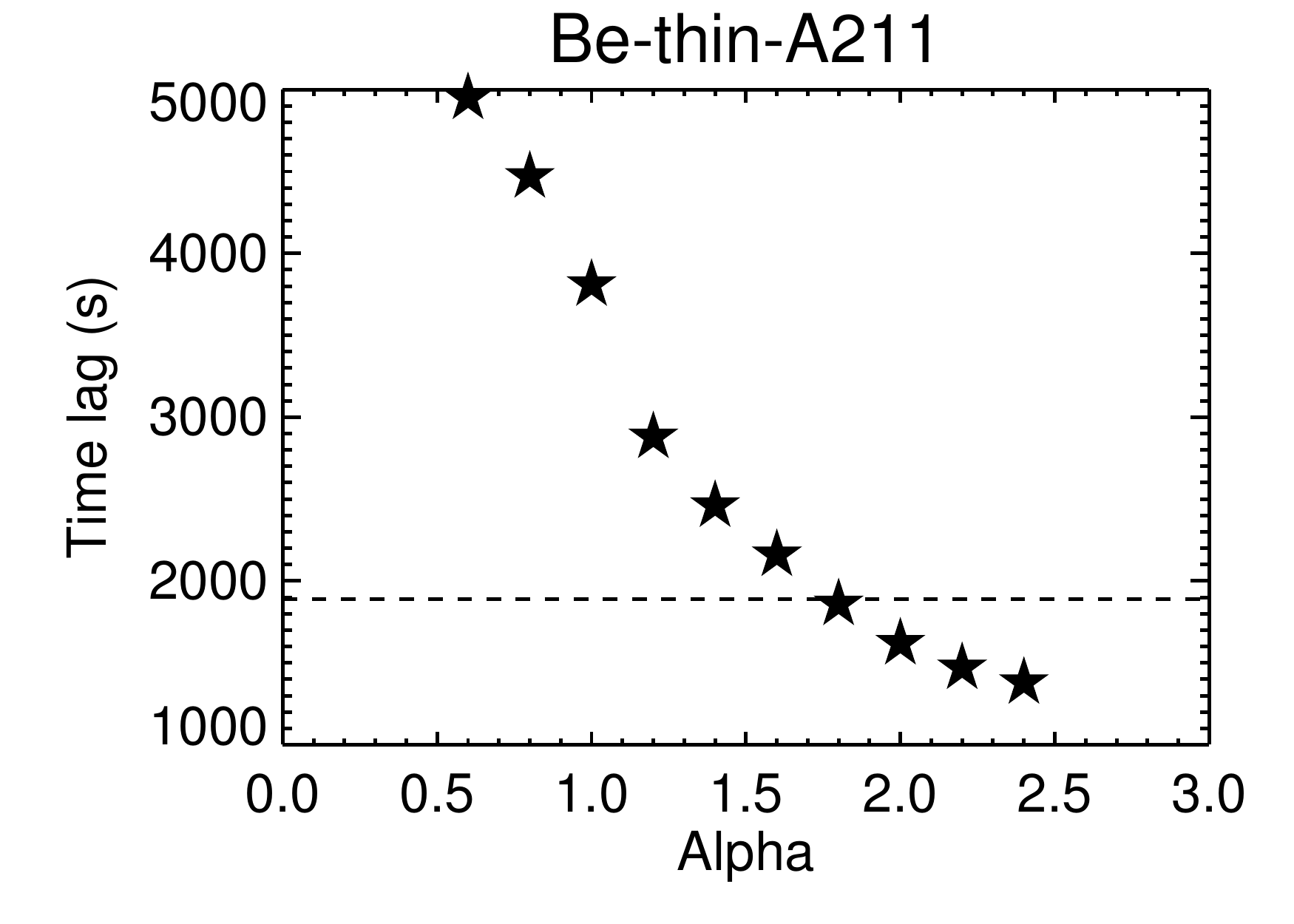}}}
\resizebox{.32\textwidth}{!}{\rotatebox{0}{\includegraphics{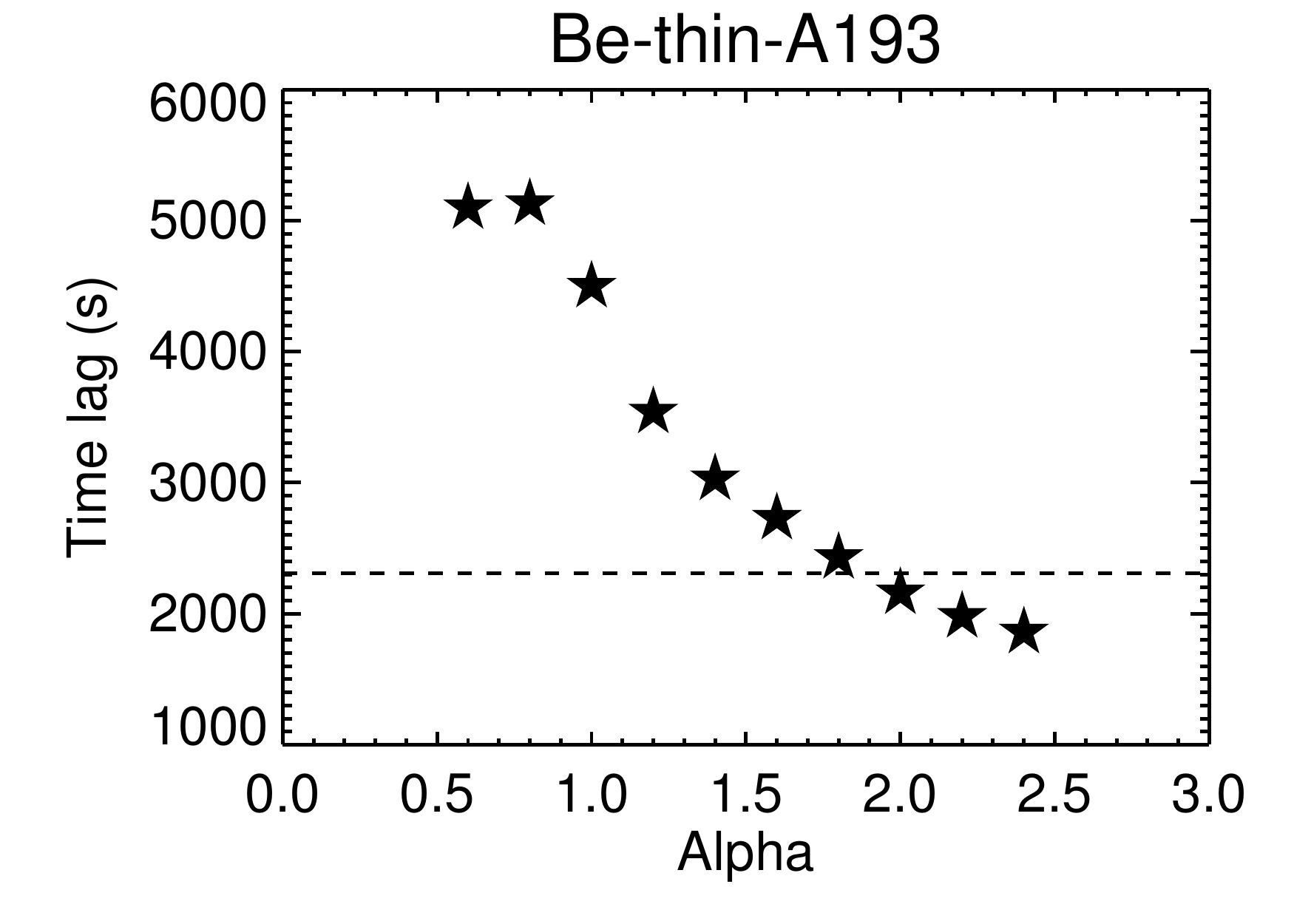}}}
\resizebox{.32\textwidth}{!}{\rotatebox{0}{\includegraphics{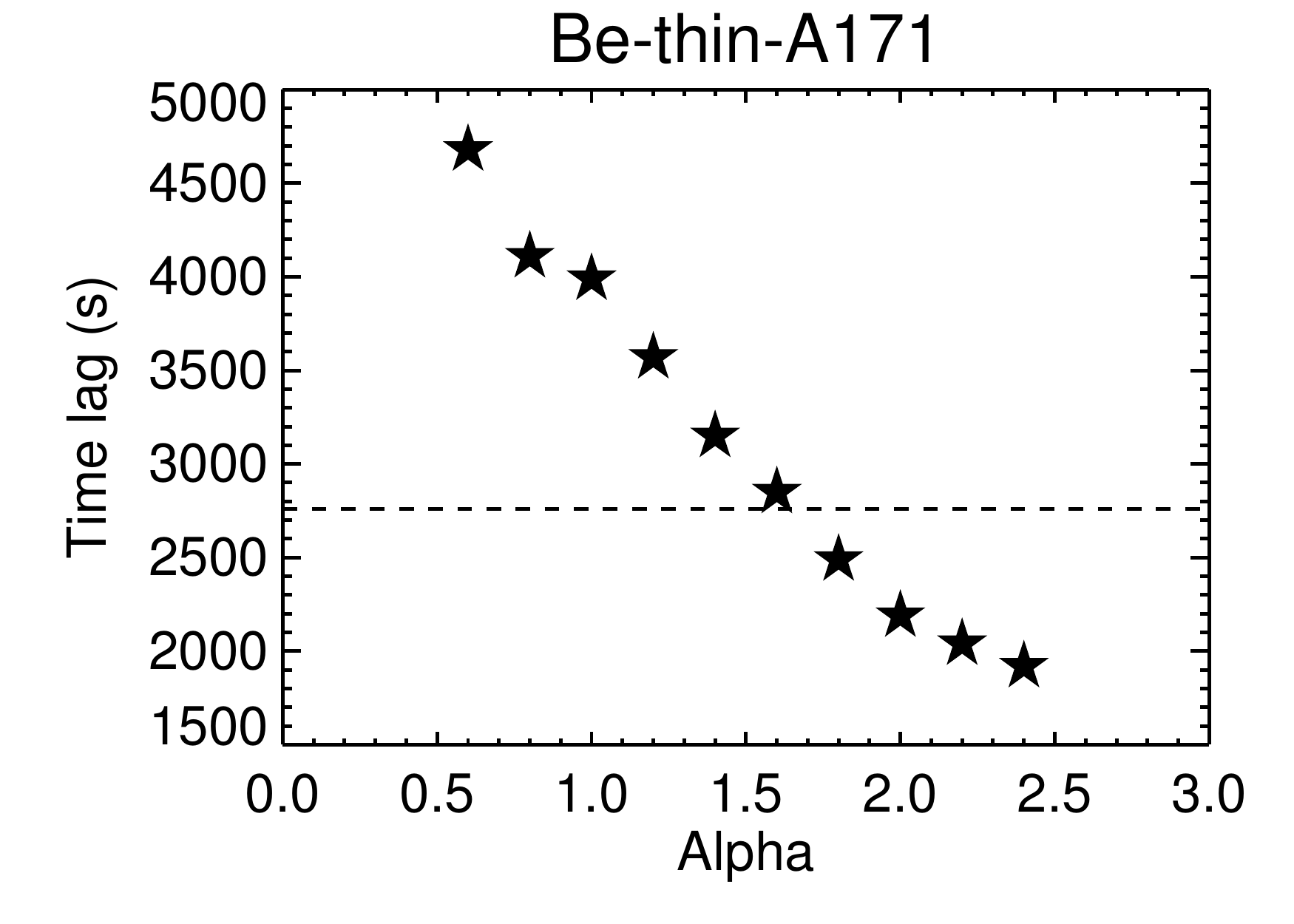}}}
\resizebox{.32\textwidth}{!}{\rotatebox{0}{\includegraphics{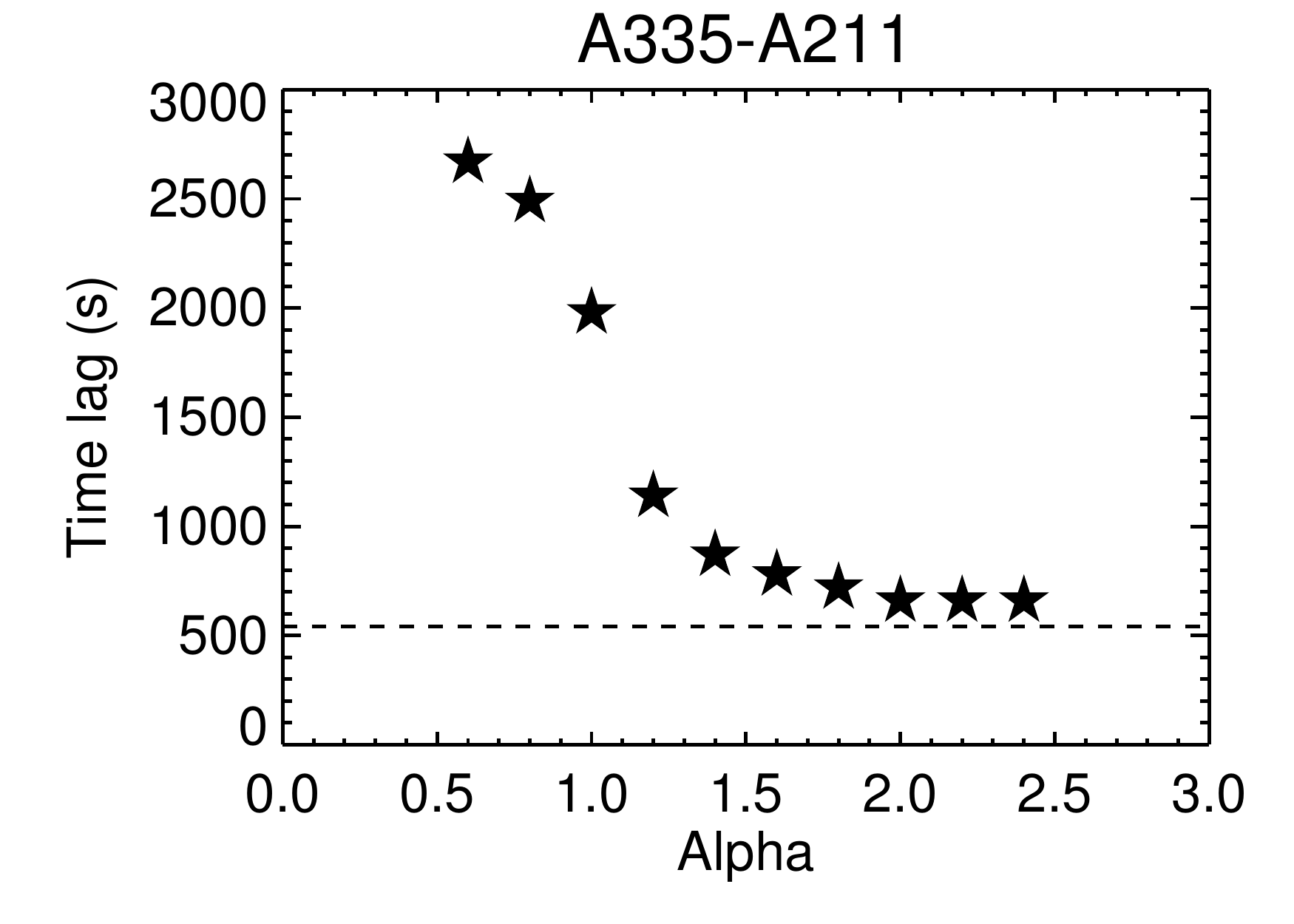}}}
\resizebox{.32\textwidth}{!}{\rotatebox{0}{\includegraphics{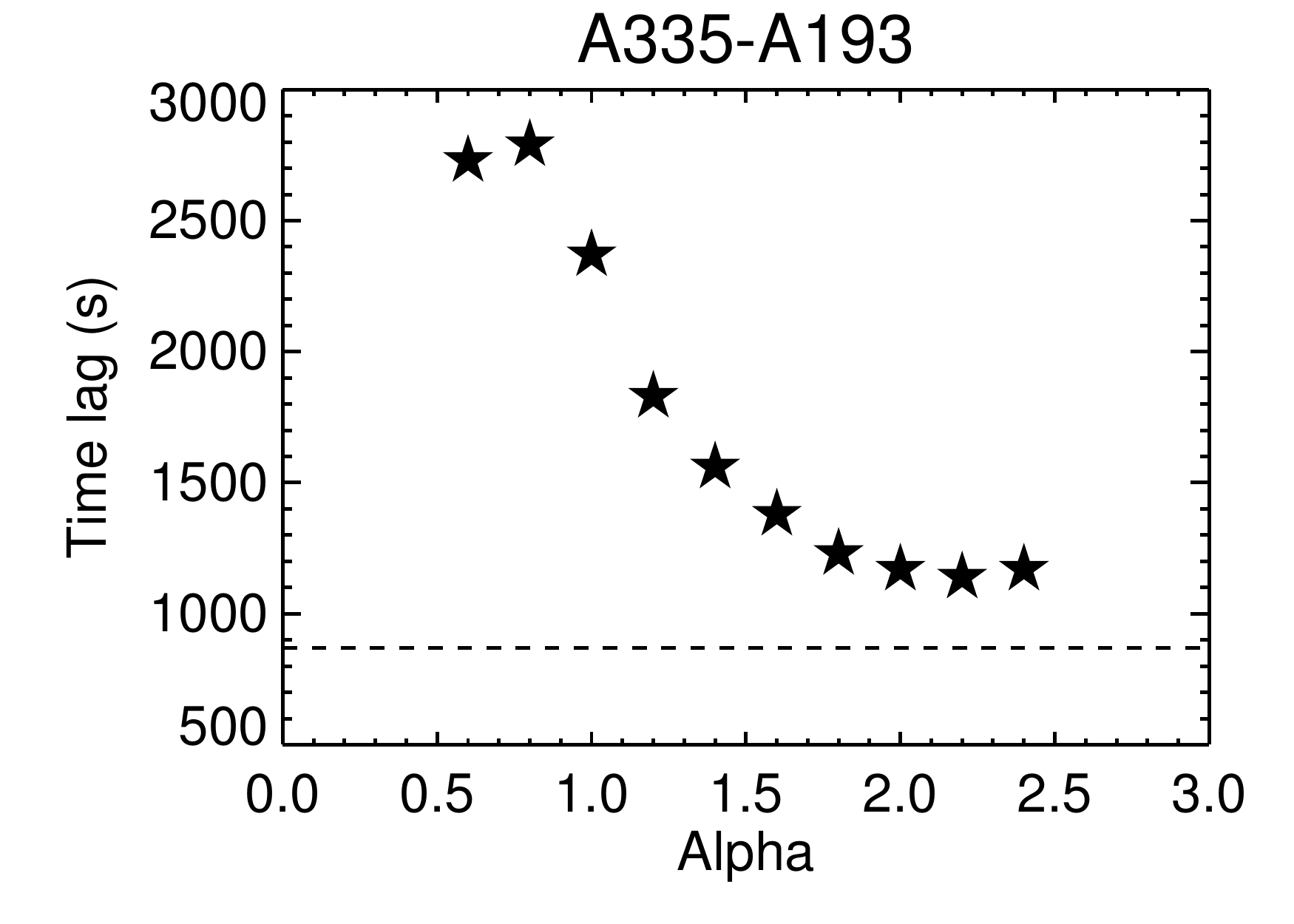}}}
\resizebox{.32\textwidth}{!}{\rotatebox{0}{\includegraphics{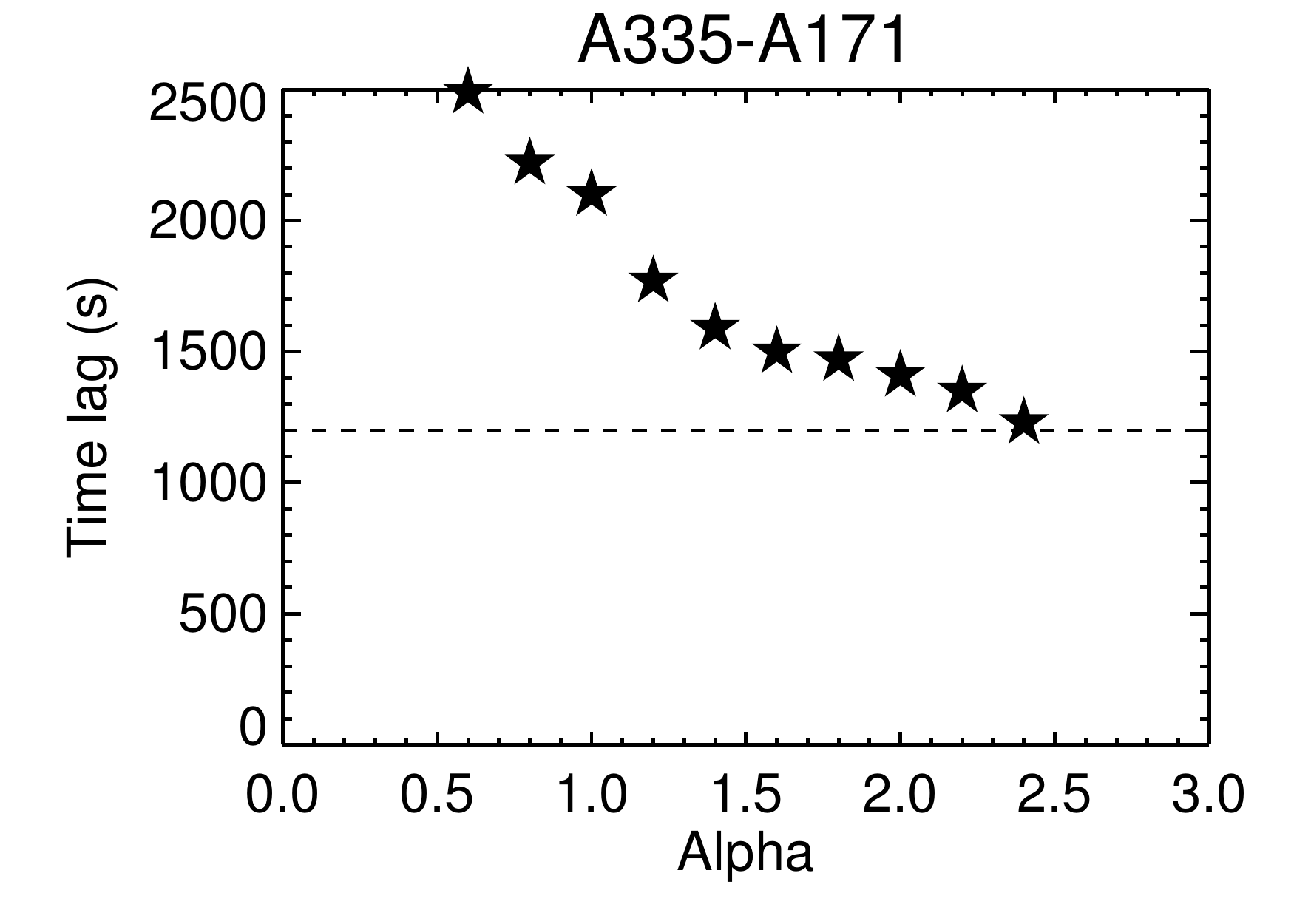}}}
\resizebox{.32\textwidth}{!}{\rotatebox{0}{\includegraphics{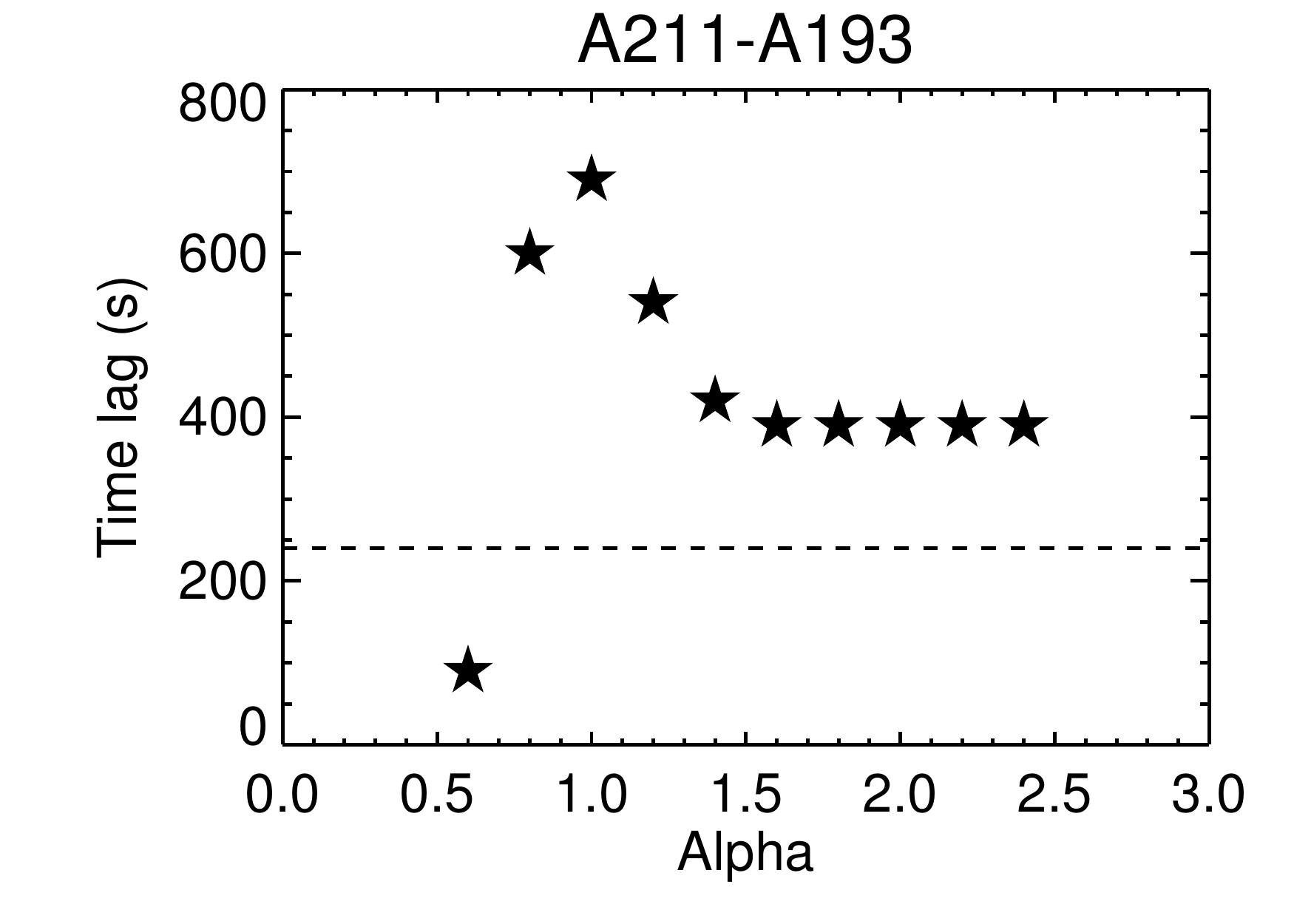}}}
\resizebox{.32\textwidth}{!}{\rotatebox{0}{\includegraphics{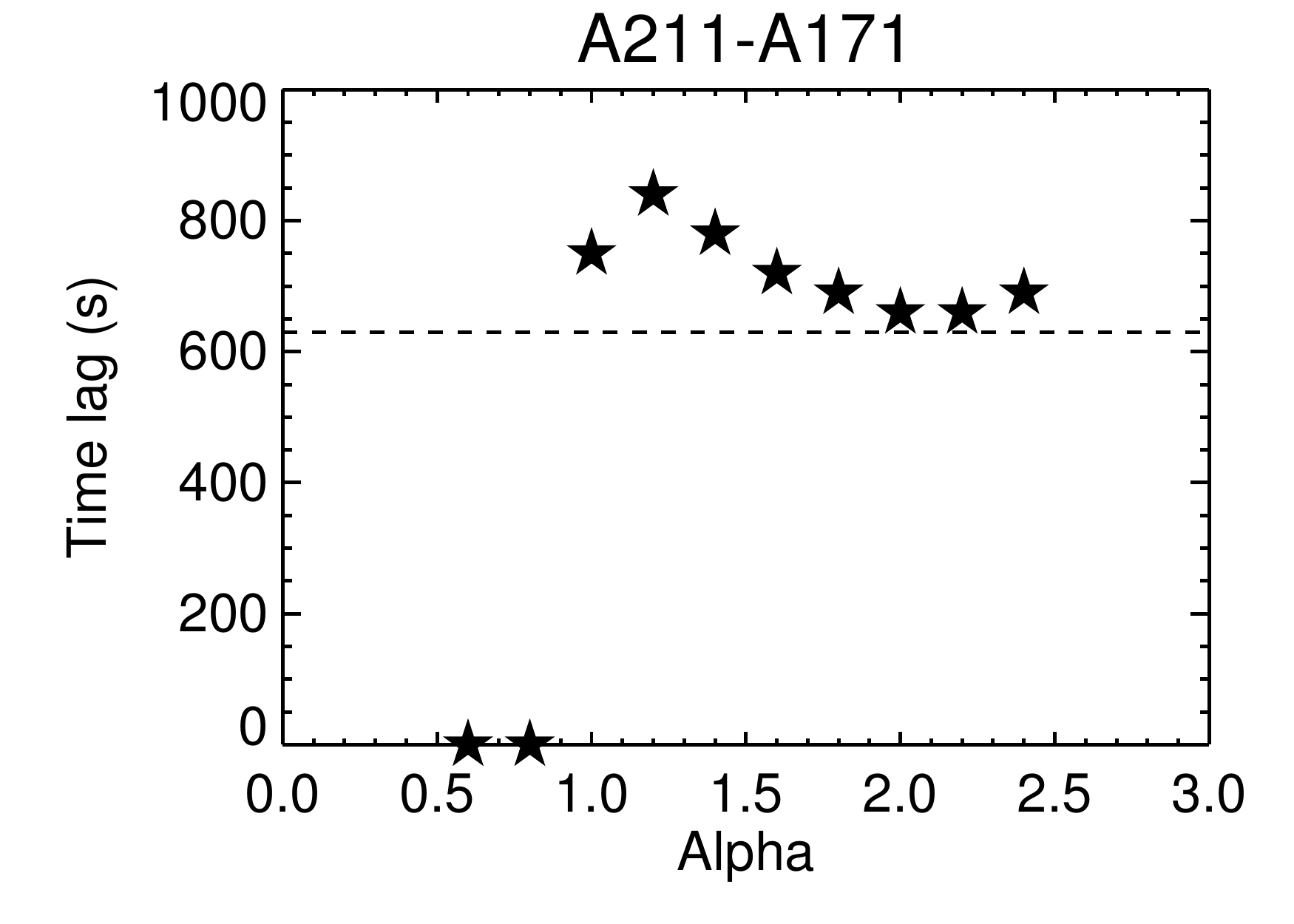}}}
\resizebox{.32\textwidth}{!}{\rotatebox{0}{\includegraphics{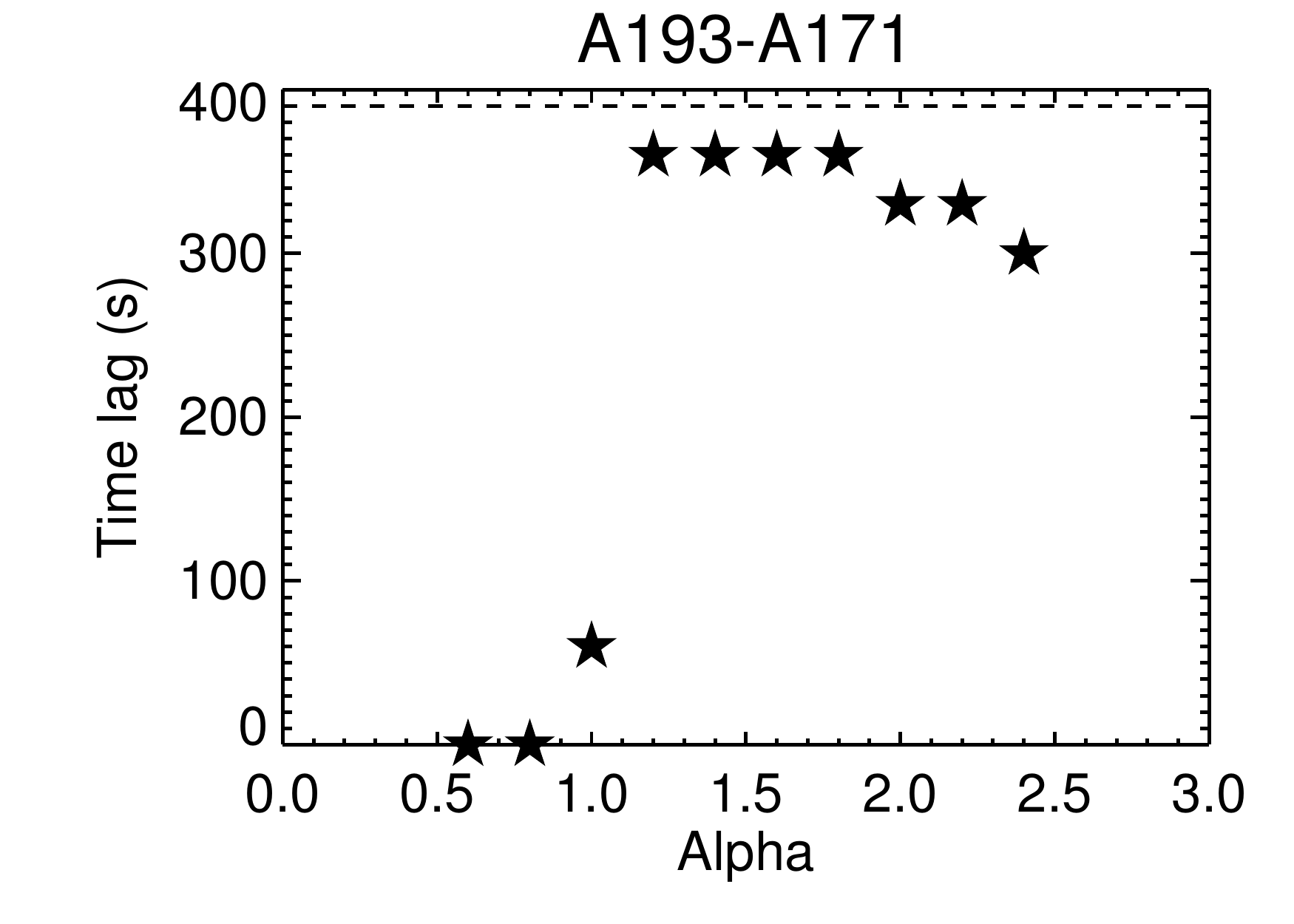}}}
\caption{Time lags for the different channel combinations as a function of alpha.  The footpoint solutions are shown with stars, the impulsive heating solution is represented as a dashed line.}
\label{fig:time_lag_exp}
\end{center}
\end{figure*}

\begin{figure*}[t!]
\begin{center}
\resizebox{.32\textwidth}{!}{\rotatebox{0}{\includegraphics{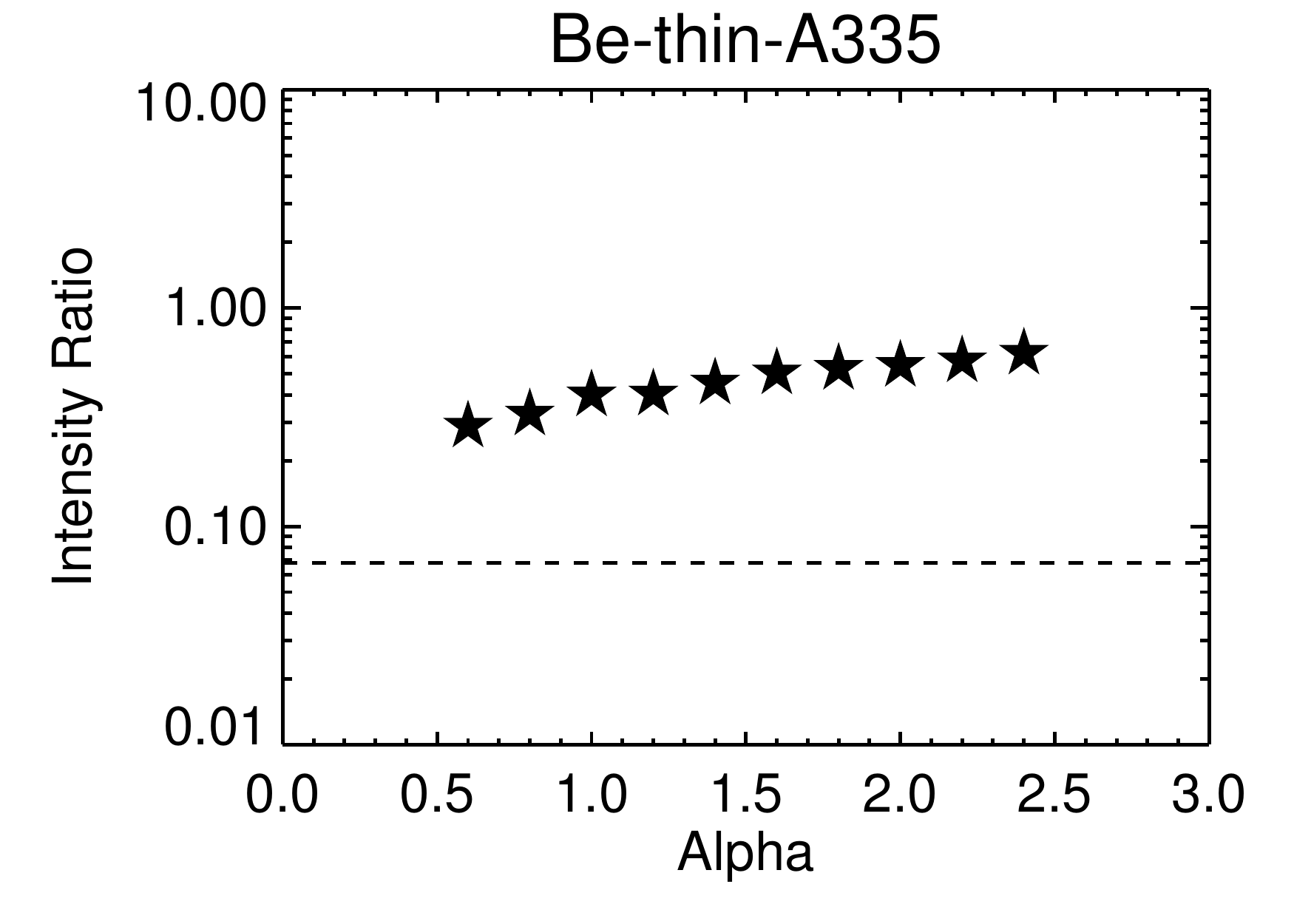}}}
\resizebox{.32\textwidth}{!}{\rotatebox{0}{\includegraphics{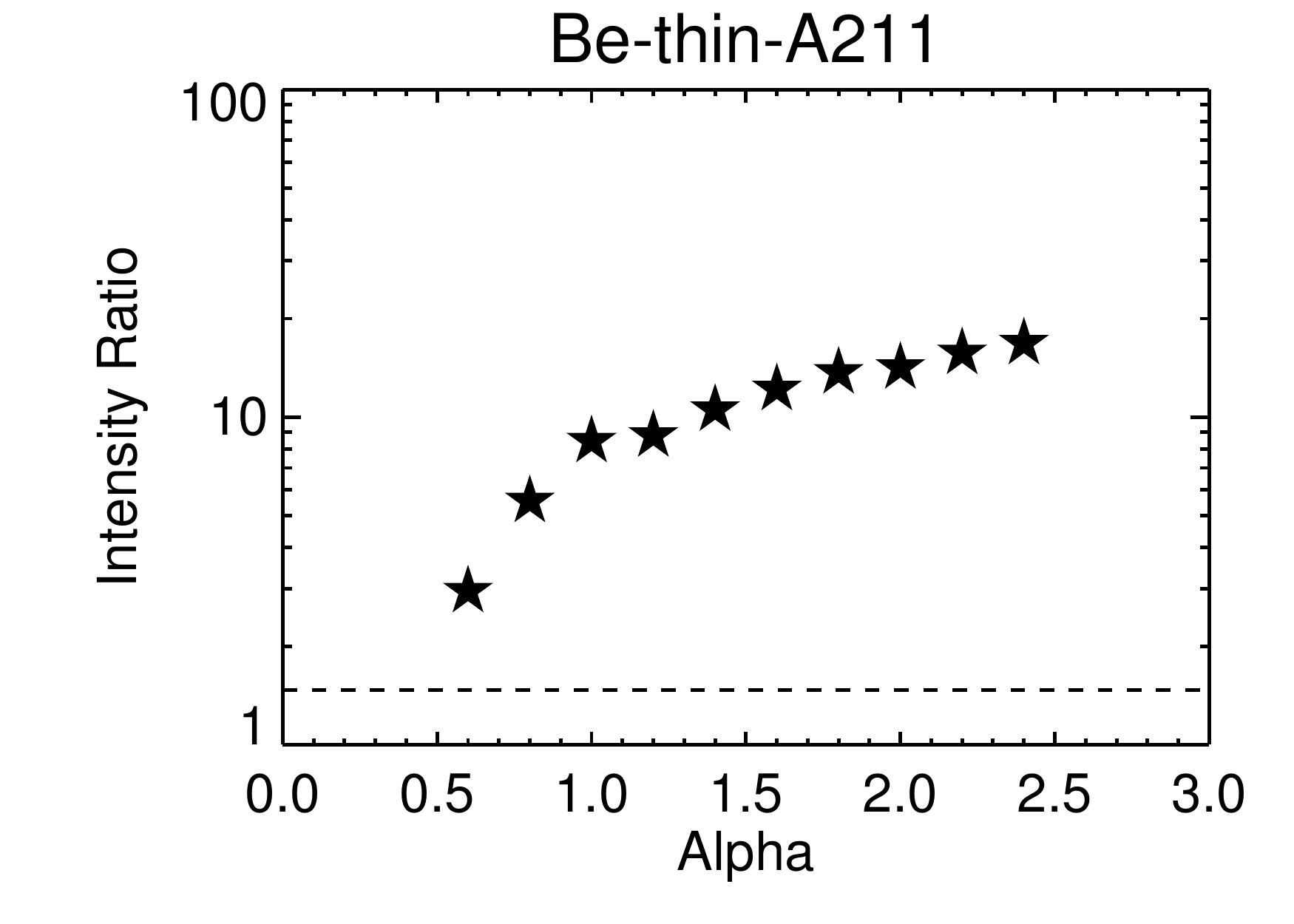}}}
\resizebox{.32\textwidth}{!}{\rotatebox{0}{\includegraphics{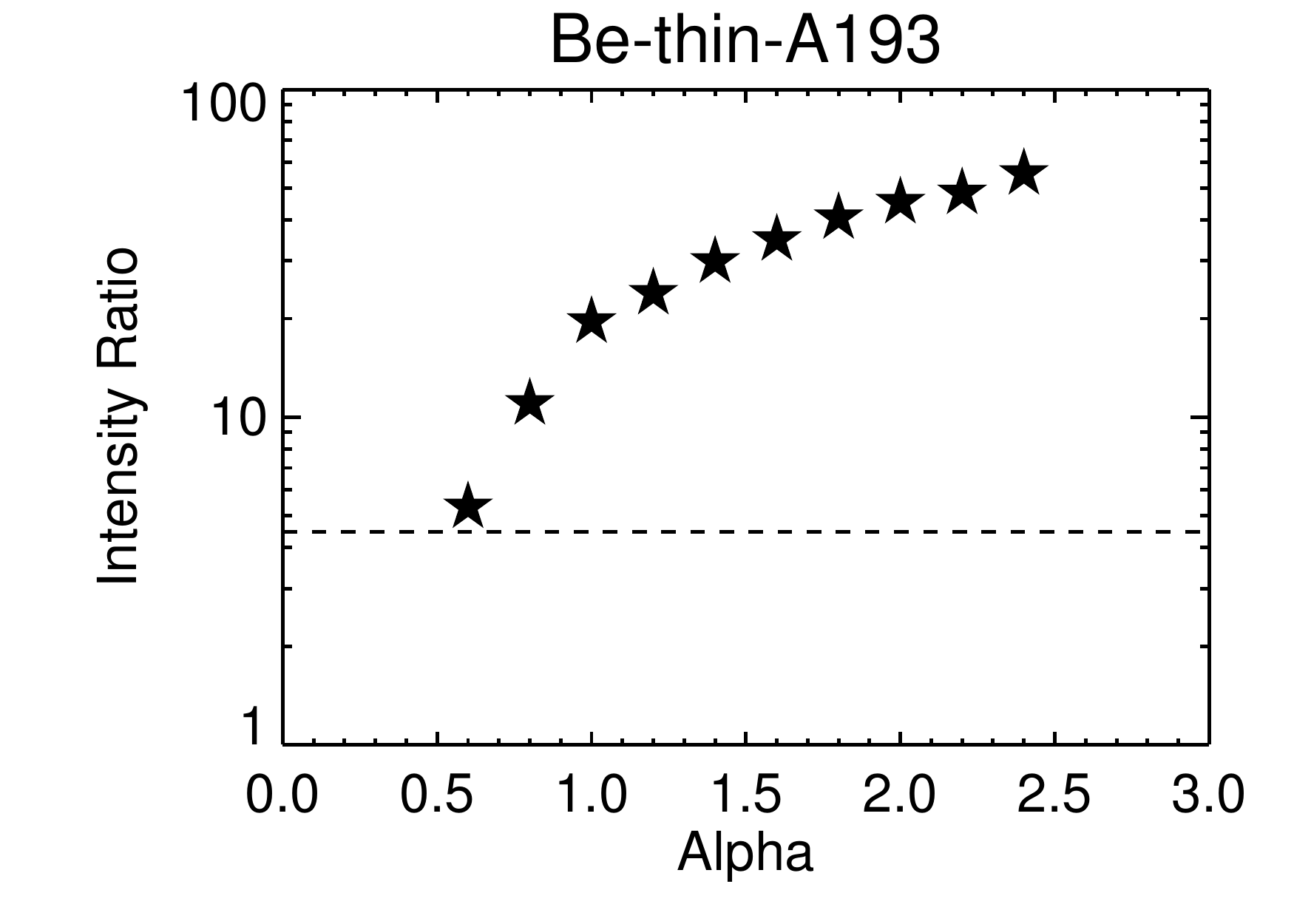}}}
\resizebox{.32\textwidth}{!}{\rotatebox{0}{\includegraphics{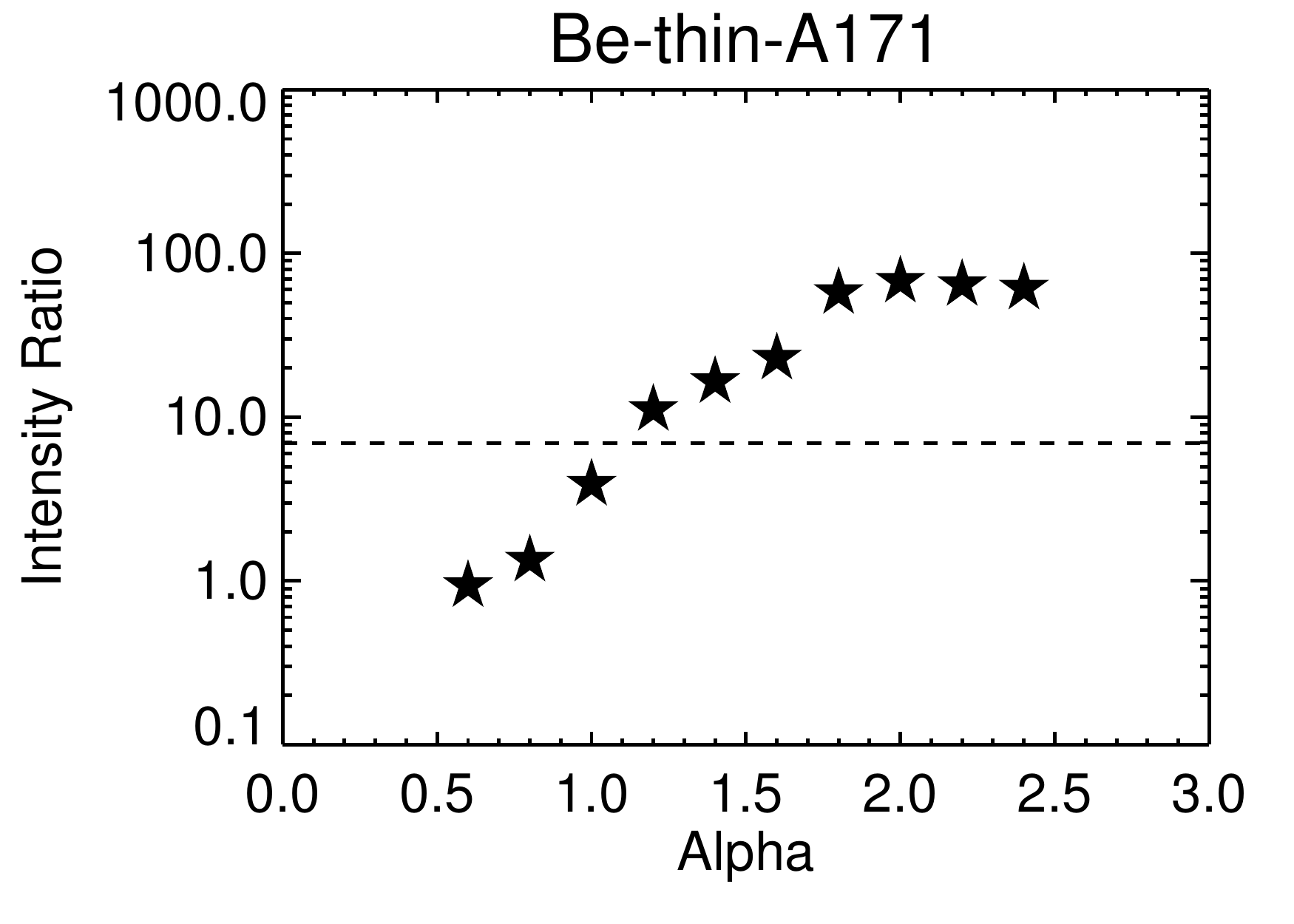}}}
\resizebox{.32\textwidth}{!}{\rotatebox{0}{\includegraphics{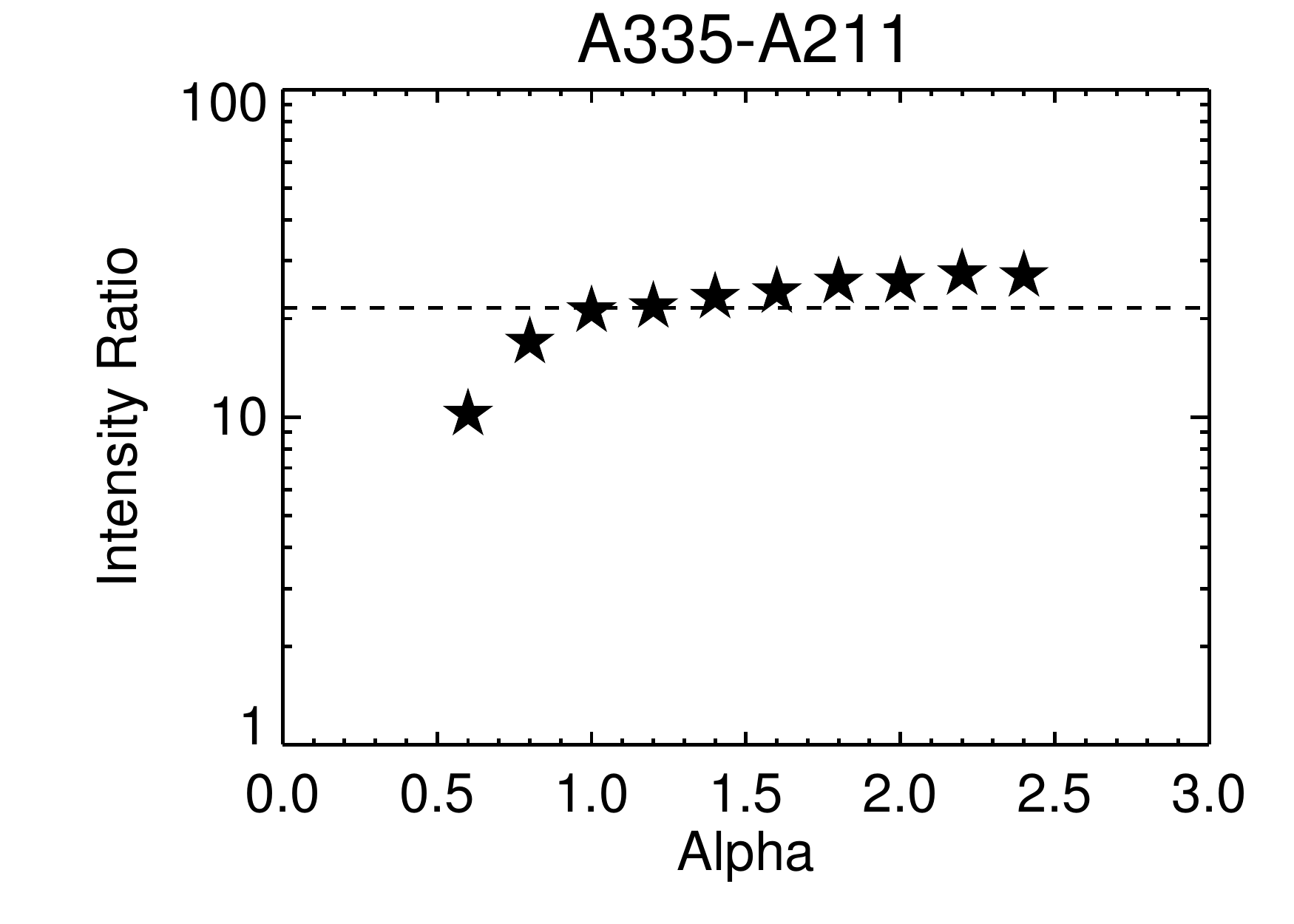}}}
\resizebox{.32\textwidth}{!}{\rotatebox{0}{\includegraphics{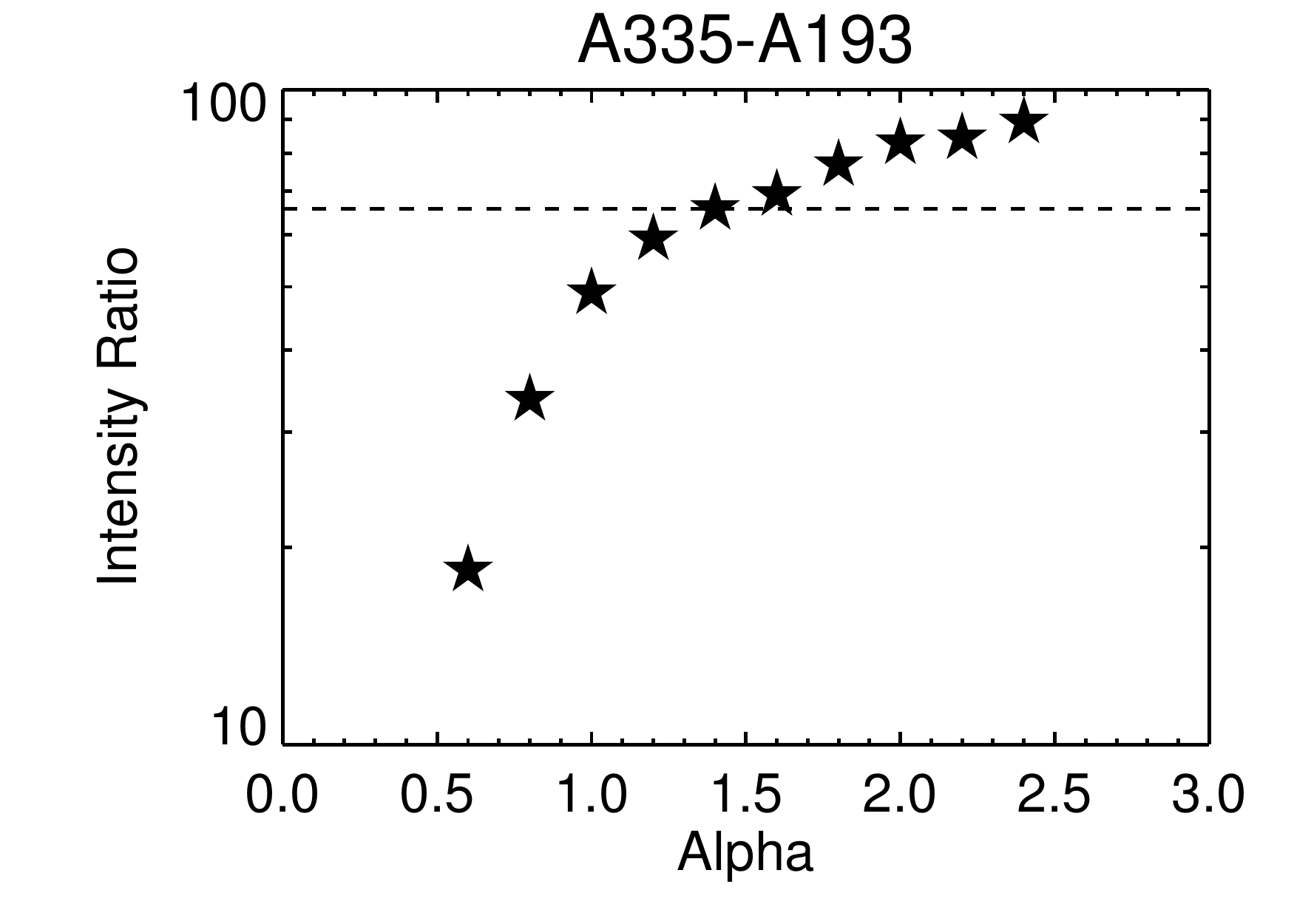}}}
\resizebox{.32\textwidth}{!}{\rotatebox{0}{\includegraphics{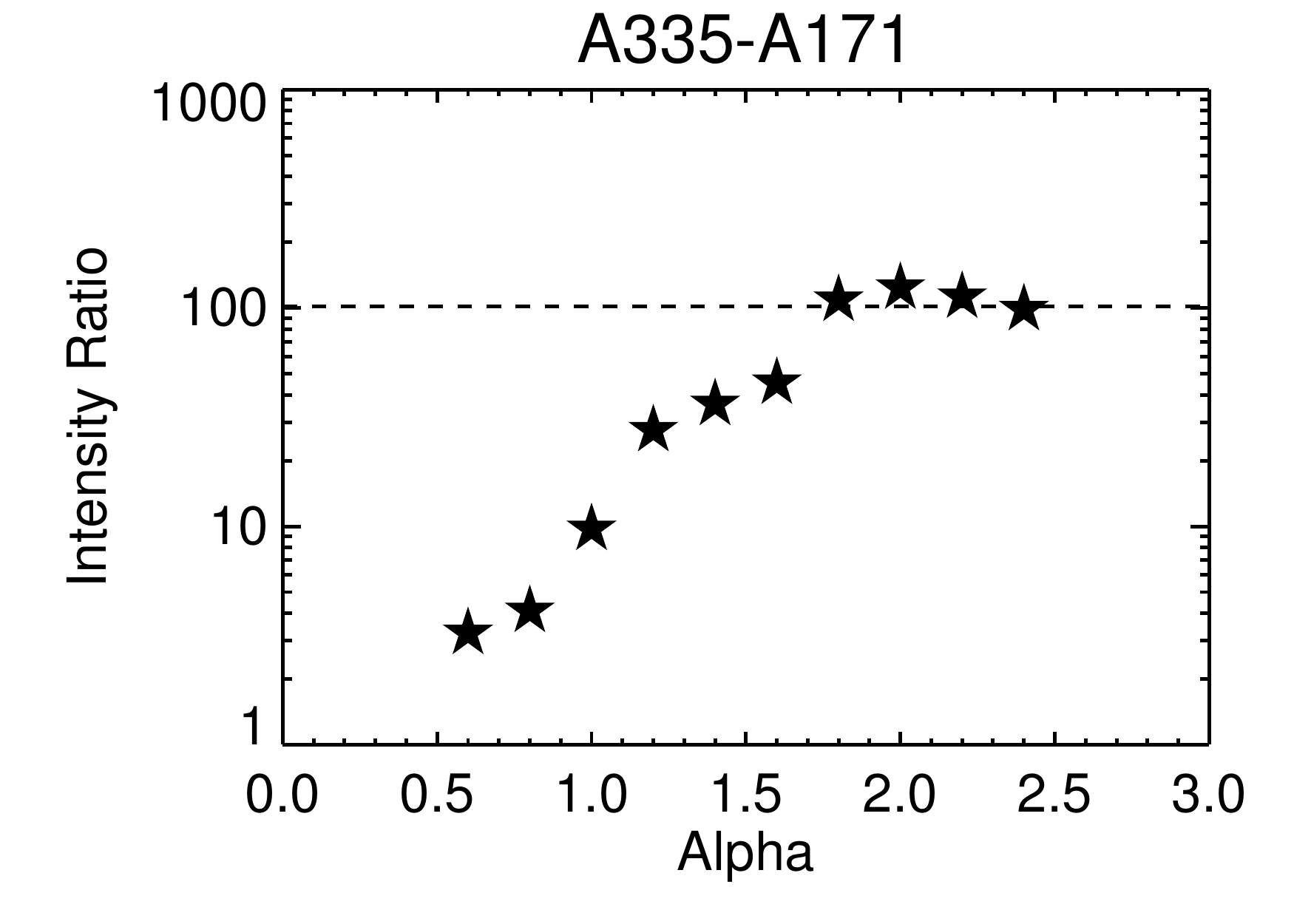}}}
\resizebox{.32\textwidth}{!}{\rotatebox{0}{\includegraphics{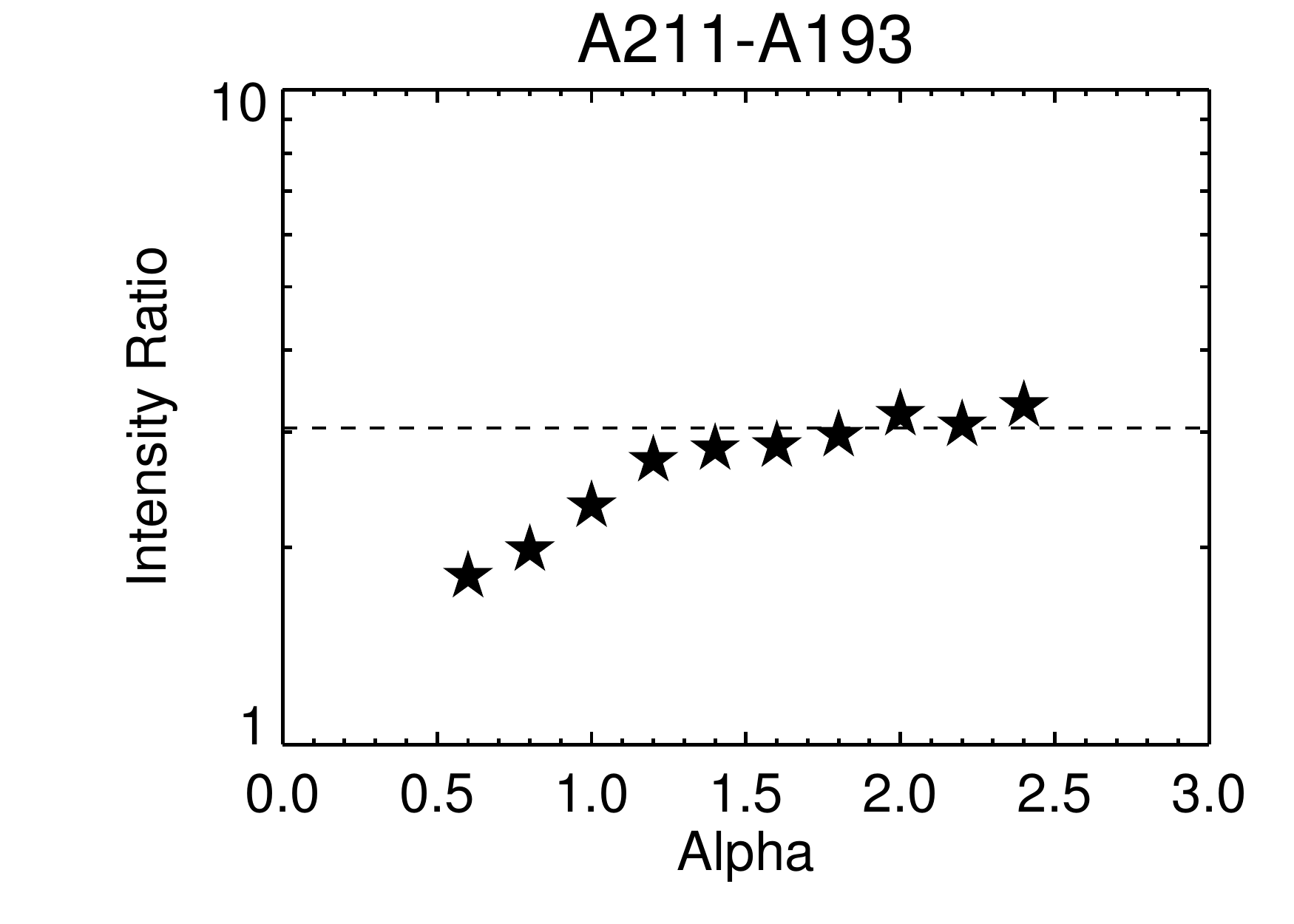}}}
\resizebox{.32\textwidth}{!}{\rotatebox{0}{\includegraphics{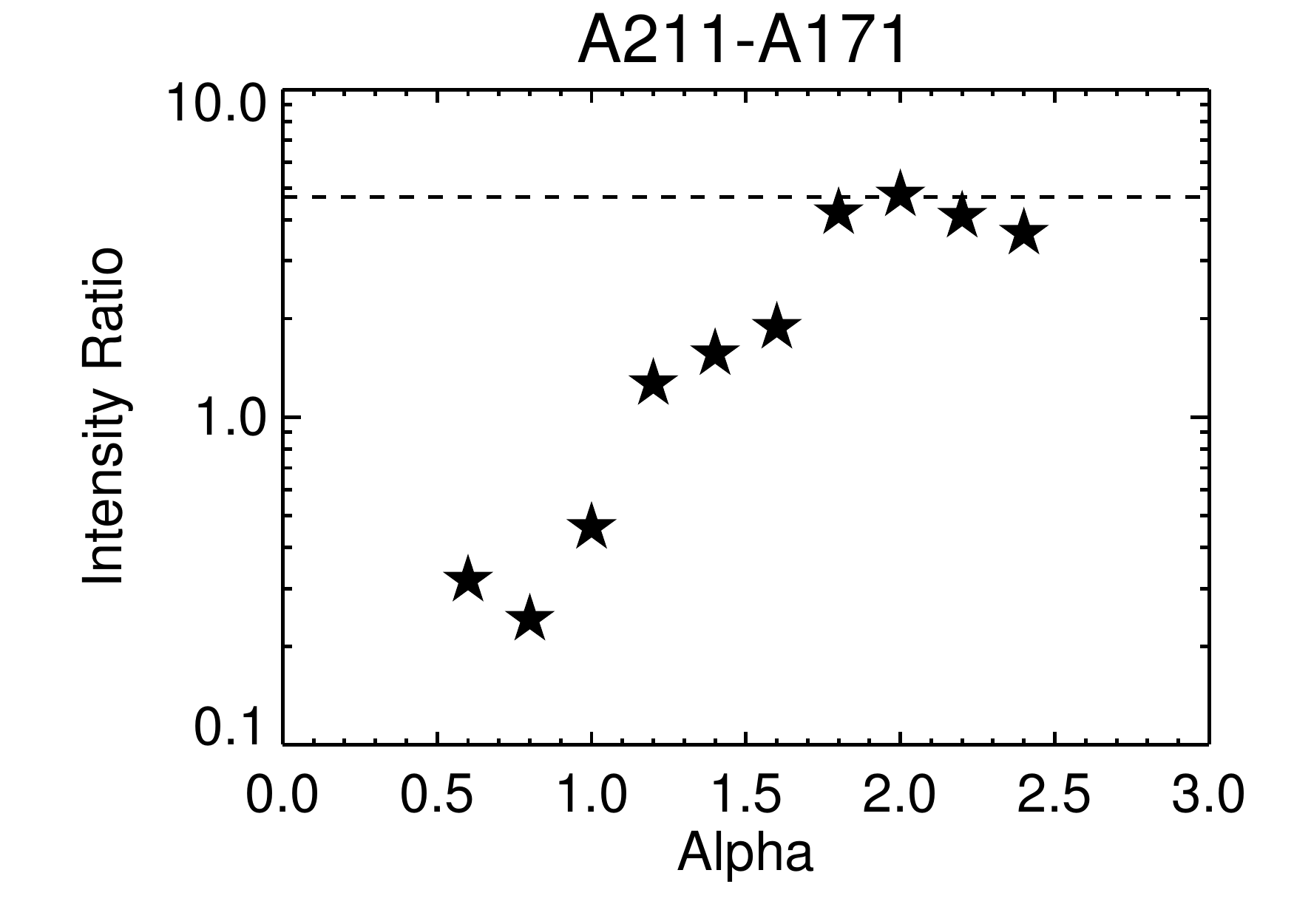}}}
\resizebox{.32\textwidth}{!}{\rotatebox{0}{\includegraphics{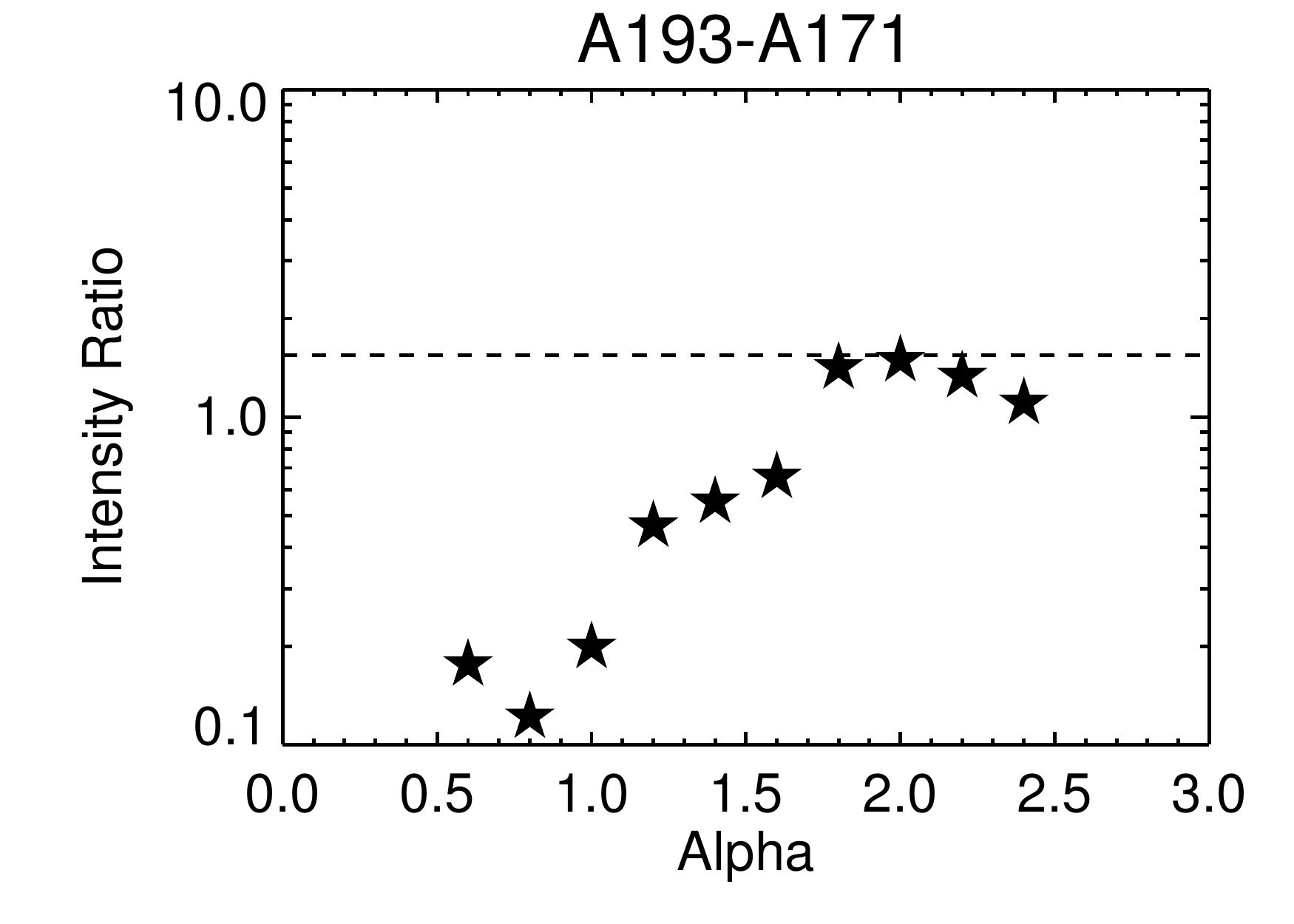}}}
\caption{The ratio of the peak intensity (Channel 2 / Channel 1) for  the different channel combinations as a function of alpha.  The footpoint solutions are shown with stars, the impulsive heating solution is represented as a dashed line.}
\label{fig:int_rat_exp}
\end{center}
\end{figure*}

Next, we take the apex temperature and density and fold it through the XRT Be-thin and four AIA channel response functions, shown in Figure~\ref{fig:resp}.  We calculate the time lag in the light curves between each channel pair and the relative intensity in the peak of the light curves.  We show these results in Figures~\ref{fig:time_lag_exp} and \ref{fig:int_rat_exp}.  We do not show time lags for intensity ratios for $\alpha < 0.5$ or $\alpha > 2.5$, as those solutions are essentially steady.  The time lags and intensity ratios are shown as a function $\alpha$.  The footpoint heating solutions are shown with stars.  The values for the comparable impulsive heated solution is shown as a horizontal line.  

In Figure~\ref{fig:time_lag_mag}, the time lags from footpoint heated solutions were typically longer than those of impulsive heating, particularly in the channels with the largest difference in the temperature response.  In this plot, we find that the time lags depend strongly on the stratification of the heating and can be both larger or smaller than predicted from impulsive heating.  As in Figure~\ref{fig:int_rat_mag}, we find there are wide range of intensity ratios predicted for footpoint heating.

\section{DISCUSSION}

There remains an ongoing debate whether cooling impulsively-heated loops or unsteady footpoint-heated loops can adequately address the observations of coronal loops.  Qualitatively, both candidate heating scenarios can predict loops that appear to cool and are overdense; however, quantitative comparisons are lacking.  In this paper, we have used a single loop geometry to help identify observables that can be used to discriminate between these two heating scenarios.    Because the nature of the solutions of the hydrodynamic equations depend strongly on the geometry of the loop in question, as well as the details of the heating function (particularly for footpoint heating), this is a preliminary study.    Regardless, several relevant indicators were found that may prove useful.  

Only impulsive heating solutions produced very hot ($> 5$ MK) plasma.  Additionally, the low-frequency, impulsively-heated loops investigated here exhibited a broader range of temperatures in general.  There may be other geometries or heating functions that could generate high temperature plasma with footpoint heating.  For instance, \cite{downs2016}  found a larger range of maximum and minimum temperatures by varying the heating parameters in a wave-turbulent model, which produces highly stratified heating similar to the heating studied in this paper.  

One potentially important observable is the time scale of the evolution of loops in different channels.  The cycle times of footpoint-heated solutions are typically longer than the cooling time of impulsively-heated plasma.  This can be understood by considering the energy equation in the hydrodynamic solution.  For impulsive heating, after the heating event concludes, the temperature change is dominated by the cooling mechanisms.  For footpoint heating, energy is always being added to the system, implying that the temperature would first change slowly during the most of the simulation, then rapidly at the end of the cycle.  This slowly-changing temperature early in the simulation implies that the time lags between high-temperature channels and cooler channels are much longer for footpoint heating than for impulsive heating.  We find the Be-thin-A193 and Be-thin-A171 time lags particularly long compared to the expected time lag for impulsive heating.  

Additionally, footpoint heating can generate very small or negative time lags in cooler channels.  This is related to the minimum temperature the plasma reaches in the simulation.  If the plasma does not cool all the way through the temperature responses of the channels, the light curves can be complicated and difficult to interpret.  Again, this is closely related to how the simulations are constructed; the minimum temperature the plasma reaches is likely highly dependent on the geometry and heating profile.
   
One new observable we identify in this paper is the ratio of the maximum intensities in the light curves in two different channels.  This is not usually measured from observations, in part due to the difficulty of subtracting the background intensity.  For impulsive heating, the intensity ratio between channel pairs generally falls within a very narrow range, almost independent of the magnitude of the heating.  This is due to the fact that the density is draining slowly in impulsive heating, so the relative intensity expected in two channel pairs is related to the relative responses in those pairs.  In contrast, there are a wider range of predicted intensity ratios in footpoint heating.  

When we changed the stratification of the heating, we found that solutions that cycle through a large range of temperature due to thermal non-equilibrium exist only for a specific range of heating stratifications.  When the heating scale height is large, the heating is essentially uniform and the solution becomes steady.  When the heating is highly stratified and asymmetric, a siphon flow is established between the footpoints and the solution returns to a near-steady-state solution.  We find that the expected time lags and intensity ratios vary as a function of the stratification when thermal non-equilibrium solutions are present.

Finally, we would like to emphasize that the limitations of this work.  First, we have only considered a single loop geometry.   Due to this limitation we cannot at present make firm conclusions that observed time lags above or below a certain threshold are not compatible with impulsive or footpoint heating.  Instead, the complete range of loop lengths and geometries in the observed active region must be considered  before making a determination.  For instance, \cite{lionello2016} argued that the longest time lags in the \cite{viall2012} active region were incompatible with low-frequency impulsive heating based on the loop lengths present in that active region.  However, that does not imply that the shorter time lags observed are compatible with impulsive heating or that footpoint heating could recreate the time lags.   Larger-scale simulations that account for multiple structures along the line of sight and have realistic loop lengths and geometries must be considered.  In the future, we plan to simulate an arcade of loops to determine how the presence of multiple structures along the line of sight impacts our  conclusions.  Additionally, we will broaden our considered observables to include the expected differential emission measure (DEM).  

Next, we have oversimplified the footpoint heating function and ignored any potential evolution of the magnetic field.  On the Sun, there is very low probability that the magnetic field or heating magnitudes will be constant for the many hours of evolution we have calculated in these simulations.  A constant geometry and heating magnitude implies the loops would reappear periodically.  Indeed, this periodicity has been observed in very localized regions on the Sun (see, for instance, \cite{froment2017}), but it cannot be expected to be the norm.  Evolution of the field and changes in the footpoint heating strength will impact the plasma evolution and disrupt the expected periodicity.  Additionally, it is likely that instead of completely steady heating, we would expect very frequent bursts of heating at the footpoints of the loops as predicted by \cite{asgaritarghi2012}.  We do not expect this simplification to impact the results except in the case where the frequency of the heating at the footpoint drops significantly.

Finally, we have only considered one (very low) frequency of heating for impulsive heating and only considered entirely impulsive heating (meaning the time scale of the heating is much shorter than the cooling time).  If the heating events were more extended, such as those found by \cite{reale2000b}, it would impact the predicted lightcurves and potentially these conclusions.  As demonstrated by \cite{barnes2016a,barnes2016b}, the modeling and detection of hot plasma associated with impulsive heating events is more complicated than we have treated it in this paper.  First it is not only the presence of high temperature plasma, but also the amount of high temperature plasma relative to moderate temperature plasma that is a constraint on heating frequency.  Second, the amount of high temperature plasma that can be observed depends on the level to which the plasma has had the opportunity to ionize which also depends on the heating duration.  Such work is outside the scope of this preliminary study.  

\acknowledgements

We would like to acknowledge the anonymous referee for helpful comments.  This  work was partially supported by NASA's Heliophysics Supporting Research Grant NNX16AH03G.


\begin{thebibliography}{68}
\expandafter\ifx\csname natexlab\endcsname\relax\def\natexlab#1{#1}\fi

\bibitem[{{Antiochos} \& {Klimchuk}(1991)}]{1991ApJ...378..372A}
{Antiochos}, S.~K., \& {Klimchuk}, J.~A. 1991, \apj, 378, 372

\bibitem[{{Antiochos} {et~al.}(1999){Antiochos}, {MacNeice}, {Spicer}, \&
  {Klimchuk}}]{1999ApJ...512..985A}
{Antiochos}, S.~K., {MacNeice}, P.~J., {Spicer}, D.~S., \& {Klimchuk}, J.~A.
  1999, \apj, 512, 985

\bibitem[{{Antolin} {et~al.}(2010){Antolin}, {Shibata}, \&
  {Vissers}}]{antolin2010}
{Antolin}, P., {Shibata}, K., \& {Vissers}, G. 2010, \apj, 716, 154

\bibitem[{{Antolin} {et~al.}(2015){Antolin}, {Vissers}, {Pereira}, {Rouppe van
  der Voort}, \& {Scullion}}]{antolin2015}
{Antolin}, P., {Vissers}, G., {Pereira}, T.~M.~D., {Rouppe van der Voort}, L.,
  \& {Scullion}, E. 2015, \apj, 806, 81

\bibitem[{{Antolin} {et~al.}(2012){Antolin}, {Vissers}, \& {Rouppe van der
  Voort}}]{antolin2012}
{Antolin}, P., {Vissers}, G., \& {Rouppe van der Voort}, L. 2012, \solphys,
  280, 457

\bibitem[{{Aschwanden}(2005)}]{2005ApJ...634L.193A}
{Aschwanden}, M.~J. 2005, \apjl, 634, L193

\bibitem[{{Aschwanden} {et~al.}(2000){Aschwanden}, {Nightingale}, \&
  {Alexander}}]{aschwanden2000a}
{Aschwanden}, M.~J., {Nightingale}, R.~W., \& {Alexander}, D. 2000, \apj, 541,
  1059

\bibitem[{{Aschwanden} {et~al.}(2008){Aschwanden}, {Nitta}, {Wuelser}, \&
  {Lemen}}]{2008ApJ...680.1477A}
{Aschwanden}, M.~J., {Nitta}, N.~V., {Wuelser}, J.-P., \& {Lemen}, J.~R. 2008,
  \apj, 680, 1477

\bibitem[{{Asgari-Targhi} \& {van Ballegooijen}(2012)}]{asgaritarghi2012}
{Asgari-Targhi}, M., \& {van Ballegooijen}, A.~A. 2012, \apj, 746, 81

\bibitem[{{Auch{\`e}re} {et~al.}(2014){Auch{\`e}re}, {Bocchialini}, {Solomon},
  \& {Tison}}]{auchere2014}
{Auch{\`e}re}, F., {Bocchialini}, K., {Solomon}, J., \& {Tison}, E. 2014, \aap,
  563, A8

\bibitem[{{Auch{\`e}re} {et~al.}(2016){Auch{\`e}re}, {Froment}, {Bocchialini},
  {Buchlin}, \& {Solomon}}]{auchere2016}
{Auch{\`e}re}, F., {Froment}, C., {Bocchialini}, K., {Buchlin}, E., \&
  {Solomon}, J. 2016, \apj, 825, 110

\bibitem[{{Barnes} {et~al.}(2016{\natexlab{a}}){Barnes}, {Cargill}, \&
  {Bradshaw}}]{barnes2016a}
{Barnes}, W.~T., {Cargill}, P.~J., \& {Bradshaw}, S.~J. 2016{\natexlab{a}},
  \apj, 829, 31

\bibitem[{{Barnes} {et~al.}(2016{\natexlab{b}}){Barnes}, {Cargill}, \&
  {Bradshaw}}]{barnes2016b}
---. 2016{\natexlab{b}}, \apj, 833, 217

\bibitem[{{Bradshaw} \& {Klimchuk}(2015)}]{bradshaw2015}
{Bradshaw}, S.~J., \& {Klimchuk}, J.~A. 2015, \apj, 811, 129

\bibitem[{{Bradshaw} {et~al.}(2012){Bradshaw}, {Klimchuk}, \&
  {Reep}}]{bradshaw2012}
{Bradshaw}, S.~J., {Klimchuk}, J.~A., \& {Reep}, J.~W. 2012, \apj, 758, 53

\bibitem[{{Bradshaw} \& {Viall}(2016)}]{bradshaw2016}
{Bradshaw}, S.~J., \& {Viall}, N.~M. 2016, \apj, 821, 63

\bibitem[{{Cargill}(2014)}]{cargill2014}
{Cargill}, P.~J. 2014, \apj, 784, 49

\bibitem[{{Cargill} \& {Klimchuk}(1997)}]{cargill1997}
{Cargill}, P.~J., \& {Klimchuk}, J.~A. 1997, \apj, 478, 799+

\bibitem[{{Cargill} \& {Klimchuk}(2004)}]{cargill2004}
---. 2004, \apj, 605, 911

\bibitem[{{Cargill} {et~al.}(1995){Cargill}, {Mariska}, \&
  {Antiochos}}]{cargill1995}
{Cargill}, P.~J., {Mariska}, J.~T., \& {Antiochos}, S.~K. 1995, \apj, 439, 1034

\bibitem[{{Cargill} {et~al.}(2015){Cargill}, {Warren}, \&
  {Bradshaw}}]{cargill2015}
{Cargill}, P.~J., {Warren}, H.~P., \& {Bradshaw}, S.~J. 2015, Philosophical
  Transactions of the Royal Society of London Series A, 373, 20140260

\bibitem[{{Dere} {et~al.}(1997){Dere}, {Landi}, {Mason}, {Monsignori Fossi}, \&
  {Young}}]{1997A&AS..125..149D}
{Dere}, K.~P., {Landi}, E., {Mason}, H.~E., {Monsignori Fossi}, B.~C., \&
  {Young}, P.~R. 1997, \aaps, 125, 149

\bibitem[{{Dere} {et~al.}(2009){Dere}, {Landi}, {Young}, {Del Zanna},
  {Landini}, \& {Mason}}]{2009A&A...498..915D}
{Dere}, K.~P., {Landi}, E., {Young}, P.~R., {Del Zanna}, G., {Landini}, M., \&
  {Mason}, H.~E. 2009, \aap, 498, 915

\bibitem[{{Downs} {et~al.}(2016){Downs}, {Lionello}, {Miki{\'c}}, {Linker}, \&
  {Velli}}]{downs2016}
{Downs}, C., {Lionello}, R., {Miki{\'c}}, Z., {Linker}, J.~A., \& {Velli}, M.
  2016, \apj, 832, 180

\bibitem[{{Feldman} {et~al.}(1992){Feldman}, {Mandelbaum}, {Seely}, {Doschek},
  \& {Gursky}}]{1992ApJS...81..387F}
{Feldman}, U., {Mandelbaum}, P., {Seely}, J.~F., {Doschek}, G.~A., \& {Gursky},
  H. 1992, \apjs, 81, 387

\bibitem[{{Froment} {et~al.}(2017){Froment}, {Auch{\`e}re}, {Aulanier},
  {Miki{\'c}}, {Bocchialini}, {Buchlin}, \& {Solomon}}]{froment2017}
{Froment}, C., {Auch{\`e}re}, F., {Aulanier}, G., {Miki{\'c}}, Z.,
  {Bocchialini}, K., {Buchlin}, E., \& {Solomon}, J. 2017, \apj, 835, 272

\bibitem[{{Froment} {et~al.}(2015){Froment}, {Auch{\`e}re}, {Bocchialini},
  {Buchlin}, {Guennou}, \& {Solomon}}]{froment2015}
{Froment}, C., {Auch{\`e}re}, F., {Bocchialini}, K., {Buchlin}, E., {Guennou},
  C., \& {Solomon}, J. 2015, \apj, 807, 158

\bibitem[{{Froment} {et~al.}(2018){Froment}, {Auch{\`e}re}, {Miki{\'c}},
  {Aulanier}, {Bocchialini}, {Buchlin}, {Solomon}, \&
  {Soubri{\'e}}}]{froment2018}
{Froment}, C., {Auch{\`e}re}, F., {Miki{\'c}}, Z., {Aulanier}, G.,
  {Bocchialini}, K., {Buchlin}, E., {Solomon}, J., \& {Soubri{\'e}}, E. 2018,
  \apj, 855, 52

\bibitem[{{Golub} {et~al.}(2007){Golub}, {Deluca}, {Austin}, {Bookbinder},
  {Caldwell}, {Cheimets}, {Cirtain}, {Cosmo}, {Reid}, {Sette}, {Weber},
  {Sakao}, {Kano}, {Shibasaki}, {Hara}, {Tsuneta}, {Kumagai}, {Tamura},
  {Shimojo}, {McCracken}, {Carpenter}, {Haight}, {Siler}, {Wright}, {Tucker},
  {Rutledge}, {Barbera}, {Peres}, \& {Varisco}}]{golub2007}
{Golub}, L., {Deluca}, E., {Austin}, G., {Bookbinder}, J., {Caldwell}, D.,
  {Cheimets}, P., {Cirtain}, J., {Cosmo}, M., {Reid}, P., {Sette}, A., {Weber},
  M., {Sakao}, T., {Kano}, R., {Shibasaki}, K., {Hara}, H., {Tsuneta}, S.,
  {Kumagai}, K., {Tamura}, T., {Shimojo}, M., {McCracken}, J., {Carpenter}, J.,
  {Haight}, H., {Siler}, R., {Wright}, E., {Tucker}, J., {Rutledge}, H.,
  {Barbera}, M., {Peres}, G., \& {Varisco}, S. 2007, \solphys, 243, 63

\bibitem[{{Grevesse} \& {Sauval}(1998)}]{1998SSRv...85..161G}
{Grevesse}, N., \& {Sauval}, A.~J. 1998, \ssr, 85, 161

\bibitem[{{Kamio} {et~al.}(2011){Kamio}, {Peter}, {Curdt}, \&
  {Solanki}}]{kamio2011}
{Kamio}, S., {Peter}, H., {Curdt}, W., \& {Solanki}, S.~K. 2011, \aap, 532,
  A96+

\bibitem[{{Karpen} {et~al.}(2001){Karpen}, {Antiochos}, {Hohensee}, {Klimchuk},
  \& {MacNeice}}]{2001ApJ...553L..85K}
{Karpen}, J.~T., {Antiochos}, S.~K., {Hohensee}, M., {Klimchuk}, J.~A., \&
  {MacNeice}, P.~J. 2001, \apjl, 553, L85

\bibitem[{{Karpen} {et~al.}(2006){Karpen}, {Antiochos}, \&
  {Klimchuk}}]{2006ApJ...637..531K}
{Karpen}, J.~T., {Antiochos}, S.~K., \& {Klimchuk}, J.~A. 2006, \apj, 637, 531

\bibitem[{{Karpen} {et~al.}(2003){Karpen}, {Antiochos}, {Klimchuk}, \&
  {MacNeice}}]{2003ApJ...593.1187K}
{Karpen}, J.~T., {Antiochos}, S.~K., {Klimchuk}, J.~A., \& {MacNeice}, P.~J.
  2003, \apj, 593, 1187

\bibitem[{{Klimchuk}(2006)}]{klimchuk2006}
{Klimchuk}, J.~A. 2006, \solphys, 234, 41

\bibitem[{{Klimchuk} {et~al.}(2010){Klimchuk}, {Karpen}, \&
  {Antiochos}}]{klimchuk2010}
{Klimchuk}, J.~A., {Karpen}, J.~T., \& {Antiochos}, S.~K. 2010, \apj, 714, 1239

\bibitem[{{Kuin} \& {Martens}(1982)}]{1982A&A...108L...1K}
{Kuin}, N.~P.~M., \& {Martens}, P.~C.~H. 1982, \aap, 108, L1

\bibitem[{{Landi} {et~al.}(2002){Landi}, {Feldman}, \&
  {Dere}}]{2002ApJS..139..281L}
{Landi}, E., {Feldman}, U., \& {Dere}, K.~P. 2002, \apjs, 139, 281

\bibitem[{{Landi} {et~al.}(2013){Landi}, {Young}, {Dere}, {Del Zanna}, \&
  {Mason}}]{2013ApJ...763...86L}
{Landi}, E., {Young}, P.~R., {Dere}, K.~P., {Del Zanna}, G., \& {Mason}, H.~E.
  2013, \apj, 763, 86

\bibitem[{{Lemen} {et~al.}(2012){Lemen}, {Title}, {Akin}, {Boerner}, {Chou},
  {Drake}, {Duncan}, {Edwards}, {Friedlaender}, {Heyman}, {Hurlburt}, {Katz},
  {Kushner}, {Levay}, {Lindgren}, {Mathur}, {McFeaters}, {Mitchell}, {Rehse},
  {Schrijver}, {Springer}, {Stern}, {Tarbell}, {Wuelser}, {Wolfson}, {Yanari},
  {Bookbinder}, {Cheimets}, {Caldwell}, {Deluca}, {Gates}, {Golub}, {Park},
  {Podgorski}, {Bush}, {Scherrer}, {Gummin}, {Smith}, {Auker}, {Jerram},
  {Pool}, {Soufli}, {Windt}, {Beardsley}, {Clapp}, {Lang}, \&
  {Waltham}}]{lemen2012}
{Lemen}, J.~R., {Title}, A.~M., {Akin}, D.~J., {Boerner}, P.~F., {Chou}, C.,
  {Drake}, J.~F., {Duncan}, D.~W., {Edwards}, C.~G., {Friedlaender}, F.~M.,
  {Heyman}, G.~F., {Hurlburt}, N.~E., {Katz}, N.~L., {Kushner}, G.~D., {Levay},
  M., {Lindgren}, R.~W., {Mathur}, D.~P., {McFeaters}, E.~L., {Mitchell}, S.,
  {Rehse}, R.~A., {Schrijver}, C.~J., {Springer}, L.~A., {Stern}, R.~A.,
  {Tarbell}, T.~D., {Wuelser}, J.-P., {Wolfson}, C.~J., {Yanari}, C.,
  {Bookbinder}, J.~A., {Cheimets}, P.~N., {Caldwell}, D., {Deluca}, E.~E.,
  {Gates}, R., {Golub}, L., {Park}, S., {Podgorski}, W.~A., {Bush}, R.~I.,
  {Scherrer}, P.~H., {Gummin}, M.~A., {Smith}, P., {Auker}, G., {Jerram}, P.,
  {Pool}, P., {Soufli}, R., {Windt}, D.~L., {Beardsley}, S., {Clapp}, M.,
  {Lang}, J., \& {Waltham}, N. 2012, \solphys, 275, 17

\bibitem[{{Lenz} {et~al.}(1999){Lenz}, {Deluca}, {Golub}, {Rosner}, \&
  {Bookbinder}}]{lenz1999}
{Lenz}, D.~D., {Deluca}, E.~E., {Golub}, L., {Rosner}, R., \& {Bookbinder},
  J.~A. 1999, \apjl, 517, L155

\bibitem[{{Lionello} {et~al.}(2016){Lionello}, {Alexander}, {Winebarger},
  {Linker}, \& {Miki{\'c}}}]{lionello2016}
{Lionello}, R., {Alexander}, C.~E., {Winebarger}, A.~R., {Linker}, J.~A., \&
  {Miki{\'c}}, Z. 2016, \apj, 818, 129

\bibitem[{{Lionello} {et~al.}(2009){Lionello}, {Linker}, \&
  {Miki{\'c}}}]{2009ApJ...690..902L}
{Lionello}, R., {Linker}, J.~A., \& {Miki{\'c}}, Z. 2009, \apj, 690, 902

\bibitem[{{Lionello} {et~al.}(2013){Lionello}, {Winebarger}, {Mok}, {Linker},
  \& {Miki{\'c}}}]{lionello2013}
{Lionello}, R., {Winebarger}, A.~R., {Mok}, Y., {Linker}, J.~A., \&
  {Miki{\'c}}, Z. 2013, \apj, 773, 134

\bibitem[{{Martens} \& {Kuin}(1983)}]{1983A&A...123..216M}
{Martens}, P.~C.~H., \& {Kuin}, N.~P.~M. 1983, \aap, 123, 216

\bibitem[{{Miki{\'c}} {et~al.}(2013){Miki{\'c}}, {Lionello}, {Mok}, {Linker},
  \& {Winebarger}}]{mikic2013}
{Miki{\'c}}, Z., {Lionello}, R., {Mok}, Y., {Linker}, J.~A., \& {Winebarger},
  A.~R. 2013, \apj, 773, 94

\bibitem[{{Mok} {et~al.}(2016){Mok}, {Miki{\'c}}, {Lionello}, {Downs}, \&
  {Linker}}]{mok2016}
{Mok}, Y., {Miki{\'c}}, Z., {Lionello}, R., {Downs}, C., \& {Linker}, J.~A.
  2016, \apj, 817, 15

\bibitem[{{Mok} {et~al.}(2005){Mok}, {Miki{\'c}}, {Lionello}, \&
  {Linker}}]{2005ApJ...621.1098M}
{Mok}, Y., {Miki{\'c}}, Z., {Lionello}, R., \& {Linker}, J.~A. 2005, \apj, 621,
  1098

\bibitem[{{Mok} {et~al.}(2008){Mok}, {Miki{\'c}}, {Lionello}, \&
  {Linker}}]{2008ApJ...679L.161M}
---. 2008, \apjl, 679, L161

\bibitem[{{M{\"u}ller} {et~al.}(2003){M{\"u}ller}, {Hansteen}, \&
  {Peter}}]{2003A&A...411..605M}
{M{\"u}ller}, D.~A.~N., {Hansteen}, V.~H., \& {Peter}, H. 2003, \aap, 411, 605

\bibitem[{{M{\"u}ller} {et~al.}(2004){M{\"u}ller}, {Peter}, \&
  {Hansteen}}]{2004A&A...424..289M}
{M{\"u}ller}, D.~A.~N., {Peter}, H., \& {Hansteen}, V.~H. 2004, \aap, 424, 289

\bibitem[{{Mulu-Moore} {et~al.}(2011){Mulu-Moore}, {Winebarger}, {Warren}, \&
  {Aschwanden}}]{mulu2011}
{Mulu-Moore}, F.~M., {Winebarger}, A.~R., {Warren}, H.~P., \& {Aschwanden},
  M.~J. 2011, \apj, 733, 59

\bibitem[{{Parker}(1972)}]{parker1972}
{Parker}, E.~N. 1972, \apj, 174, 499

\bibitem[{{Parker}(1983)}]{parker1983b}
---. 1983, \apj, 264, 642

\bibitem[{{Peter} {et~al.}(2012){Peter}, {Bingert}, \& {Kamio}}]{peter2012}
{Peter}, H., {Bingert}, S., \& {Kamio}, S. 2012, \aap, 537, A152

\bibitem[{{Porter} \& {Klimchuk}(1995)}]{porter1995}
{Porter}, L.~J., \& {Klimchuk}, J.~A. 1995, \apj, 454, 499

\bibitem[{{Reale}(2014)}]{reale2014}
{Reale}, F. 2014, Living Reviews in Solar Physics, 11, 4

\bibitem[{{Reale} {et~al.}(2000){Reale}, {Peres}, {Serio}, {Betta}, {DeLuca},
  \& {Golub}}]{reale2000b}
{Reale}, F., {Peres}, G., {Serio}, S., {Betta}, R.~M., {DeLuca}, E.~E., \&
  {Golub}, L. 2000, \apj, 535, 423

\bibitem[{{Rosner} {et~al.}(1978){Rosner}, {Tucker}, \& {Vaiana}}]{rosner1978}
{Rosner}, R., {Tucker}, W.~H., \& {Vaiana}, G.~S. 1978, \apj, 220, 643

\bibitem[{{Schrijver}(2001)}]{schrijver2001}
{Schrijver}, C.~J. 2001, \solphys, 198, 325

\bibitem[{{Ugarte-Urra} {et~al.}(2006){Ugarte-Urra}, {Winebarger}, \&
  {Warren}}]{ugarte2006}
{Ugarte-Urra}, I., {Winebarger}, A.~R., \& {Warren}, H.~P. 2006, \apj, 643,
  1245

\bibitem[{{Viall} \& {Klimchuk}(2012)}]{viall2012}
{Viall}, N.~M., \& {Klimchuk}, J.~A. 2012, \apj, 753, 35

\bibitem[{{Warren} {et~al.}(2010){Warren}, {Kim}, {DeGiorgi}, \&
  {Ugarte-Urra}}]{warren2010}
{Warren}, H.~P., {Kim}, D.~M., {DeGiorgi}, A.~M., \& {Ugarte-Urra}, I. 2010,
  \apj, 713, 1095

\bibitem[{{Warren} {et~al.}(2002){Warren}, {Winebarger}, \&
  {Hamilton}}]{warren2002b}
{Warren}, H.~P., {Winebarger}, A.~R., \& {Hamilton}, P.~S. 2002, \apjl, 579,
  L41

\bibitem[{{Warren} {et~al.}(2003){Warren}, {Winebarger}, \&
  {Mariska}}]{warren2003}
{Warren}, H.~P., {Winebarger}, A.~R., \& {Mariska}, J.~T. 2003, \apj, 593, 1174

\bibitem[{{Winebarger} \& {Warren}(2005)}]{winebarger2005}
{Winebarger}, A.~R., \& {Warren}, H.~P. 2005, \apj, 626, 543

\bibitem[{{Winebarger} {et~al.}(2003{\natexlab{a}}){Winebarger}, {Warren}, \&
  {Mariska}}]{winebarger2003}
{Winebarger}, A.~R., {Warren}, H.~P., \& {Mariska}, J.~T. 2003{\natexlab{a}},
  \apj, 587, 439

\bibitem[{{Winebarger} {et~al.}(2003{\natexlab{b}}){Winebarger}, {Warren}, \&
  {Seaton}}]{winebarger2003a}
{Winebarger}, A.~R., {Warren}, H.~P., \& {Seaton}, D.~B. 2003{\natexlab{b}},
  \apj, 593, 1164

\end{thebibliography}
\end{document}